\documentstyle[psfig,epsfig,lscape]{mn}
\newcommand{\mnras}{MNRAS}

\newcommand{\aap}{AAP}
\newcommand{\apj}{ApJ}
\newcommand{\be}{\begin{equation}}
\newcommand{\ee}{\end{equation}}
\newcommand{\ba}{\begin{eqnarray}}
\newcommand{\ea}{\end{eqnarray}}

\newcommand{\nn}{\nonumber \\}

\newcommand{\n}{\mbox{\boldmath $n$}}

\newcommand{\bell}{\mbox{\boldmath $\ell$}}
\newcommand{\bL}{\mbox{\boldmath $L$}}

\def\gs{\mathrel{\raise1.16pt\hbox{$>$}\kern-7.0pt %
\lower3.06pt\hbox{{$\scriptstyle \sim$}}}}         %
\def\ls{\mathrel{\raise1.16pt\hbox{$<$}\kern-7.0pt %
\lower3.06pt\hbox{{$\scriptstyle \sim$}}}}         %

\title[GREAT10 Galaxy Challenge]{Image Analysis for Cosmology:\\ Results from the
  GREAT10 Galaxy Challenge}

\author[T. D. Kitching et al.]
       {T. D. Kitching$^1$\thanks{tdk@roe.ac.uk}, 
         S. T. Balan$^2$, S. Bridle$^3$, N. Cantale$^4$,
         F. Courbin$^4$, T. Eifler$^5$,
         \newauthor
         M. Gentile$^4$, M. S. S. Gill$^{5,6,7}$, S. Harmeling$^8$, C. Heymans$^1$,
         M. Hirsch$^{3,8}$, 
         \newauthor
         K. Honscheid$^5$, T. Kacprzak$^3$, D. Kirkby$^9$, D. Margala$^9$, R. J. Massey$^{10}$, 
         \newauthor
         P. Melchior$^5$, G. Nurbaeva$^4$, K. Patton$^5$, J. Rhodes$^{11,12}$, B. T. P. Rowe$^{3,11,12}$,  
         \newauthor
         A. N. Taylor$^1$, M. Tewes$^4$, M. Viola$^1$,
         D. Witherick$^3$, L. Voigt$^3$, J. Young$^5$, 
         \newauthor 
         J. Zuntz$^{3,13,14}$
         \\
$^1$SUPA, Institute for Astronomy, University of
Edinburgh, Royal Observatory, Blackford Hill, Edinburgh, EH9 3HJ,
U.K.\\
$^2$Astrophysics Group, Cavendish Laboratory, JJ Thomson Avenue, Cambridge,  CB3 0HE, U.K.\\
$^3$Department of Physics and Astronomy, University College London, Gower Street, London, WC1E 6BT, U.K.\\
$^4$Laboratoire d'Astrophysique, Ecole Polytechnique Federale de
Lausanne (EPFL), Switzerland\\
$^5$Center for Cosmology and AstroParticle Physics Physics Dept, The
Ohio State University, USA\\
$^6$Kavli Institute for Particle Astrophysics \& Cosmology, Stanford, USA\\
$^7$Centro Brasileiro de Pesquisas FÃ­sicas,Rio de Janeiro, RJ, Brazil\\
$^8$Department of Empirical Inference, Max Planck Institute for
Intelligent Systems, T\"ubingen, Germany\\
$^9$Department of Physics and Astronomy, UC Irvine, 4129 Frederick
Reines Hall, Irvine, CA 92697-4575, USA\\
$^{10}$Institute for Computational Cosmology, Durham University, South
Road, Durham, DH1 3LE, U.K.\\
$^{11}$Jet Propulsion Laboratory, California Institute of Technology, 4800 Oak Grove Drive, Pasadena, CA 91109, USA\\
$^{12}$California Institute of Technology, 1200 East California
Boulevard, Pasadena, CA 91106, USA\\
$^{13}$Astrophysics Group, University of Oxford, Denys Wilkinson
Building, Keble Road, Oxford OX1 3RH, U.K\\
$^{14}$Oxford Martin School, University of Oxford, Old Indian
Institute, 34 Broad Street, Oxford OX1 3BD, U.K
}
\pagerange{\pageref{firstpage}--\pageref{lastpage}}
\pubyear{2012}
\begin{document}
\maketitle
\begin{abstract}
In this paper we present results from the weak lensing shape
measurement GRavitational lEnsing
Accuracy Testing 2010 (GREAT10) Galaxy Challenge. This marks an order
of magnitude step change in the level of scrutiny employed in weak lensing
shape measurement analysis. We provide descriptions of each
method tested and include 10 evaluation metrics over 24
simulation branches. 

GREAT10 was the first
shape measurement challenge to include variable fields; both
the shear field and the Point Spread Function (PSF) vary across the images
in a realistic manner. The variable fields enable a variety of metrics that are
inaccessible to constant shear simulations including a direct measure of
the impact of shape measurement inaccuracies, and the impact of PSF size and ellipticity, 
on the shear power spectrum.
To assess the impact of shape measurement bias for
cosmic shear we present a general pseudo-Cl formalism, that propagates 
spatially varying systematics in cosmic shear through to power
spectrum estimates. We also show how one-point estimators of 
bias can be extracted from variable shear simulations. 

The GREAT10 Galaxy Challenge received $95$ 
submissions and saw a factor of $3$ improvement in the 
accuracy achieved by shape measurement methods. 
The best methods achieve sub-percent average biases. 
We find a strong dependence on 
accuracy as a function of signal-to-noise, and indications of a weak dependence on
galaxy type and size. Some requirements for the most ambitious 
cosmic shear experiments are met above a signal-to-noise ratio of
$20$. 
These results have the caveat that the simulated PSF was a ground-based PSF. 
Our results are a snapshot of the accuracy of current
shape measurement methods and are a benchmark upon which 
improvement can continue. This provides a foundation for a better
understanding of the strengths and limitations of shape measurement
methods. 
\end{abstract}
\begin{keywords}
Cosmology: observations, gravitational lensing: weak,
methods: statistical, techniques: image processing
\\
\noindent{\footnotesize $^*$tdk@roe.ac.uk}
\end{keywords}

\section{Introduction}
\label{Introduction}
In this paper we present the results from the GRavitational lEnsing
Accuracy Testing 2010 (GREAT10) Galaxy Challenge. GREAT10 was an image
analysis challenge for cosmology that focused on the task of measuring the weak
lensing signal from galaxies. 
Weak lensing is the effect whereby the image of a source galaxy is
distorted by intervening massive structure along the line-of-sight.
In the weak field limit this distortion is a change in the observed
ellipticity of the object, and this change in ellipticity is called
shear. Weak lensing is particularly important for
understanding the nature of dark energy and dark matter, because 
it can be used to
measure the cosmic growth of structure and the expansion history of
the Universe (see reviews by e.g. Albrecht et al., 2001; Massey, Kitching,
Richards, 2010; Hoekstra \& Jain, 2008; Bartelmann \& Schneider, 2001; Weinberg et al., 2012). 
In general, by measuring the ellipticities of distant galaxies --
hereafter denoted ``shape measurement'' -- we can make statistical
statements about the nature of the intervening matter.  
The full process through which photons propagate from galaxies to detectors is
described in a previous companion paper, the GREAT10 Handbook (Kitching et al.,
2011).

There are a number of features, in the physical processes and optical
systems, through which the photons we ultimately use for weak lensing
pass. These features must be accounted for when designing shape
measurement algorithms. These are primarily the
convolution effects of the atmosphere and the telescope optics,
pixelisation effects of the detectors used and the presence of noise
in the images.                   
The simulations in GREAT10 aimed to address each of these complicating factors. 
GREAT10 consisted of two concurrent challenges as described in
Kitching et al. (2011): the Galaxy Challenge, where entrants were
provided with $50$ million simulated galaxies and asked to measure their
shapes and spatial variation of the shear field with a known Point
Spread Function (PSF) and the Star Challenge wherein entrants were
provided with an unknown PSF, sampled by stars, and asked to
reconstruct the spatial variation of the PSF across the field.

In this paper we present the results of the GREAT10 Galaxy Challenge. 
The challenge provided a controlled simulation development environment
in which shape measurement methods could be tested, and was run as a
blind competition for 9 months from December 2010 to September 2011. 
Blind analysis of shape measurement algorithms began with the Shear
TEsting Programme (STEP; Heymans et al., 2006; Massey et al., 2007) and
GREAT08 (Bridle et al., 2009, 2010). The blindness of these competitions is critical
in testing methods under circumstances that will
be similar to those encountered in real astronomical data. This
is because for weak lensing, unlike photometric redshifts for example,
we cannot observe a training set from which we know the shear
distribution (we can however observe a subset of galaxies at 
high signal-to-noise to train upon, which is something we address in this paper). 

The GREAT10 Galaxy Challenge is the first
shape measurement analysis that includes \emph{variable fields}. Both
the shear field and the PSF vary across the images
in a realistic manner. This enables a variety of metrics that are
inaccessible to constant shear simulations (where the fields are a
single constant value across the images), including a direct measure of
the impact of shape measurement inaccuracies on the inferred shear power
spectrum and a measure of the correlations between shape measurement
inaccuracies and the size and ellipticity of the PSF. 

We present a general pseudo-Cl formalism for
a flat-sky shear field in Appendix A, which we use to show how to propagate
general spatially varying shear measurement biases through to the
shear power spectrum. This has a more general application in cosmic
shear studies. 
 
This paper summarises the results of the GREAT10 Galaxy Challenge. 
We refer the reader to a companion paper 
that discusses the GREAT10 Star challenge (Kitching et al., in prep).
Here we summarise the results that we show, distilled from the
wealth of information that we present in this paper: 
\begin{enumerate}
\item 
Signal-to-noise: We find a strong dependence of the metrics below
S/N$=10$. However we find methods that meet bias requirements for the most ambitious
experiments when S/N$ > 20$. We note that methods tested here have
been optimised for use on ground based data in this regime. 
\item
Galaxy type: We find marginal evidence 
that model fitting methods have a relatively low
dependence on galaxy type compared to model-independent methods.
\item 
PSF dependence: We find 
contributions to biases from PSF size, but less so from PSF
ellipticity. 
\item 
Galaxy Size: For large galaxies well sampled by the PSF, with scale
radii $\gs 2$ times the mean PSF size we find that methods meet
requirements on bias parameters for the most ambitious
experiments. However if galaxies are unresolved, with radii $\ls 1$
times the mean PSF size, biases become significant.
\item 
Training: We find that calibration on a high signal-to-noise sample can
significantly improve a method's average biases. 
\item  
Averaging Methods: We find that averaging ellipticities over several
methods is clearly beneficial, but
that the weight assigned to each method will need to be correctly
determined. 
\end{enumerate} 

In Section \ref{Description of the Competition} we describe the 
Galaxy Challenge structure, in Section \ref{Description of the
  Simulations} we describe the simulations. Results are summarised in Section \ref{Results}
and we present conclusions in Sections \ref{astrocrowdsourcing} 
and \ref{Conclusions}. We make
extensive use of Appendices that contain technical information on the
metrics and a more detailed breakdown of individual shape measurement
method's performance. 

\section{Description of the Competition}
\label{Description of the Competition}
The GREAT10 Galaxy Challenge was run as an open competition for 9
months between $3^{\rm rd}$ December 2010 and $2^{\rm nd}$ September
2011\footnote{Between $2^{\rm nd}$ September 2011 and $8^{\rm th}$ September
2011 we extended the challenge to allow submissions from those
participants who had not met the deadline; those submissions will be
labelled in Section \ref{Results}.}. The challenge was open for
participation from anyone, the website\footnote{{\tt
  http://www.greatchallenges.info}} served as the portal for
participants, and data could be freely downloaded.

The challenge was to reconstruct the shear power spectrum from
subsampled images of sheared galaxies (Kitching et
al. 2011). All shape measurement methods 
to date do this by measuring the ellipticity from each galaxy in
an image, although scope for alternative approaches
was allowed. Participants in the challenge were asked to submit either 
\begin{enumerate}
\item 
\emph{Ellipticity catalogues} that contained an estimate of the
ellipticity for each object in each image, or
\item 
\emph{Shear power spectra}, that consisted of an estimate of the
shear power spectrum for each simulation set.
\end{enumerate}
For ellipticity catalogue submissions all objects were required
to have an ellipticity estimate, and no galaxies were removed or
down-weighted in the power spectrum calculation; if such weighting
functions were desired by a participant then a shear power spectra
submission was encouraged.

Participants were required to access $1$ TB of imaging data in
the form of {\tt FITS} images. Each image contained $10$,$000$
galaxies arranged on a $100$x$100$ grid. 
Each galaxy was captured in a single postage stamp of
48x48 pixels (to incorporate the largest galaxies in the simulation
with no truncation), and the grid was arranged so that each neighbouring postage
stamp was positioned contiguously i.e. there were no 
gaps between postage stamps and no overlaps. Therefore each image was 
$4800$x$4800$ pixels in size. The simulations were divided into $24$ sets (see Section
\ref{Simulation Structure}) and each set contained $200$ images. 
For each galaxy in each image participants were provided with a
functional description of the PSF (described in Section \ref{The
  PSF Models}) and an image showing a pixelised realisation of the
PSF. In addition a suite of development code was provided to help read in
the data and perform a simple analysis\footnote{{\tt http://great.roe.ac.uk/data/code/}}.

\subsection{Summary of metrics}
The metric with which the live leaderboard was scored during the
challenge was a Quality factor $Q$, defined as 
\be
\label{Qe}
Q\equiv 1000\frac{5\times 10^{-6}}{\int \!{\rm d}\!\ln \ell |\widetilde C^{EE}_{\ell}-C^{EE,\gamma\gamma}_{\ell}|\ell^2}, 
\ee
averaged over all sets, a quantity that relates the reconstructed shear power
spectrum $\widetilde C^{EE}_{\ell}$ with the true shear power spectrum
$C^{EE,\gamma\gamma}_{\ell}$. We describe this metric in more detail in
Appendices A and B. This is a general integral expression for the
Quality factor, in the simulations we use discrete bins in $\ell$
that are defined in Appendix C. By evaluating this metric for each submission,  
results were posted to a live leaderboard that ranked
methods based on the value of $Q$. We will also
investigate a variety of alternative metrics extending the STEP $m$
and $c$ bias formalism to variable fields. 

The measured ellipticity
of an object at position $\btheta$ can be related to the true ellipticity and shear, 
\ba
&&e_{\rm measure}(\btheta)=\gamma(\btheta)+e_{\rm intrinsic}(\btheta)\nn
&+&c(\btheta)+m(\btheta)[\gamma(\btheta)+e_{\rm intrinsic}(\btheta)]+ \nn
&+&q(\theta)[\gamma(\btheta)+e_{\rm intrinsic}(\btheta)]|\gamma(\btheta)+e_{\rm intrinsic}(\btheta) |\nn
&+&e_{\rm n}(\btheta),
\ea
with a multiplicative bias $m(\btheta)$, an offset
$c(\btheta)$, and a quadratic term $q(\btheta)$ 
(this is $\gamma|\gamma|$, not $\gamma^2$, since we may expect divergent behaviour to more
positive and more negative shear values for each domain
respectively), that in general are functions of position due to PSF
and galaxy properties. $e_{\rm n}(\btheta)$ is a potential stochastic
noise contribution. 
For spatially variable shear fields, biases between measured and true
shear can vary as a function of position, mixing angular modes and
power between E and B-modes. In Appendix A, we present a general
formalism that allows for the propagation of biases into shear power
spectra using a pseudo-Cl methodology; this approach has applications
beyond the treatment of shear systematics. 
The full set of metrics are described in detail in Appendix B and are 
summarised in Table \ref{metrics}.
\begin{table*}
\begin{center}
\begin{tabular}{|l|c|c|}
\hline
{\bf Metric}&{\bf Definition}&{\bf Features}\\
\hline
$m$, $c$,
$q$&$\hat\gamma=(1+m)\gamma^t+c+q\gamma^t|\gamma^t|$&One-point estimators
of bias. Links to STEP\\
\hline
$Q$&$1000\frac{5\times 10^{-6}}{\int \!{\rm d}\!\ln \ell |\widetilde
  C^{EE}_{\ell}-C^{EE,\gamma\gamma}_{\ell}|\ell^2}$&Numerator relates to
bias on $w_0$\\
\hline
$Q_{\rm dn}$&$1000\frac{5\times 10^{-6}}{\int \!{\rm d}\!\ln \ell |C^{EE}_{\ell}-C^{EE,\gamma\gamma}_{\ell}-\frac{\langle\sigma^2_{\rm n}\rangle}{N_{\rm realisation}N_{\rm
    object}}|\ell^2}$&Corrects Q for pixel noise\\
\hline 
${\mathcal M}\simeq m^2+2m$, ${\mathcal A}\propto \sigma(c)^2$&$\widetilde
C^{EE}_{\ell}=C^{EE,\gamma\gamma}_{\ell}+{\mathcal A}+
{\mathcal M}C^{EE,\gamma\gamma}_{\ell}$&Power spectrum relations.\\ 
\hline
$\alpha_X$&$m(\btheta)=m_0+\alpha [X(\btheta)/X_0]$&Variation of $m$ with PSF ellipticity/size\\
$\beta_X$&$c(\btheta)=c_0+\beta [X(\btheta)/X_0]$&Variation $c$ with PSF ellipticity/size\\
\hline
\end{tabular}
\caption{A summary of the metrics used to evaluate shape measurement
  methods for GREAT10. These are defined in detail in Appendices A and
  B. We refer to $m$ and $c$ as the one-point estimators
  of bias, and make the distinction between these and spatially constant terms
  ($m_0$, $c_0$) and correlations ($\alpha$, $\beta$) only where clearly stated. } 
\label{metrics}
\end{center}
\end{table*}

The metric with which the live leaderboard was scored was the $Q$
value, and the same metric was used for ellipticity catalogue
submissions and power spectrum submissions. 
However in this paper we will introduce and focus on $Q_{\rm dn}$ (see Table
\ref{metrics}) that for ellipticity
catalogue submissions removes any residual pixel-noise error
(nominally associated with biases caused by finite signal-to-noise or 
inherent shape measurement method noise). For details see Appendix B. Note that 
this is not a correction for ellipticity (shape) noise which is removed in GREAT10
through the implementation of a B-mode only intrinsic ellipticity
field. 

The metric $Q$ takes into account scatter between the estimated shear
and the true shear due to stochasticity in a method or spatially
varying quantities, such that a small $m(\btheta)$ and $c(\btheta)$ do not necessarily
correspond to a large $Q$ value (see Appendix B). This is discussed within the context
of previous challenges in Kitching et al. (2008). 
Spatial variation is important because the shear and PSF fields vary, 
so that there may be scale-dependent correlations between them, and
stochasticity is important because we wish
methods to be accurate (such that errors do not dilute cosmological or
astrophysical constraints) as well as being unbiased. 

For variable fields we can complement the linear biases, $m(\btheta)$ and $c(\btheta)$, with
a component that can be correlated with any spatially varying quantity
$X(\btheta)$, for example PSF ellipticity or size; 
\be 
m(\btheta)=m_0+\alpha \left[\frac{X(\btheta)}{X_0}\right], \,\,\,\,\,\,\,\,\,\,\, c(\btheta)=c_0+\beta \left[\frac{X(\btheta)}{X_0}\right],
\ee
with spatially constant terms $m_0$ and $c_0$ and 
correlation coefficients $\alpha$ and $\beta$; $X_0$
  is a constant reference value that ensures that the units of $\alpha$ and $\beta$
are dimensionless: for ellipticity this is set to unity $X_0=1$, for PSF size squared
this is the mean PSF size squared $X_0=\langle r_{\rm PSF}^2\rangle$.
Only ellipticity catalogue submissions can have $m_0$, $c_0$, $\alpha$ and $\beta$ values
calculated because these parameters require individual galaxy ellipticity estimates (in order
to calculate the required mixing matrices, see Appendices A and B). 
Throughout we will refer to $m$ and $c$ as the one-point estimators
of bias and make the distinction between spatially constant terms
$m_0$ and $c_0$ and
correlations $\alpha$ and $\beta$ only where clearly stated. 
Finally we also include a non-linear shear response (see Table
\ref{metrics}), we do not include a discussion of this in the main results, 
because $q\gamma|\gamma|\approx 0$ for most methods, but show the results in Appendix E. 

To measure biases at the power spectrum level we define constant
linear bias parameters (see Appendix A equation \ref{genCl3})
\be
\widetilde
C^{EE}_{\ell}=C^{EE,\gamma\gamma}_{\ell}+{\mathcal A}+
{\mathcal M}C^{EE,\gamma\gamma}_{\ell},
\ee
that relate the measured power spectrum to the true power
spectrum. These are approximately related to one-point shear bias 
$m$, and the variance of $c$, by ${\mathcal M}/2\simeq m$ for values of $m\ll 1$ and
$\sqrt{\mathcal A}\simeq \sigma(c)$. These parameters can
be calculated for both ellipticity and power spectrum submissions.

\section{Description of the Simulations}
\label{Description of the Simulations}

In this Section we describe the overall structure of the simulations. 
For details on the local modelling of the
galaxy and star profiles and the spatial variation of the PSF and
shear fields we refer to Appendix C. 

\subsection{Simulation structure}
\label{Simulation Structure}
The structure of the simulations was engineered such that, in the final 
analysis, the various aspects of performance for a given shape measurement 
method could be gauged. The competition was split into sets of
images, where one set was a `fiducial' set and the remaining 
sets represented perturbations about the parameters in that set. 
Each set consisted of $200$ images. This number was justified by calculating
the expected pixel-noise effect on shape measurement methods (see
Appendix B) such that when averaging over all $200$ images this effect
should be suppressed (however, see also Section \ref{Results} where we
investigate this noise term further). 

Participants were provided with a functional description and a
pixelated realisation of the PSF at
each galaxy position. The task of estimating the PSF itself was set a
separate `Star Challenge' that is described in a companion paper 
(Kitching et al. in prep). 

The variable shear field was
constant in each of the images within a set, but the PSF field and
intrinsic ellipticity could vary such that there were three kinds of
set 
\begin{itemize}
\item 
{\bf Type 1}, `Single Epoch', fixed $C^{EE}_{\ell}$, variable PSF, variable intrinsic ellipticity. 
\item 
{\bf Type 2}, `Multi-Epoch', Fixed $C^{EE}_{\ell}$, variable PSF, fixed intrinsic ellipticity. 
\item 
{\bf Type 3}, `Stable Single Epoch', Fixed $C^{EE}_{\ell}$, fixed PSF, variable intrinsic ellipticity. 
\end{itemize}
The default, fiducial, type being one in which both PSF and intrinsic
ellipticity vary between images in a set.
This was designed in part to test the ability of any method which took
advantage of stacking procedures, where galaxy images are averaged over some
population, by testing whether stacking worked when either the galaxy
or PSF were fixed across images within a set or not. 
Stacking methods achieved high scores in GREAT08, Bridle et
al. (2010), but in actuality were not submitted for GREAT10.
For each type of set the PSF and intrinsic
ellipticity fields are \emph{always spatially varying} but this variation did not 
change \emph{within a set}; when we refer to a quantity being
`fixed' this means that its spatial variation does not vary between images
within a set. 

Type 1 (variable PSF and intrinsic field) sets test the 
ability of a method to reconstruct the shear field in the presence of both a
variable PSF field and variable intrinsic ellipticity between images. This nominally
represents a sequence of observations of different patches of sky but
with the same underlying shear power spectrum. Type 2 sets 
(variable PSF and fixed intrinsic field) represent an 
observing strategy where the PSF is different in each exposure of the
same patch of sky (a typical ground based observation); so called
`multi-epoch' data. Type 3 sets (fixed PSF) 
represent `single-epoch' observations with a highly stable PSF. These were only 
simple approximations to reality because, for example, properties in
the individual exposures for the `multi-epoch' sets were not 
correlated (as they may be in real data), and the signal-to-noise was constant in all images for the 
single and multi-epoch sets. 
Participants were aware of the PSF variation from image to image within a set but 
not of the intrinsic galaxy properties or shear.  Thus the conclusions
drawn from these tests will be conservative with regard to the testing between the different set types, 
relative to real data; where in fact this kind of observation is 
known to the observer \emph{ab initio}. In
subsequent challenges this hidden layer of complexity could be removed.

In Appendix D we list in detail the parameter values that
define each set, and the parameters themselves are described in the
sections below. In Table \ref{shortsets} we summarise each set by
listing its distinguishing feature and parameter value. 
\begin{table*}
\begin{center}
\begin{tabular}{|l|c|c|c|}
\hline
Set Number&Set Name&Fixed PSF/Intrinsic Field&Distinguishing Parameter\\
\hline
$1$ & Fiducial & --  & -- \\
$2$ & Fiducial & PSF & -- \\
$3$ & Fiducial & Int& -- \\
$4$ & Low S/N & -- &  S/N$=10$ \\
$5$ & Low S/N & PSF &  S/N$=10$\\
$6$ & Low S/N & Int & S/N$=10$\\
$7$ & High S/N {\bf Training Data}& -- & S/N$=40$ \\ 
$8$ & High S/N  & PSF& S/N$=40$\\ 
$9$ & High S/N  & Int& S/N$=40$\\ 
$10$ & Smooth S/N & --& S/N distribution Rayleigh\\ 
$11$ & Smooth S/N & PSF &S/N distribution Rayleigh\\ 
$12$ & Smooth S/N & Int &S/N distribution Rayleigh\\ 
$13$ & Small Galaxy & -- &$r_b=1.8$ $r_d=2.6$\\ 
$14$ & Small Galaxy & PSF & $r_b=1.8$ $r_d=2.6$\\
$15$ & Large Galaxy & -- & $r_b=3.4$ $r_d=10.0$\\ 
$16$ & Large Galaxy & PSF& $r_b=3.4$ $r_d=10.0$\\ 
$17$ & Smooth Galaxy & -- & Size distribution Rayleigh\\ 
$18$ & Smooth Galaxy & PSF & Size distribution Rayleigh\\ 
$19$ & Kolmogorov & -- & Kolmogorov PSF\\ 
$20$ & Kolmogorov & PSF & Kolmogorov PSF\\ 
$21$ & Uniform b/d & --& b/d fraction $[0.3,0.95]$\\ 
$22$ & Uniform b/d  & PSF &b/d fraction $[0.3,0.95]$\\ 
$23$ & Offset b/d & --& b-d offset variance $0.5$\\
$24$ & Offset b/d  & PSF& b-d offset variance $0.5$\\
\hline
\end{tabular}
\caption{A summary of the simulations sets with the parameter or
  function that distinguishes each set from the fiducial one. In the
  third column we list whether either the PSF or intrinsic ellipticity
  field (Int) were kept fixed between images within a set. $r_b$
  and $r_d$ are the scale radii of the bulge and disk components of
  the galaxy models in pixels, $b/d$ is the ratio between the
  integrated flux in the bulge to disk components of the galaxy
  models. See Appendix C and D for more details.}
\label{shortsets}
\end{center}
\end{table*}
There were two additional sets that used a pseudo-Airy PSF which we do
not include in this paper because of technical reasons (see Appendix F).

Training data was provided in the form of a set with exactly the same
size and form as the other sets. In fact the training set was a copy
of Set $7$, a set which
contained high signal-to-noise galaxies. In this way the structure was
set up to enable an assessment of whether training on high signal to
noise data is useful when extrapolating to other domains, in
particular low galaxy signal-to-noise regime. This is similar to
being able to observe a region of sky with deeper exposures than a
main survey.

\subsection{Variable shear and intrinsic ellipticity fields}
\label{Variable shear and intrinsic ellipticity fields}
In the GREAT10 simulations the key and unique aspect was that the shear
field was a variable quantity and not a static scalar value (as for
all previous shape measurement simulations; STEP1, STEP2, GREAT08). 
To make a variable shear field we
generated a spin-2 Gaussian random field from a $\Lambda$CDM weak
lensing power spectrum (Hu, 1999)
\be 
C^{\gamma\gamma}_{\ell}=\int_0^{r_H}\! {\rm d}r
    \, W^{\rm GG}_{ii}(r)
    P_{\delta\delta}\!\left(\frac{\ell}{r};r \right),
 \ee
where $P_{\delta\delta}$ is the matter power spectrum, and the lensing weight can be expressed as
 \be
 W^{\rm GG}_{ii}(r)=\frac{q_i(r)q_i(r)}{r^2},
\ee
where the kernel is 
\be
q_i(r) = \frac{3 H^2_0 \Omega_m r}{2 a(r)}
\int_r^{r_H}\!  dr' \, p_i(r')
\frac{(r'-r)}{r'}.
\ee
We have assumed a flat Euclidean geometry throughout and $r_H$
is the horizon size. $p_i(r)$ refers to the redshift distribution of the
lensed sources in redshift bin $i$; this expression can be generalised to an
arbitrary number (even a continuous set) of redshift bins (see Kitching, Heavens \& Miller,
2011). For these simulations we have a single redshift bin with a
median redshift of $z_m=1.0$ and a delta-function probability
distribution $p_i(r')=\delta^D(r-r_i)$. We assume
an Eisenstein \& Hu (1999) linear matter power spectrum with a
Smith et al. (2003) non-linear correction. The cosmological parameter
values used were $\Omega_m = 0.25$, $h=H_0/100=0.75$, $n_s= 0.95$ and $\sigma_8=0.78$.
In order to add a random component to the shear power spectrum, so 
that participants could not guess the functional form, we added a 
series of Legendre polynomials $P_n(x)$ up to $5^{\rm th}$ order, such that 
\be 
C^{EE,\gamma\gamma}_{\ell}\rightarrow C^{EE,\gamma\gamma}_{\ell}+2\times 10^{-9}\sum_{n=1}^5 c_n P_n(x_L)
\ee
where the variable $x_L=-1+2(\ell-1)/(\ell_{\rm max}-1)$ is contained within the 
range $[-1$,$1]$ as $\ell$ varies from $\ell_{\rm min}$ to $\ell_{\rm max}$. 
The shear field generated has an E-mode power
spectrum only. The size of the shear field was $\theta_{\rm
  image}=2\pi/\ell_{\rm min}$ and to generate the shear field we set $\theta_{\rm
  image}=10$ degrees, such that the range in $\ell$ we used to
generate the power was
$\ell=[36$, $3600]$ from the fundamental mode to the grid separation
cut-off; the exact $\ell$-modes used are shown in Appendix
C. Note that the Legendre polynomials add fluctuations to the
power spectra, this is benign in the calculation of the
evaluation metrics but would not be expected from real data.

The shear field is generated on a grid of $100$x$100$,
which is then converted into an image of galaxy objects via an image
generation code\footnote{To generate the image simulations we used a Monte
Carlo code that simulates the galaxy model and PSF stages at a photon
level; this code is a modified version of that used for the GREAT08
simulations (Bridle et al., 2010). The modified code is
  available here {\tt http://great.roe.ac.uk/data/code/image\_code},
  the original code is by Konrad Kuijken, modified by SB 
  and SBr for GREAT08, and modified by TDK for GREAT10.} with galaxy
properties described in Appendix C. When postage stamps of objects are generated they 
point-sample the shear field at each position, and a postage stamp is
generated. The postage stamps are then combined to form an
image. 

Throughout, the intrinsic ellipticity field had a variation that 
contained B-mode power only (in every image and when
also averaged over all images in a set) , 
as described in the GREAT10 Handbook. This meant that the contribution from 
intrinsic ellipticity correlations, as well from intrinsic shape
noise, to the lensing shear power spectra was zero. 

\section{Results}
\label{Results}

In total the challenge received $95$ 
submissions from $9$ separate teams and $12$ different methods in total, these were 
\begin{itemize}
\item 
82 submissions before the deadline, 
\item 
13 submissions in the post challenge period,
\end{itemize}
split into 
\begin{itemize}
\item 
85 ellipticity catalogue submissions,
\item
10 power spectra submissions.
\end{itemize} 
We summarise the methods that analysed the GREAT10 Galaxy Challenge in
detail in Appendix E. The method that won the challenge, with the
highest $Q$ value at the end of the challenge period, was 'fit2unfold'
submitted by the \emph{DeepZot} team, authors D. Kirkby and D. Margala.

During the challenge a number of aspects of the 
simulations were corrected (we list these in Appendix
F). 
Several methods generated low scores due to misunderstanding of
simulation details, and in this paper we summarise only those
results for which these errata did not occur. 
In the following we choose the best performing entry for each of the
$12$ shape measurement method entries. 
\begin{table*}
\begin{center}
\begin{tabular}{|l|c|c|c|r|r|r|r|}
\hline
Method & $Q$ & $Q_{\rm dn}$& $Q_{\rm dn\,\, \&
  \,\,trained}$&$m\,\,\,\,\,\,\,\,\,\,$&$c/10^{-4}$&${\mathcal M}/2\,\,\,\,\,\,$&$\sqrt{\mathcal A}/10^{-4}$\\
\hline
$^{\dagger}$ARES 50/50         & $ 105.80 $ & $ 163.44 $ & $ 277.01 $  &$-0.026483$&$ 0.35\,\,\,$& $ -0.018566 $ & $ 0.0728$\\ 
$^{\dagger}$cat7unfold2 (ps)   & $ 152.55 $ &  & $ 150.37 $ & & & $ 0.021409 $ & $ 0.0707$\\ 
 DEIMOS C6                   & $ 56.69 $ & $ 103.87 $ & $ 203.47 $ &$0.006554$&$ 0.08\,\,\,$& $ 0.004320 $ & $ 0.6329$\\ 
 fit2-unfold (ps)            & $ 229.99 $ &  & $ 240.11 $ & & & $ 0.040767 $ & $ 0.0656$\\ 
 gfit                        & $ 50.11 $ & $ 122.74 $ & $ 249.88 $ &$0.007611$&$0.29\,\,\,$& $ 0.005829 $ & $ 0.0573$\\ 
$^*$im3shape NBC0            & $ 82.33 $ & $ 114.25 $ & $ 167.53 $ &$-0.049982$&$ 0.12\,\,\,$& $ -0.053837 $ & $ 0.0945$\\ 
 KSB                         & $ 97.22 $ & $ 134.42 $ & $ 166.96 $&$-0.059520$&$ 0.86\,\,\,$& $ -0.037636 $ & $ 0.0872$\\ 
$^*$KSB f90                  &  $ 49.12 $ & $ 102.29 $ & $ 202.83 $ &$-0.008352$&$ 0.19\,\,\,$& $ 0.020803 $ & $ 0.0789$\\ 
$^{\dagger}$MegaLUTsim2.1 b20  & $ 69.17 $ & $ 75.30 $ & $ 52.62 $ &$-0.265354$&$ -0.55\,\,\,$& $ -0.183078 $ & $ 0.1311$\\ 
 method 4                    & $ 83.52 $ & $ 92.66 $ & $ 116.02 $ &$-0.174896$&$-0.12\,\,\,$& $ -0.090748 $ & $ 0.0969$\\ 
$^{\dagger}$NN23 func          & $ 83.16 $ & $ 60.92 $ & $ 17.19 $ &$-0.239057$&$ 0.47\,\,\,$& $ -0.015292 $ & $ 0.0982$\\ 
 shapefit                    &$ 39.09 $ & $ 63.49 $ & $ 84.68 $  &$0.108292$&$ 0.17\,\,\,$& $ 0.049069 $ & $ 0.8686$\\ 
\hline
\end{tabular}
\caption{The Quality factors, $Q$, with denoising and training, and
  the $m$ and $c$ values for each method (not available for power
  spectrum submissions) that we explore in detail in this
  paper, in alphabetical order of the
  methods name. A ``(ps)'' indicates a power spectrum submission, in
  these cases $Q_{\rm dn\,\, \& \,\,trained}=Q_{\rm trained}$, all others
  were ellipticity catalogue submissions. An $^*$ indicates that this
  team had knowledge of the internal parameters of the simulations, and
  access to the image simulation code. A
  $^{\dagger}$ indicates that this submission was made in the
  post-challenge time period.}
\label{avscores}
\end{center}
\end{table*}

\subsection{One-point estimators of bias: m and c values}
\begin{figure*}
  {\includegraphics[angle=0,clip=]{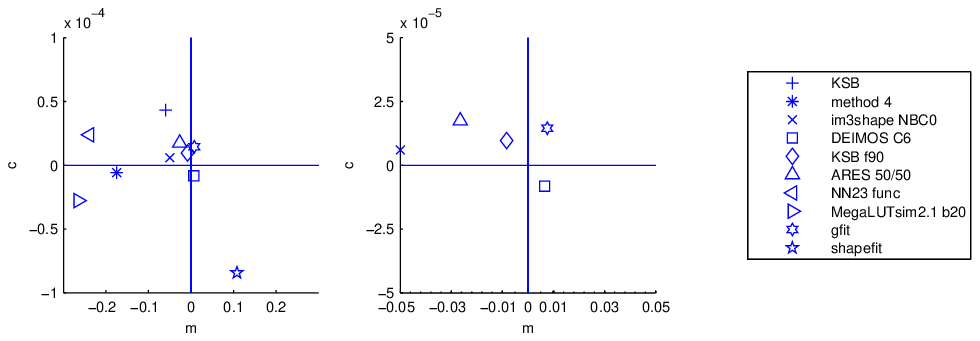}}
 \caption{In the lefthand panel we show the multiplicative $m$ and additive $c$ biases for each 
   ellipticity catalogues method, for which one-point estimators can
   be calculated, see Appendix B. The symbols indicate the method with a legend in the
   righthand panel. The central panel expands the x- and y-axes to show the
   best performing methods.}
 \label{69}
\end{figure*}

In Appendix B we describe how the estimators for shear biases on a
galaxy-by-galaxy basis in the simulations -- what we refer to as
`one-point estimators' of biases -- can be derived, and how these relate
to the STEP $m$ and $c$ parameters (Heymans et al. 2006).
In Figure \ref{69} and in Table \ref{avscores} we show the $m$
and $c$ biases for the best performing entries for each method (those
with the highest quality factors). 
In Appendix E we show how the $m$ and $c$ parameters,
and the difference of the measured and true shear
$\hat\gamma-\gamma^t$, vary for each method as a function
of several quantities: PSF ellipticity, PSF size, galaxy size, galaxy
bulge to disk fraction and galaxy bulge to disk angle offset. 
We show in Appendix E that some methods have a strong $m$ dependence on
PSF ellipticity and size (e.g. TVNN and method04). Model fitting methods (gfit, im3shape) tend
to have fewer model-dependent biases, whereas the KSB-like methods
(DEIMOS, KSB f90) have the smallest average biases. 

\subsection{Variable shear}
\begin{figure*}
  {\includegraphics[width=0.99\columnwidth,angle=0,clip=]{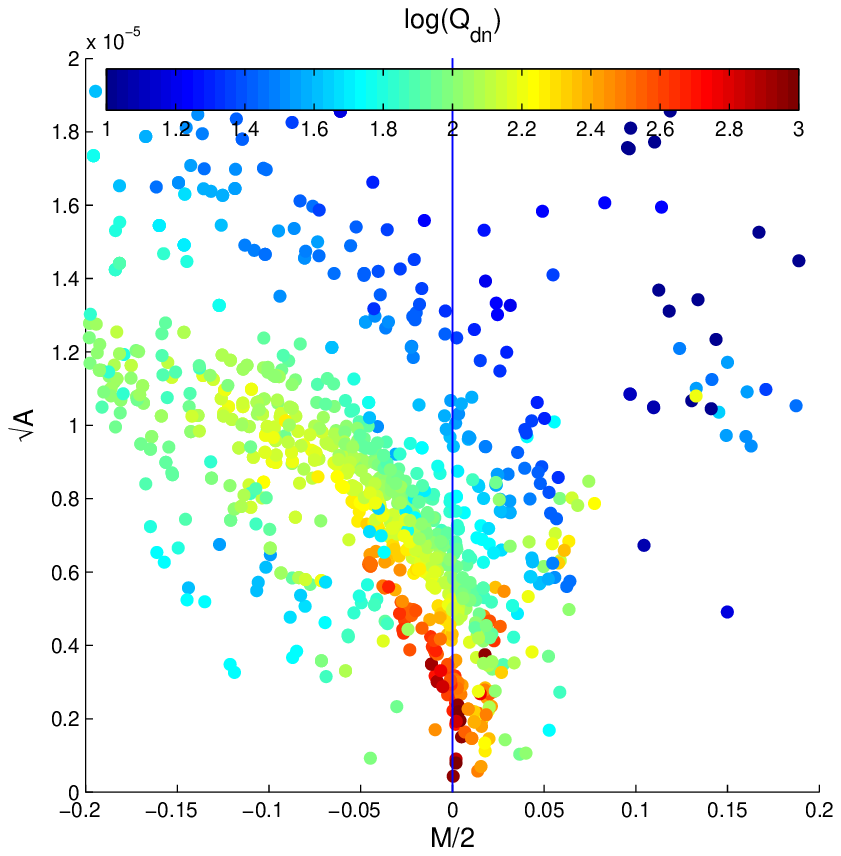}}
  {\includegraphics[width=0.99\columnwidth,angle=0,clip=]{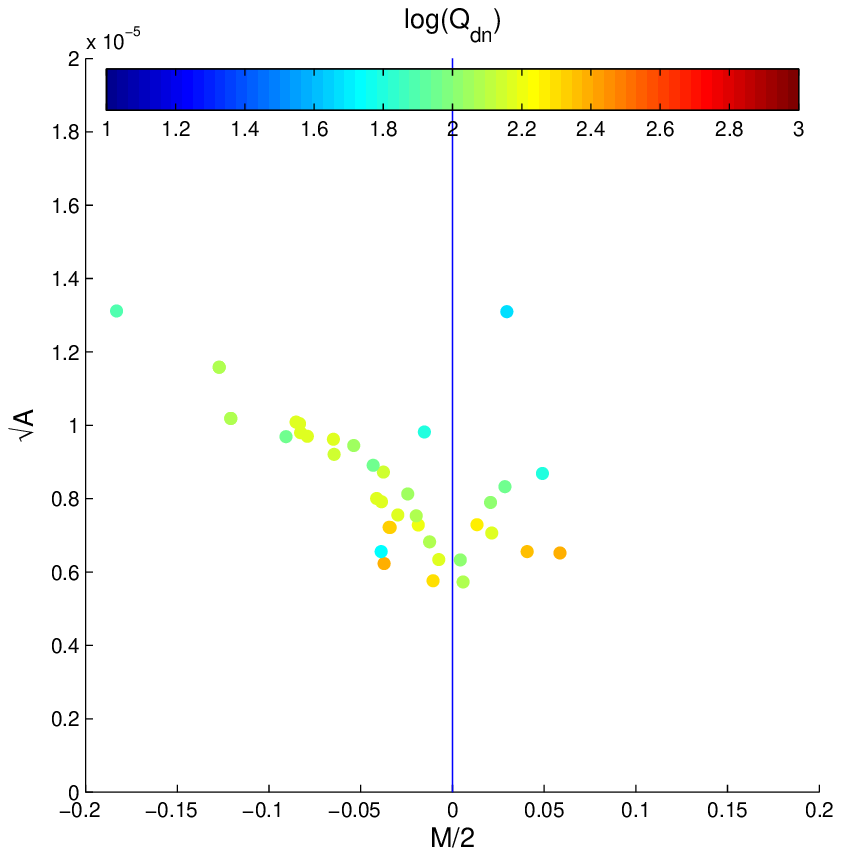}}
 \caption{In the lefthand panel we show  ${\mathcal M}$ and ${\mathcal A}$ for each 
   method for each set. The colour scale represents the logarithm of the quality factor $Q_{\rm dn}$. In the righthand 
   panel we show the metrics ${\mathcal M}$, ${\mathcal A}$ and
   $Q_{\rm dn}$ for each method averaged over all sets. For a
   breakdown of these into dependence on set type see Figure \ref{Leps}.}
 \label{avmc}
\end{figure*}

In the lefthand panel of Figure \ref{avmc} we show the values of the linear
power spectrum parameters ${\mathcal M}$ and ${\mathcal A}$ for each method for each set, and
display by color code the Quality factor $Q_{\rm dn}$. In
Table \ref{avscores} we show the mean values of these parameters
averaged over all sets. 
We find a clear anti-correlation between ${\mathcal M}$ and ${\mathcal
  A}$ and $Q_{\rm dn}$, with higher Quality factors corresponding to
smaller ${\mathcal M}$ and ${\mathcal A}$ values. 
We will explore this further in the subsequent sections. 
We refer the reader to Appendix B where we show how the 
${\mathcal M}$, ${\mathcal A}$ and $Q_{\rm dn}$ parameters are
expected to be related in an ideal case.
In the righthand panel of Figure \ref{avmc} we also show the ${\mathcal M}$,
${\mathcal A}$ and $Q_{\rm dn}$ values for each method averaged over
all sets.

\begin{figure*}
  {\includegraphics[width=\columnwidth,angle=0,clip=]{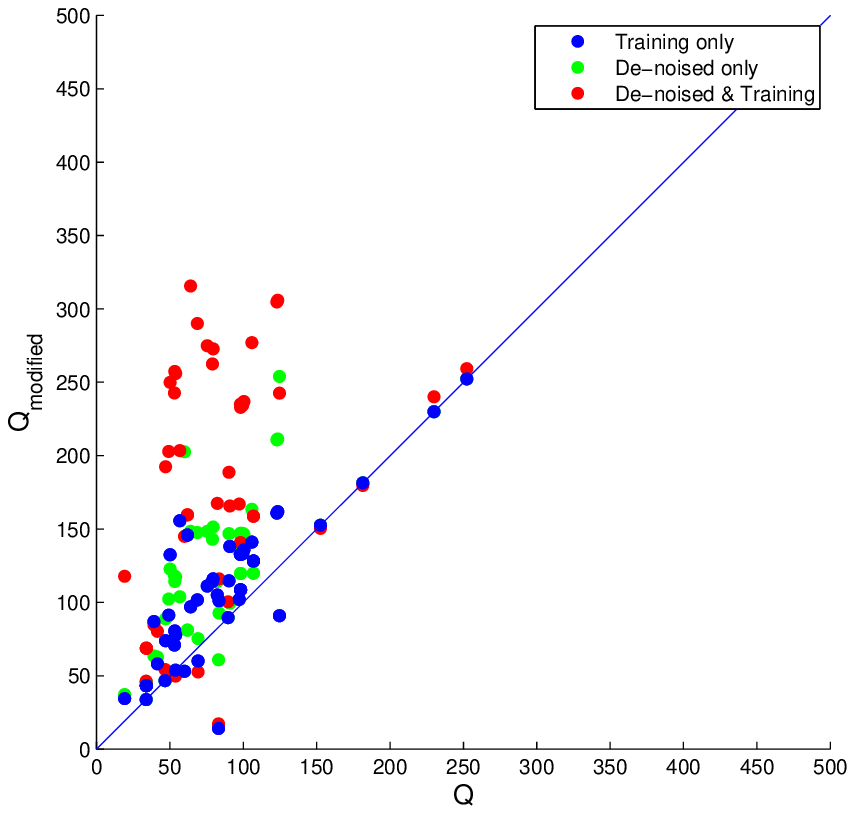}}
  {\includegraphics[width=\columnwidth,angle=0,clip=]{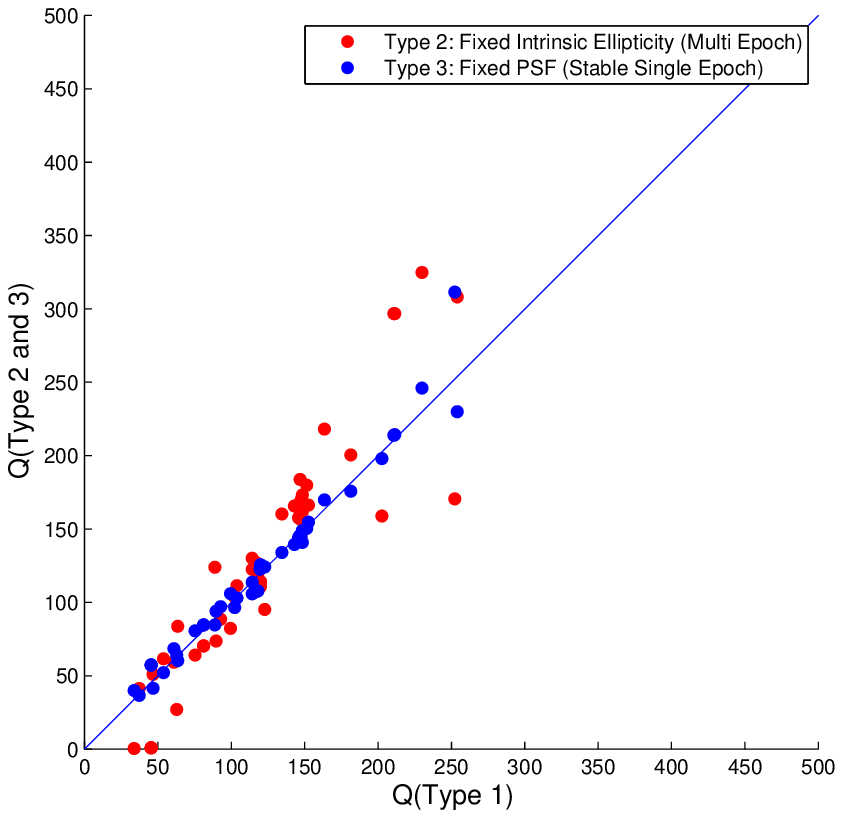}}
 \caption{In the lefthand panel we show the un-modified quality factor
   $Q$ (equation \ref{Qe}) and how this relates 
   to the quality factor with pixel (shape measurement) noise removed
   $Q_{\rm dn}$ and the quality factor obtained when high signal to
   noise training is applied to each submission (equation \ref{tr}). 
   Methods that submitted power
   spectra could not be modified to remove the denoising in this way,
   so only show the training values are shown. 
   The righthand panel shows the $Q_{\rm dn}$ for those sets with
   fixed intrinsic 
   ellipticities (`multi-epoch'; Type 2) or a fixed PSF (`stable 
   single epoch'; Type 3) over all images compared to the
   quality factor in the variable PSF and intrinsic ellipticity  
   case (`single epoch'; Type 1).}
 \label{Q_mod}
\end{figure*}

In the lefthand panel of Figure \ref{Q_mod} we show the effect that the pixel
noise denoising step has on the Quality factor, $Q$.
Note that the way that the denoising step is implemented here uses the
variance of the true shear values (but not the true shear values
themselves). This is a method that was not
available to power spectrum submissions and indeed part of the
challenge was to find optimal ways to account for this in power
spectrum submissions.  
The final layer used to generate the `fit2-unfold' submission performed power-spectrum
estimation and used the model-fit errors themselves to determine and subtract the
variance due to shape measurement errors, including pixel noise.
We find as expected that $Q$ in
general increases for all methods when pixel noise is removed, 
by a factor of $\ls 1.5$, such that a method that has $Q\simeq 100$ 
has a $Q_{\rm dn}\simeq 150$. 
When this correction is applied the method `fit2-unfold' still obtains
the highest Quality factor, and the ranking of the top five methods is
unaffected. 

\subsubsection{Training}
Several of the methods used the training data to help 
debug and test code. For example, and in particular, `fit2-unfold' used the data to
help build the galaxy models used and to set initial parameter values
and ranges in the maximum likelihood fits. This meant that
`fit2-unfold' performed particularly well in sets similar to the
training data (sets 7, 8, and 9) at high signal-to-noise; for details see Appendix D Figure
\ref{fit2unfold}, where `fit2-unfold' has smaller combined 
${\mathcal M}$ and ${\mathcal A}$ values than any other method for
some sets.

To investigate whether using high signal-to-noise training data is useful for methods
we investigate a scenario that
training on the power spectra had been used for all methods. This
modification was potentially available to all participants if they
chose to implement it. To do this we measure the ${\mathcal M}$ and ${\mathcal
  A}$ values from the high signal-to-noise 
Set 7 (see Table \ref{shortsets}) and apply the transformation to the
power spectra, which is to first order equivalent to an $m$ and $c$
correction,  
\be 
\label{tr}
C_{\ell}\rightarrow \frac{C_{\ell}-{\mathcal A}_{{\rm set}=7}}{
1+{\mathcal M}_{{\rm set}=7}}
\ee
to calibrate the method using the training data. In Figure \ref{Q_mod} we
show the resulting Quality factors where we apply both a denoising
step and a training step and when we apply a training step only. 
When both steps are applied we find that Quality factor improves by a
factor $\gs 2$ and some methods perform as well as the 
`fit2-unfold' method (if not better).  In particular `DEIMOS C6'
achieves an average Quality factor of $316$ (see Table 3). 
We find that the increase in the quality 
factor is uniform over all sets, including the low signal-to-noise sets.

We conclude that it was a combination of model calibration 
on the data, and using a denoised power spectrum, 
that enabled `fit2-unfold' to win the challenge. 
We also conclude that calibration of measurements on high
signal-to-noise samples, i.e. those that could be observed using a deep survey
within a wide/deep survey strategy, is an approach that can improve
shape measurement accuracy by about a factor of two. Note that using
this approach is not doing shear calibration as it is practised
historically because the true shear is not known. 
This holds as long as the deep survey is a representative
sample and the PSF of the deep data has similar properties to the PSF
in the shallower survey.

\subsubsection{Multi-epoch data}
In  Figure \ref{Q_mod} 
we show how $Q_{\rm dn}$ varies for
each submission averaged over all those sets that had a fixed
intrinsic ellipticity field (Type 2) or a fixed PSF
(Type 3), described in 
Section \ref{Simulation Structure}. 
Despite the simplicity of this implementation we find that for the
majority of methods, this variation, corresponding to multi-epoch
data, results in an improvement of approximately $1.1$ to $1.3$ in
$Q_{\rm dn}$, although there is large scatter in the relation.  
In GREAT10 the coordination team made a decision to keep the labelling
of the sets private, so that participants were not explicitly aware
that these particular sets had the same PSF (although the functional
PSFs were available) or the same intrinsic ellipticity field. 
These were designed to test stacking methods, however no such methods
were submitted. The 
approach of including this kind of subset can form a basis for
further investigations.
\begin{figure*}
  \centering{\bf Signal to Noise}
  {\includegraphics[width=2.\columnwidth,angle=0,clip=]{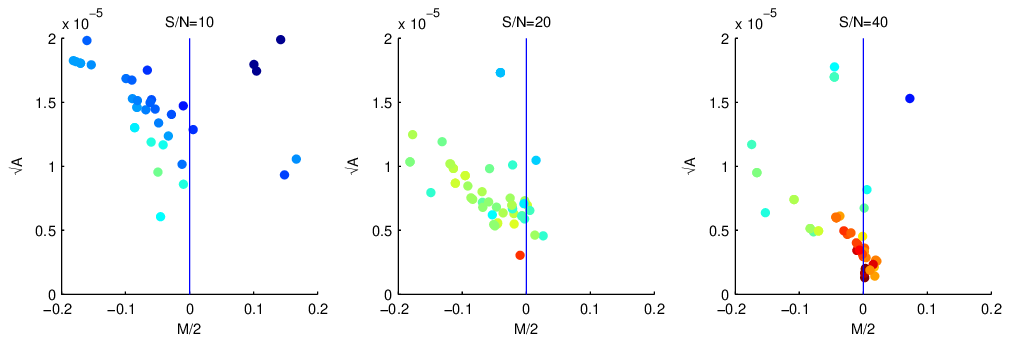}}
 \centering {\bf Galaxy Size/PSF Size}
  {\includegraphics[width=2.\columnwidth,angle=0,clip=]{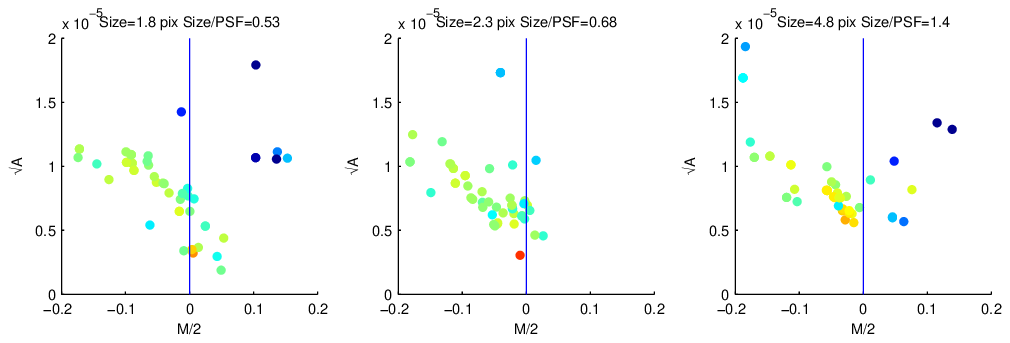}}
  \centering{\bf Galaxy Model}
  {\includegraphics[width=2.\columnwidth,angle=0,clip=]{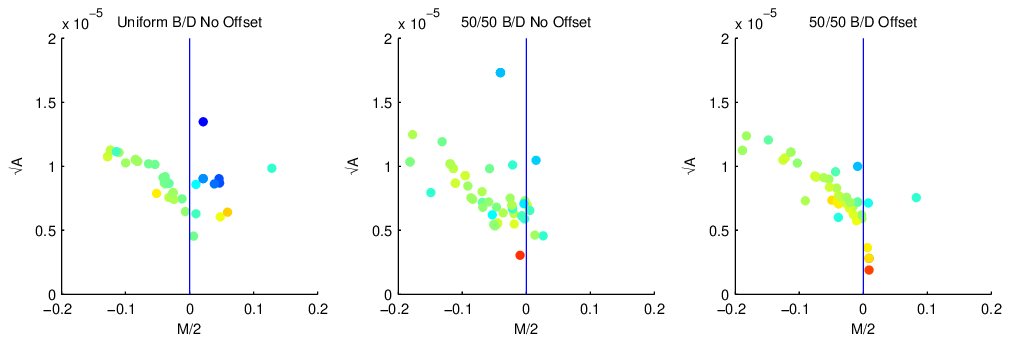}}
  \\{\bf PSF Variation}\\
  {\includegraphics[width=1.333\columnwidth,angle=0,clip=]{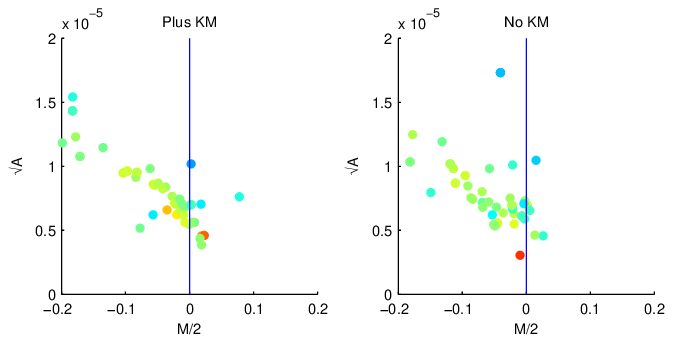}
   \includegraphics[width=0.2\columnwidth,angle=0,clip=]{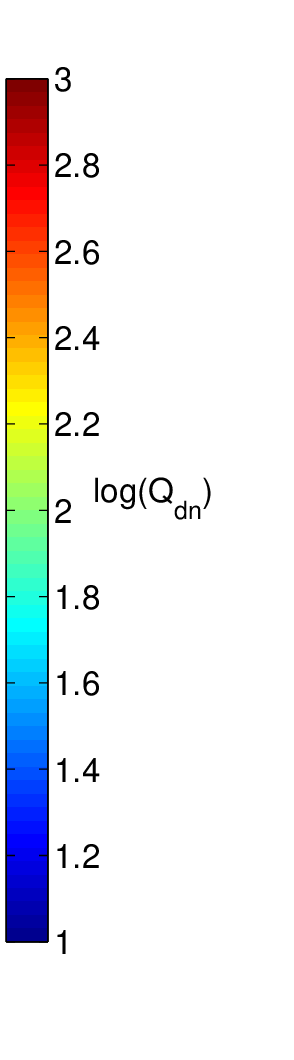}}
 \caption{In each panel we show the metrics, ${\mathcal
     M}$, ${\mathcal A}$ and $Q_{\rm dn}$, for each of the parameter 
   variations between sets, for each submission; 
   the colour scale labels the logarithm of $Q_{\rm
     dn}$ as show in the lower right. 
   The first row shows the signal-to-noise variation, the second row
   shows the galaxy size variation, the third row shows the galaxy
   model variation (the galaxy models are: uniform bulge-to-disk
   fractions where each galaxy has a b/d ratio randomly sampled from
   the range b/d$=[0.3$, $0.95]$ with no offset (Uniform B/D No Offset),
   a 50\% bulge-to-disk fraction b/d$=0.5$ with no offset (50/50 B/D No
   Offset) and a 50\% bulge-to-disk fraction b/d$=0.5$ with a
   bulge/disk centroid offset (50/50 B/D Offset)), the fourth
   row shows PSF variation with and without Kolmogorov (KM) PSF
   variation.}
 \label{Leps}
\end{figure*}
\\

As a summary we show in Figure \ref{Leps} how the population of ${\mathcal M}$,  ${\mathcal A}$
and $Q_{\rm dn}$ parameters for each of the quantities that were 
varied between the sets, for all methods (averaging over all the other
properties of the sets that are kept constant between these
variations). In the following Sections we will analyse each behaviour in detail. 

\subsubsection{Galaxy signal-to-noise}

\begin{figure*}
  {\includegraphics[width=2.\columnwidth,angle=0,clip=]{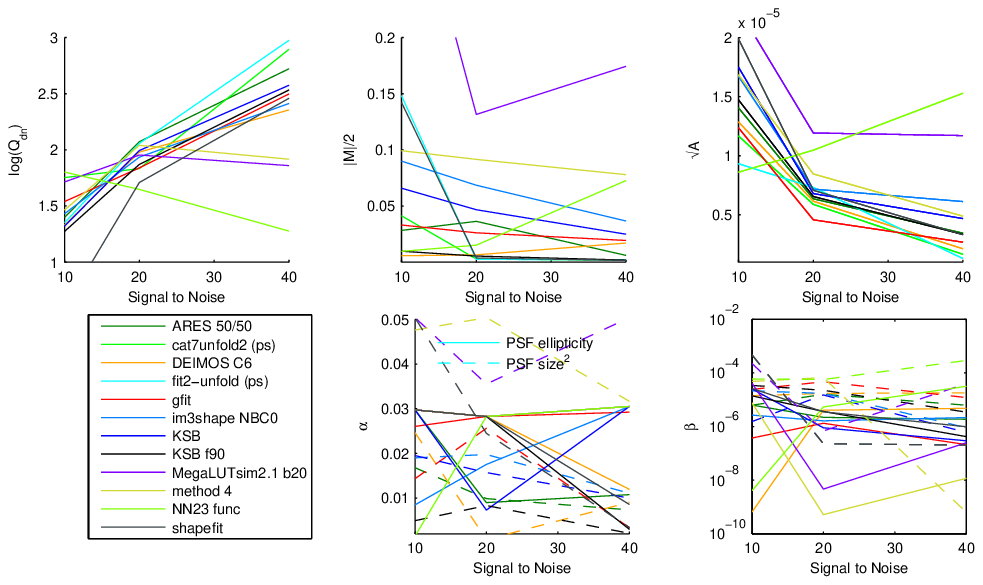}}
 \caption{In the top panels we show how the metrics, ${\mathcal M}$,
   ${\mathcal A}$ and $Q_{\rm dn}$ for submissions change as the signal-to-noise
   increases; the colour scale labels the logarithm of $Q_{\rm
     dn}$. In the lower
   panels we show the PSF size and ellipticity contributions $\alpha$
   and $\beta$. In the bottom-lefthand panel we show the key
   that labels each method.}
 \label{SN}
\end{figure*}

In the top row of Figure \ref{SN} we show how the metrics for each method
change as a function of the galaxy signal-to-noise. We find a clear trend for all
methods to achieve better measurements on higher signal-to-noise
galaxies; with higher $Q$ values and a smaller 
additive biases ${\mathcal A}$. 
In particular `fit2-unfold', `cat2-unfold', `DEIMOS', `shapefit' and `KSB f90'
have a close to zero multiplicative bias for $S/N>20$. Because
signal-to-noise has a particularly strong impact we tabulate the 
${\mathcal M}$ and ${\mathcal A}$ values in Table \ref{sntable}.
We also show in the lower row of Figure \ref{SN} the breakdown of the
multiplicative and additive biases into the components that are
correlated with the PSF size and ellipticity (see Table \ref{metrics}).
We find that for the methods
with the smallest biases at high signal-to-noise (e.g. 
`DEIMOS', `KSB f90', `ARES') the contribution from the PSF size is
also small. For all methods we find that the contribution from PSF
ellipticity correlations is subdominant for ${\mathcal A}$.
\begin{table*}
\begin{center}
\begin{tabular}{|l|r|c|r|c|r|c|}
\hline
 & S/N=10 & & S/N=20 & &S/N=40\\
\hline
Method & ${\mathcal M}/2\,\,\,\,\,\,$&$\sqrt{\mathcal A}/10^{-4}$& ${\mathcal M}/2\,\,\,\,\,\,$&$\sqrt{\mathcal A}/10^{-4}$& ${\mathcal M}/2\,\,\,\,\,\,$&$\sqrt{\mathcal A}/10^{-4}$ \\
\hline
 $^{\dagger}$ARES 50/50         & $ -0.028320 $ & $ 0.140511 $ & $ -0.036322 $ & $ 0.063551 $ & $ -0.006060 $ & $ 0.034517 $\\ 
 $^{\dagger}$cat7unfold2 (ps)   & $ -0.041280 $ & $ 0.116732 $ & $ -0.002803 $ & $ 0.058890 $ & $ 0.001880 $ & $ 0.016527 $\\ 
 DEIMOS C6                      & $ 0.005676 $ & $ 0.128678 $ & $ -0.006533 $ & $ 0.061440 $ & $ 0.017020 $ & $ 0.021269 $\\ 
 fit2-unfold (ps)               & $ 0.148242 $ & $ 0.093275 $ & $ -0.002501 $ & $ 0.073071 $ & $ 0.002228 $ & $ 0.012961 $\\ 
 gfit                           & $ -0.033046 $ & $ 0.123692 $ & $ 0.026172 $ & $ 0.045710 $ & $ 0.019359 $ & $ 0.026773 $\\ 
 $^*$im3shape NBC0              & $ -0.089984 $ & $ 0.167280 $ & $ -0.068486 $ & $ 0.071842 $ & $ -0.036627 $ & $ 0.061176 $\\ 
 KSB                            & $ -0.065856 $ & $ 0.175017 $ & $ -0.046715 $ & $ 0.068038 $ & $ -0.024967 $ & $ 0.046845 $\\ 
 $^*$KSB f90                    & $ -0.009688 $ & $ 0.147320 $ & $ 0.005480 $ & $ 0.065486 $ & $ -0.001810 $ & $ 0.033502 $\\ 
 $^{\dagger}$MegaLUTsim2.1 b20  & $ -0.380576 $ & $ 0.224465 $ & $ -0.131563 $ & $ 0.119239 $ & $ -0.174472 $ & $ 0.117005 $\\ 
 method 4                       & $ -0.099330 $ & $ 0.168536 $ & $ -0.091481 $ & $ 0.084571 $ & $ -0.077907 $ & $ 0.048824 $\\ 
 $^{\dagger}$NN23 func          & $ -0.009595 $ & $ 0.086018 $ & $ 0.015145 $ & $ 0.104664 $ & $ 0.072641 $ & $ 0.152932 $\\ 
 shapefit                       & $ 0.142251 $ & $ 0.198852 $ & $-0.003768 $ & $ 0.070808 $ & $ 0.001568 $ & $ 0.033164 $\\ 
\hline
\end{tabular}
\caption{The metrics ${\mathcal M}/2\simeq m$ and $\sqrt{\mathcal
    A}\simeq \sigma(c)$ for each of the signal-to-noise values used in
  the simulations.}
\label{sntable}
\end{center}
\end{table*}

\subsubsection{Galaxy size} 
\begin{figure*}
  {\includegraphics[width=2.\columnwidth,angle=0,clip=]{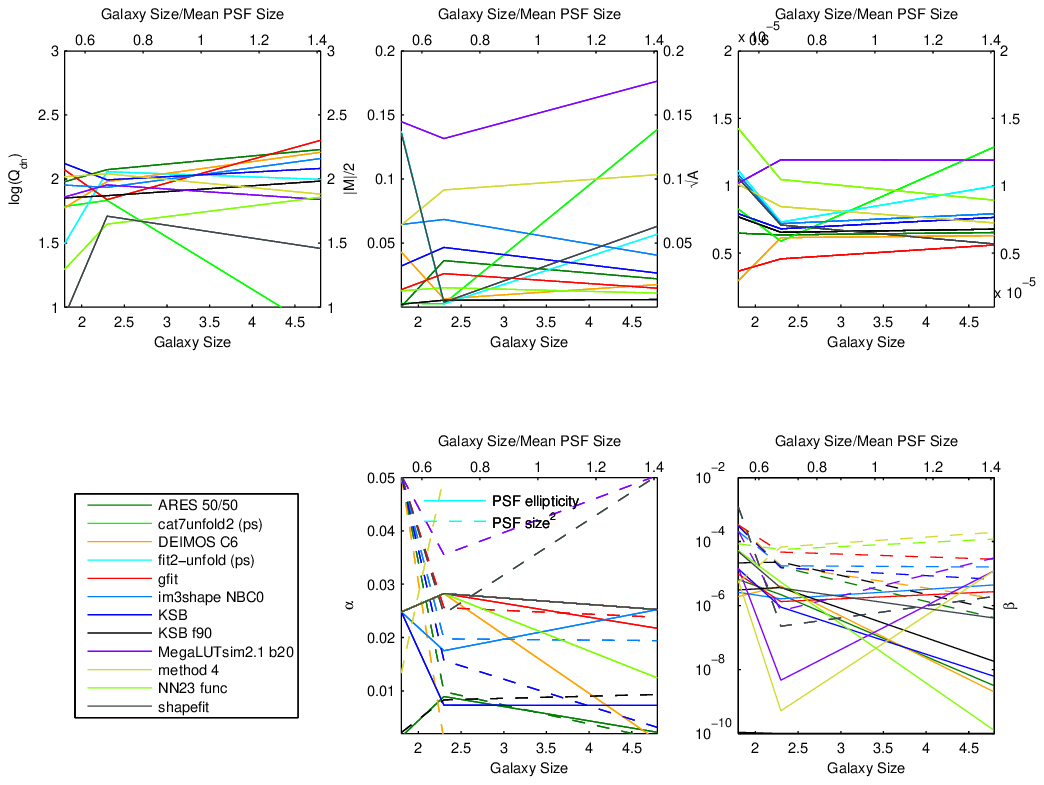}}
 \caption{In the top panels we show how the metrics,
   ${\mathcal M}$,
   ${\mathcal A}$ and $Q_{\rm dn}$ for submissions change as the galaxy size 
   increases; the colour scale labels the logarithm of $Q_{\rm
     dn}$. In the lower
   panels we show the PSF size and ellipticity contributions $\alpha$
   and $\beta$. In the bottom-lefthand panel we show the key
   that labels each method. The mean PSF is the mean within an image
   not between all sets.}
 \label{size}
\end{figure*}

In Figure \ref{size} we show how the metrics of each method change as a
function of the galaxy size -- the mean PSF size was $\simeq 3.4$ pixels.
Note that the PSF size is statistically the same in each set, such
that a larger galaxy size corresponds to either a case where the
galaxies are larger in a given survey or where observations are taken
where the pixel size and PSF size are relatively smaller for the same
galaxies. 

We find that the majority of methods have
a weak dependency on the galaxy size, but that at scales of $\ls 2$
pixels, or size/mean PSF size $\simeq 0.6$, the accuracy decreases (larger ${\mathcal M}$ and ${\mathcal
  A}$ and smaller $Q_{\rm dn}$). This weak dependence is partly due to
the small (but realistic) dynamical range in size, compared to a
larger dynamical range in signal-to-noise. 
The exceptions are `cat7unfold2', `fit2unfold' and `shapefit'
that appear 
to perform very well on the fiducial galaxy size and less well on the small and
large galaxies -- this is consistent with the model calibration approach of these
methods, which was done on Set 7 that used the fiducial galaxy type. 
The PSF size appears to have a small contribution at large
galaxy sizes, as one should expect, but a large contribution to the
biases at scales smaller than the mean PSF size. We find that the
methods with largest biases have a strong PSF size contribution. Again the PSF
ellipticity has a subdominant contribution to the biases for all
galaxy sizes for ${\mathcal A}$.  

\subsubsection{Galaxy model} 

\begin{figure*}
  {\includegraphics[width=2.\columnwidth,angle=0,clip=]{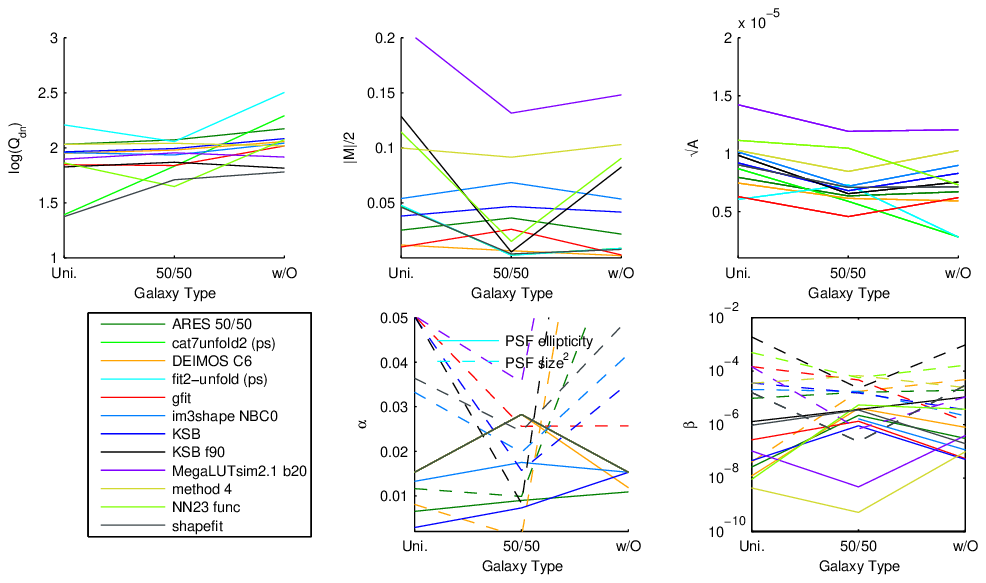}}
 \caption{In the top panels we show how the metrics,
   ${\mathcal M}$, ${\mathcal A}$ and $Q_{\rm dn}$ for submissions change as the galaxy model changes; 
   the colour scale labels the logarithm of $Q_{\rm dn}$, the galaxy
   models are: uniform bulge-to-disk fractions each galaxy has a b/d
   ratio randomly sampled from the range $b/d=[0.3,0.95]$ with no
   offset (Uni.), a 50\% bulge-to-disk fraction $b/d=0.5$ with no
   offset (50/50.) and a 50\% bulge-to-disk fraction $b/d=0.5$ with
   a bulge/disk centroid offset (w/O). In the lower
   panels we show the PSF size and ellipticity contributions $\alpha$
   and $\beta$. In the bottom-lefthand panel we show the key
   that labels each method.}
 \label{model}
\end{figure*}

In Figure \ref{model} we show how each method's metrics change as a
function of the galaxy type. The majority of methods have a weak
dependency on the galaxy model. The exceptions, similar to the galaxy
size dependence, are `cat7unfold2', `fit2unfold' and `shapefit'
that appear 
to perform very well on the fiducial galaxy model and less so on the small and
large galaxies -- this again is consistent with model calibration approach of these
methods. Again the
contribution to ${\mathcal A}$ from the PSF size dependence is dominant over the PSF
ellipticity dependence, and is consistent with no model dependency for
the majority of methods, except those highlighted here. We refer to
Section \ref{Discussion} and Appendix E for a breakdown of $m$ and $c$
behaviour as a function of galaxy model for each method.

\subsubsection{PSF model} 

\begin{figure*}
  {\includegraphics[width=2.\columnwidth,angle=0,clip=]{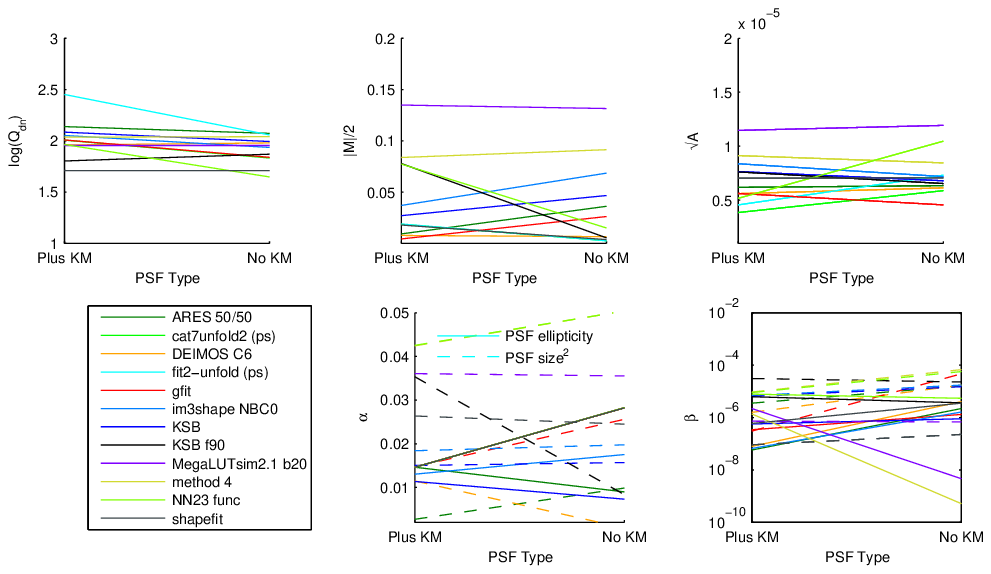}}
 \caption{In the top panels we show how the metrics,
   ${\mathcal M}$,
   ${\mathcal A}$ and $Q_{\rm dn}$ for submissions change as the PSF model changes; 
   the colour scale labels the logarithm of $Q_{\rm dn}$, the PSF
   models are the fiducial PSF, and the same PSF except with a
   Kolmogorov power spectrum in ellipticity added. In the lower
   panels we show the PSF size and ellipticity contributions $\alpha$
   and $\beta$. In the bottom-lefthand panel we show the key
   that labels each method.}
 \label{PSF}
\end{figure*}

In Figure \ref{PSF} we show the impact of changing the PSF spatial variation on the 
metrics for each method. We show results for the fiducial PSF, which does not
include a Kolmogorov (turbulent atmosphere) power spectrum, and one which includes a
Kolmogorov power spectrum in PSF ellipticity. We find that the
majority of methods have a weak dependence on the inclusion of the
Kolmogorov power. But it should be noted that participants knew the
local PSF model exactly in all cases.   

\subsection{Averaging methods}
\label{Averaging methods}
In order to reduce shape measurement biases one may also wish to average
together a number of shape measurement methods. In this way any random
component, and any biases, in the ellipticity estimates may be
reduced. In fact the `ARES' method (see Appendix E) averaged
catalogues from DEIMOS and KSB and attained better quality metrics. 
Doing this exploited the fact that DEIMOS had in some sets a strong
response to the ellipticity whereas KSB had a weak response.
\begin{figure*}
  {\includegraphics[width=2\columnwidth,angle=0,clip=]{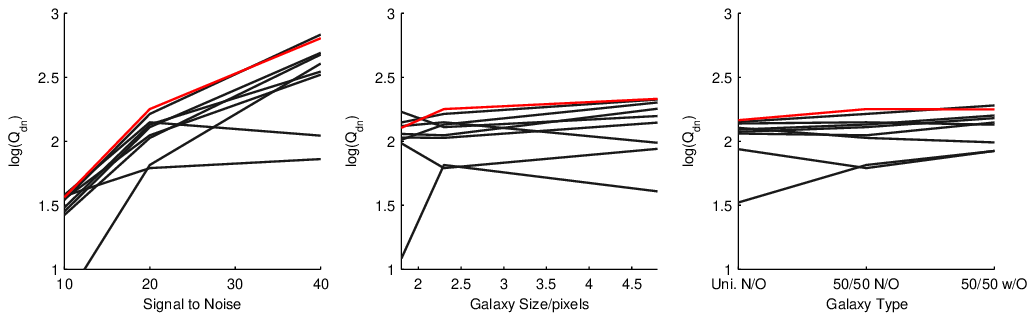}}
\caption{The Quality factor as a function of signal-to-noise (left
  panel), galaxy size (middle panel) and galaxy type (right panel) for
  an averaged ellipticity catalogue submission (red, using the
  averaging described in Section \ref{Averaging methods}); compared to the
  methods used to average (black).}
 \label{avP}
\end{figure*}

To test this we averaged the ellipticity catalogues
from the entries with the best metrics for each method that submitted
an ellipticity catalogue (ARES 50/50, DEIMOS C6,
gfit, im3shape NBC0, KSB, KSB f90, MegaLUTsim2.1 b20, method 4,
shapefit) as so:
\be 
\langle e_i\rangle=\frac{\sum_{\rm methods}e_{m,i}w_{m,i}}{\sum_{\rm
    methods}w_{m,i}}
\ee
where $i$ labels each galaxy and in general $w_{m,i}$ is some weight
that depends on the method, galaxy and PSF properties. We wish to
weight methods that perform better, and so choose the Quality factor
from the high signal-to-noise training set (set 7) as the weight $w_{m,i}=Q_{\rm
  dn,m}($set 7$)$ applied over all other sets. This is close to an inverse variance weight on the
noise induced on the shear power spectrum ($\propto 1/\sigma_{\rm
  sys}^2$). We leave the determination of optimal weights for future
investigation. 

We find that the average Quality factors over all sets for this approach are 
$Q=131$ and $Q_{\rm dn}=210$, which are slightly smaller
on average than some of the individual methods. However we find
that for the fiducial signal-to-noise and large galaxy size the
Quality factor increases, see Figure \ref{avP}. This suggests that such
an averaging approach can improve the accuracy of an ellipticity
catalogue but that a weight function should be optimised to be a
function of signal-to-noise, galaxy size and type; however averaging
many methods with a similar over or under estimation of the shear would
not improve in the combination. If we take the highest quality factors 
in each set, as an optimistic case that a weight function had been 
found that could identify the best shape measurement in each regime 
we find an average $Q_{\rm dn}=393$. 

\subsection{Overall performance}
\label{Discussion}
We now list some 
observations of method accuracy for each method by
commenting on the behaviour of the metrics and dependencies shown in
Section \ref{Results} and Appendix E. Words such as
`relative' are with respect to the other methods analysed here. This
is a snapshot of methods performance as submitted for GREAT10 blind
analysis.
\begin{itemize}
\item 
{\bf KSB}: has low PSF ellipticity correlations, and a small galaxy
morphology dependence, however it has a relatively large absolute $m$ bias
value.
\item 
{\bf KSB f90}: has small relative $m$ and $c$ biases on average, but a relatively strong
 PSF size and galaxy morphology dependence, in particular on the galaxy bulge fraction.
\item 
{\bf DEIMOS}: has small $m$ and $c$ biases on average, but a relatively strong
dependence on galaxy morphology again in particular on the bulge fraction, similar to KSB
f90. Dependence on galaxy size is low except for small galaxies with
size smaller than the mean PSF.
\item 
{\bf im3shape}: has a relatively large PSF ellipticity and size
correlation, a small galaxy size dependence for $m$ and $c$ but a 
stronger bulge fraction dependence. 
\item 
{\bf gfit}: has relatively small average $m$ and $c$ biases, and a small galaxy morphology 
dependence, there is a relatively large correlation with PSF
ellipticity. This was the only method to employ a denoising step at
the image level, suggesting that this may be partly responsible for the small biases. 
\item 
{\bf method 4}: has relatively strong PSF ellipticity, size and galaxy type dependence.
\item 
{\bf fit2unfold}: has strong model dependence,
but relatively small $m$ and $c$ biases for the fiducial model type, and also a
relatively low PSF ellipticity correlation.
\item 
{\bf cat2unfold}: has strong model dependence in particular on galaxy size,
but relatively small $m$ and $c$ biases for the fiducial model type, and also a
relatively low PSF ellipticity correlation.
\item 
{\bf shapefit}: has a relatively low quality factor, and a strong
dependence on model types and size that are not the fiducial values,
but small $m$ and $c$ biases for the fiducial model type.
\end{itemize}

\noindent To make some general conclusions, we find that 
\begin{enumerate}
\item 
Signal-to-noise: We find a strong dependence of the metrics below
S/N$=10$ especially for additive biases, however
we find methods that meet bias requirements for the most ambitious
experiments when S/N$ > 20$. 
\item
Galaxy type: We find marginal evidence 
that model fitting methods have a relatively low
dependence on galaxy type compared to KSB-like methods, 
but that this is only true if the model matches the
underlying input model (note that GREAT10 used simple models). 
We find evidence that if one trains on a
particular model then biases are small for this subset of galaxies.
\item 
PSF dependence: Despite the PSF being known exactly we find 
contributions to biases from PSF size, but less so from PSF
ellipticity. The methods with the largest biases have a strong PSF
size correlation. 
\item 
Galaxy Size: For large galaxies well sampled by the PSF, with scale
radii $\gs 2$ times the mean PSF size we find that methods meet
requirements on bias parameters for the most ambitious
experiments. However if galaxies are unresolved with radii $\ls 1$
time the PSF size biases become significant.
\item 
Training: We find that calibration on a high signal-to-noise sample can
significantly improve a method's average biases. This is true whether
training is a model calibration, or a more direct form of training on the ellipticity
values of power spectra themselves. 
\item  
Averaging Methods: We find that averaging methods is clearly beneficial, but
that the weight assigned to each method will need to be correctly
determined. An individual entry (ARES) found that this was the case,
and we find similar conclusions when averaging over all methods.
\end{enumerate} 
Note that statements on required accuracy are only
on bias and not on the statistical accuracy, that a selection in objects
with a particular property (e.g. high signal-to-noise) would achieve. 
Such selection is dependent on the observing conditions and survey design 
for a particular experiment, so we leave such an investigation for future work.

\section{Astrocrowdsourcing}
\label{astrocrowdsourcing}
The GREAT10 Galaxy Challenge was an example of `crowdsourcing' 
astronomical algorithm development (`astrocrowdsourcing'). This was part of a wider effort
during this time period, that included the GREAT10 Star Challenge and
the sister project Mapping Dark Matter\footnote{Run in conjunction
  with Kaggle {\tt http://www.kaggle.com/c/mdm}} (see companion papers
for these challenges). In this Section we discuss this aspect
of the challenge and list some observations 
\begin{figure}
  {\includegraphics[width=0.7\columnwidth,angle=-90,clip=]{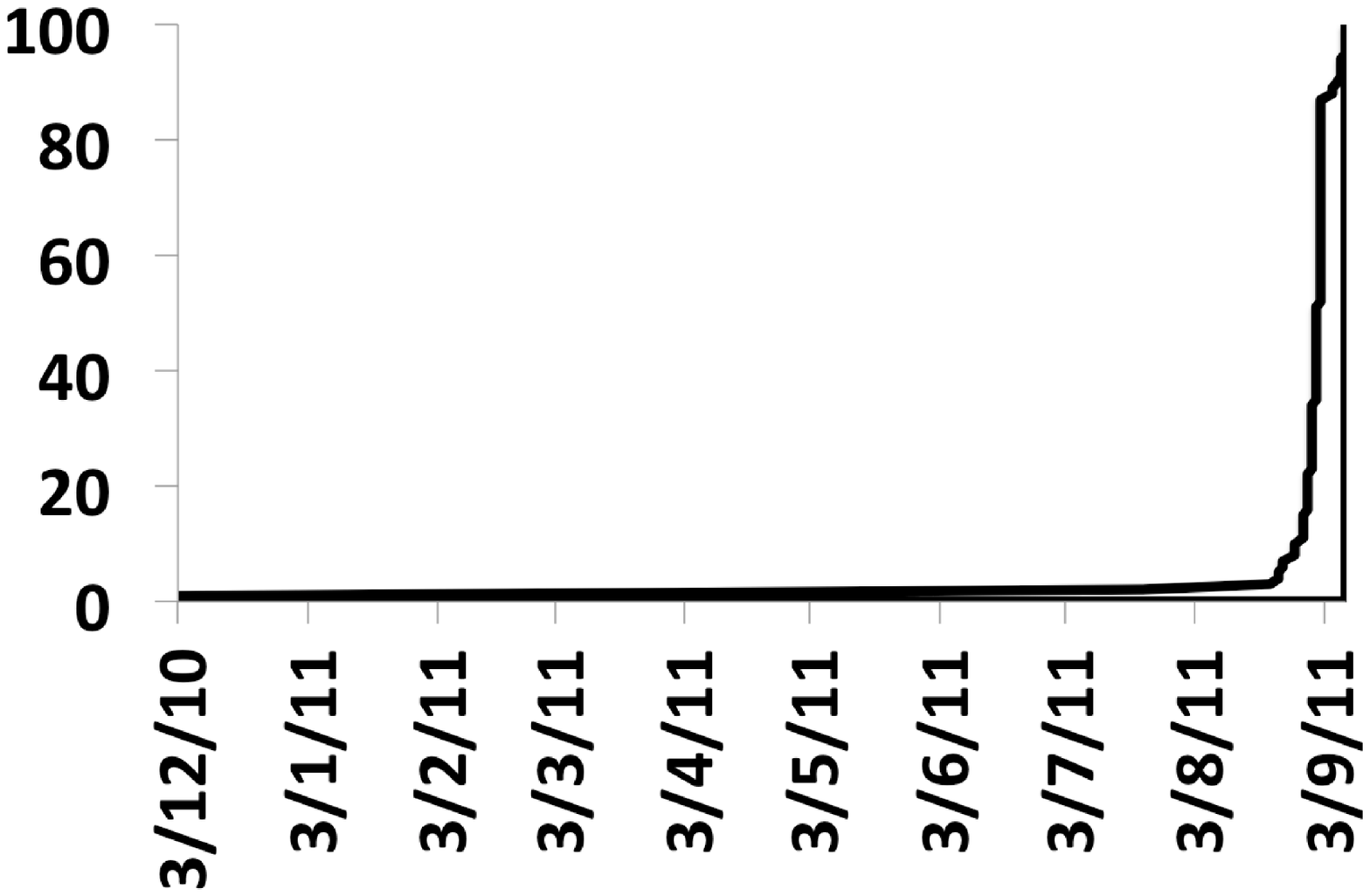}}
\caption{The cumulative submission number as a function of the challenge
time, which started on 3$^{\rm rd}$ December 2010 and ran for 9 months.}
 \label{g10time}
\end{figure}

GREAT10 was a major success in its effort to generate new ideas and attract new 
people into the field. For example, the winners of the challenge
(authors D. Kirkby and D. Margala), were new
to the field of gravitational lensing. A variety of entirely new
methods have also been attempted for the first time on blind data, including
the Look Up Table (MegaLUT) approach, an auto-correlation approach 
(method 4 and TVNN), and the use of training data. Furthermore the TVNN
method is a real pixel-level deconvolution method, which is the first
time a genuine deconvolution of the data has been used in  
shape measurement. 

The limiting factor in designing the scope of the GREAT10 Galaxy Challenge was the size of
the simulations, that was kept below 1TB for ease of distribution; a
larger challenge could have addressed even more observational
regimes. In the future executables could be distributed that locally
generate the data. However in this case 
participants may still need to store the data. Another approach might
be to host challenges on a remote server where participants can upload and run
algorithms. Care should be taken however to retain the integrity of
the blindness of a challenge, without which results become largely
meaningless as methods could be tuned to the parameters or functions of specific 
solutions if those solutions are known \emph{a priori}. We require algorithms to be of high 
fidelity and to be useful on large amounts of data, which requires them to be fast: an algorithm 
that takes a second per galaxy needs $\simeq 50$ CPU years to run on
$1.5$x$10^9$ galaxies (the number observable by the most ambitious
lensing experiments e.g. Euclid\footnote{\tt
  http://www.euclid-ec.org}, Laureijs et. al., 2011), 
a large simulation generates innovation in this direction.

In Figure \ref{g10time} we show the cumulative submission of the
GREAT10 Galaxy Challenge as a function of time, from the beginning of
the challenge to the end and in the post-challenge submission period. All
submissions (except one made by the GREAT10 coordination team) 
were made in the last 3 weeks of the 9 month period. For
future challenges intra-challenge milestones could be used to encourage
early submissions. This submission profile also reflects the size and
complexity of the challenge; it took time for participants to
understand the challenge and to run
algorithms over the data to generate a submission. For future
challenges submissions on smaller subsets of the
data could be enabled, with submission over the entire data set being
optional.  

We note that the winning team (Kirkby and Margala) made $18$
submissions during the challenge, compared to the mean
submission number of $9$. The winners also recognised from the information
provided that the submission
procedure was open to power spectrum and ellipticity catalogue
submissions. The leaderboard was designed such that accuracy
was reported in a manner that was indicative of performance, but such that this information 
could not be trivially used to directly calibrate methods (for example if
$m$ and $c$ were provided a simple ellipticity catalogue correction
could have been made). 

Many of these issues were overcome in the sister Mapping
Dark Matter challenge (see the Mapping Dark Matter results paper, 
Kitching et al., in prep) that received over $700$ entries, over $2000$
downloads of the data and a constant rate of submission. It also used an
alternative model for leaderboard feedback where the simulated data
was split into public and private sets, and useful feedback only
provided for the public sets.  

For a discussion of the simplifications present in GREAT10 we also
refer the reader to Section 5 of the GREAT10 Handbook (Kitching et al., 2011).

\section{Conclusions}
\label{Conclusions}

The GREAT10 Galaxy Challenge was the first weak lensing shear simulation to include 
variable fields: both the PSF and the shear field varied as a function of position. 
It was also the largest shear simulation to date, consisting of over 50 million simulated galaxies, 
and a total of 1TB of data. The challenge ran for 9 months from December 2010 to 
September 2011, and during that time approximately $100$ submissions
were made. 

In this paper we define a general pseudo-Cl methodology for propagating shape
measurement biases into cosmic shear power spectra and use this to
derive a series of metrics that we use to investigate methods. We present a 
quality factor $Q$ that relates the inaccuracy in shape measurement methods to the shear power spectrum itself. 
A $Q=1000$ denotes a method that could measure the dark energy equation of state parameter $w_0$ with a bias 
less than or equal to the predicted statistical error from the most
ambitious planned weak lensing experiments (for a more general
expression we refer to Massey et al., in prep). We show 
how one can correct such a metric to account for pixel noise in a shape
measurement method. During the challenge, submissions were publicly
ranked on a live leaderboard and ranked by this metric $Q$.

We show how a variable shear simulation can be used to determine 
$m$ and $c$ parameters (Heymans et al., 2006) that are a measure of bias between the measured and true shear 
(those parameters used in constant shear simulations: STEP and GREAT08) on an object by object basis. 
We link the quality factor to linear power spectrum biases 
including a multiplicative ${\mathcal M}\approx 2m$ and additive bias ${\mathcal
  A}\propto \sigma(c)^2$ that are approximately related to the STEP one-point estimators
of shape measurement bias. The equality is only approximate because in
general ${\mathcal M}$ and ${\mathcal A}$ are a measure of spatially 
varying method biases.  
We introduce further metrics that allow an assessment of the contribution to 
the multiplicative and additive biases from correlations between the
biases and any spatially variable quantity (in this paper we focus
on PSF size and ellipticity). 
 
The simulations were divided into sets of
$200$ images each containing a grid of $10$,$000$ galaxies. In each
set the shear field was spatially varying
but constant between images. The challenge was to reconstruct the shear \emph{power spectrum}
for each set. Participants could submit either catalogues of  
ellipticities one per image or power spectra one for each set, and were provided with an 
exact functional description of the PSF and the positions of all objects to within half a pixel.

The simulations were structured in such a way that conclusions could
be made about a shape measurement method's accuracy  
as a function of galaxy signal-to-noise, galaxy size, galaxy
model/type and the PSF type. The simulations also contained  
some `multi-epoch' sets in which the shear and intrinsic ellipticities
were fixed between images in a set but where the PSF varied between
images, and some `static single-epoch' sets where the PSF was fixed between images
in a set but the intrinsic ellipticity field varied between images. All fields were
always spatially varying. Participants were  
provided with true shears for one of the high signal-to-noise sets
that they could use as a training set. 

Despite the simplicity of the
challenge, making conclusions about which aspects of which algorithm generated
accurate shape measurement is difficult due to the complexity of the
algorithms themselves (see Appendix E). We leave investigations 
into tunable aspects of each method to future work. We can however make some statements about the
regimes in which methods perform well or poorly.

The best methods submitted to GREAT10 scored an average $Q\simeq 300$ 
with $m\simeq 7\times 10^{-3}$ and $c\simeq 10^{-5}$.  
The best performing non-stacking method at a 
signal-to-noise $20$, using the GREAT10/Sextractor definition, in
GREAT08 was KSBf90 (CH) 
which had an $m=0.0095\pm 0.003$ and $c\simeq 8$x$10^{-4}$, and we
find a similar performance on GREAT10. Comparing this benchmark
against methods here we find at least a factor $3$
improvement in performance by methods tested on blind simulations (we
refer to Table 3 where the mean improvement over KSBf90 is $2.6\pm 1.6$ over
all metrics).
The methods that won the challenge (scoring the highest $Q$ on the leaderboard)
employed a maximum likelihood
model-fitting method. 
Several methods used the training data to test code, and 
we find that by directly training on a high signal-to-noise set the
majority of methods achieve a factor of $2$ increase in the average value of $Q$. 
We find some evidence that shape measurement 
inaccuracies can be reduced by averaging methods together, but conclude that for such a method to
be usable an optimal weight for each method as a function of
signal-to-noise and galaxy properties would have to be found.

For a signal-to-noise of $40$ the best methods achieved a $Q\gs 1000$, 
$m<1$x$10^{-3}$ and $c<1$x$10^{-5}$; the majority of methods 
have an accuracy that is strongly dependent on signal-to-noise with 
$Q\simeq 100$ and $\simeq 50$ for signal-to-noise of $20$ and $10$ 
respectively. However the dependence on galaxy model (bulge-to-disk
ratio or bulge-to-disk offset) 
and size is not strong. 
There is a contribution to the multiplicative bias $m$ from PSF size correlations for the majority of 
methods over all sets, but a smaller contribution 
from PSF ellipticity dependence (as expected from theoretical
calculations, e.g. Massey et al., in prep).

The testing of shape measurement methods by GREAT10 suggests methods 
now exist that can be used for cosmic shear surveys covering up to a few thousand square degrees ($\ls
3000$ square degrees, that require $m\ls 6$x$10^{-3}$; Kitching et
al., 2008\footnote{The scaling formula from this paper can be
  rewritten for the maximum applicable area of a survey for a given
  bias $m$ as $A_{\rm max}\ls
  20$,$000[(0.001/m)^{2.4}/0.17/10^{\beta}]^{1/1.5}$ square degrees,
  assuming that the redshift behaviour is $m\propto (1+z)^{\beta}$.}) to measure 
cosmological parameters in an unbiased fashion.
We find that on the additive bias $c$ methods already meet requirements for 
even the most ambitious surveys ($c< 1$x$10^{-3}$) over all simulated conditions, 
and that in the high signal-to-noise regime ($\gs 40$) methods 
already meet the most ambitious requirements on the multiplicative bias
($m < 2$x$10^{-3}$; Kitching et al., 2008). Now that 
such accuracy has been demonstrated in the high signal-to-noise regime, it is now plausible that such
accuracy may be possible at lower signal-to-noise, in principle. However we note that the
requirements are on all galaxies in a survey and that the demonstration
here is averaged over a simulation with particular properties, in
particular the fiducial signal-to-noise is $20$.
Therefore these conclusions have a caveat that the GREAT10 simulations 
were intentionally simplistic in some respects, so 
that clear statements about methods could be made, but they provide a foundation for shape measurement 
development to continue increasing in realism and complexity. 
\\
\\
\noindent{\em Acknowledgements:} We thank the GREAT10 Advisory team for discussions
before and after the challenge. TDK is supported by a Royal Society
University Research Fellowship, and was supported by a Royal
Astronomical Society 2010 Fellowship for the majority of this
work. This work was funded by a EU FP7 PASCAL 2 Challenge
Grant. Workshops for the GREAT10 challenge were funded by the eScience
STFC Theme, PASCAL 2 and by JPL, run under a contract for NASA by Caltech, 
and hosted at the eScience Institute
Edinburgh, by UCL and by IPAC at Caltech. We thank Bob Mann, Francesca
Ziolkowska, Harry Teplitz and Helene Seibly for local organisation of
the workshops. We thank Mark Holliman for system administrator tasks
for the GREAT10 web server. We thank Whitney Clavin for assistance on
the NASA press release for GREAT10. We thank Lance Miller
comments on a first draft and throughout the challenge. We also thank 
Alan Heavens, Alina Keissling, Benjamin Joachimi, Marina Shmakova, 
Gary Bernstein, Konrad Kuijken, Yannick Mellier, Mark
Cropper, Malin Velander, Elisabetta Semboloni, Henk Hoekstra, Karim
Benabed for useful discussions. CH acknowledges support from the ERC
under grant EC FP7 240185. SBr, MH, TKa, BR, JZ acknowledge 
support from ERC Starting Grant with number 240672. BR acknowledges
support from the NASA WFIRST Project Office. DK and DM acknowledge the
support of the US DOE. RM acknowledges support from a Royal Society
University Research Fellowship. We thank an anonymous referee for
helpful comments that improved the analysis and clarity of the paper. 

{\small {\em Contributions:} All authors contributed to the development
  of this paper. TDK was PI of GREAT10, created the
  simulations, and wrote this paper.
  CH, MG, RM, BR  were active members of the GREAT10 coordination board during the
  challenge (12/2010 to 10/2011). 
  SBr, FC, MG, SH, CH, MH, TK, DK, DM, LM, PM, GN, KP, BR, MT, LV, MV, JY,
  JZ submitted entries to the GREAT10 galaxy challenge. 
  JR hosted and ran the GREAT10 challenge final workshop, 
  TDK, CH, SBr, MH, TKa, RM, BR, LV, JZ were on the LOC for the
  mid-challenge workshops. SBa and SBr created 
  the image simulation code for GREAT08 that was extended by TDK for
  the GREAT10 challenge. AT contributed to the a code on which the
  spatially varying field code was based, and provided consultation with
  regard to the pseudo-Cl formalism. DW  
  maintained the GREAT10 leaderboard and processed submissions with
  TDK during the challenge.}



\onecolumn
\newpage
\section*{Appendix A: Pseudo-Cl Estimators for Weak Lensing}
In this Section we describe a formalism for the evaluation of
variable shear systematics in weak lensing. We note that this has a
more general application to that described here, such that any mask in
general could be accounted for in weak lensing power spectrum 
estimation. This closely follows the pseudo-Cl formalism described
in Memari (2010) and Brown et al. (2005) that has been applied in CMB
studies, for survey masks.  

We start by defining a generalised shear systematic response where 
\be 
\label{genmc}
e_{\rm measure}(\btheta)=\gamma(\btheta)+e_{\rm intrinsic}(\btheta)+
c(\btheta)+m(\btheta)[\gamma(\btheta)+e_{\rm intrinsic}(\btheta)]
+q(\btheta)[\gamma(\btheta)+e_{\rm intrinsic}(\btheta)]|\gamma(\btheta)+e_{\rm intrinsic}(\btheta)|
\ee
where all variables are a function of position on the sky, and all are
complex quantities (e.g. $\gamma(\btheta)=\gamma_1(\btheta)+{\rm
  i}\gamma_2(\btheta)$). We expect that $m(\btheta)$ will in general
depend on spatially varying quantities including PSF ellipticity
and size or galaxy properties such as signal-to-noise, so that 
one could write $m(\btheta)\rightarrow m({\rm PSF}(\btheta),{\rm
  Galaxy}(\btheta))$ or $m(e_{\rm PSF}(\btheta), r_{\rm PSF}(\btheta),
{\rm S/N}(\btheta),\dots)$ for example, but this does not qualitatively
change the following treatment. We note also that in general the
systematic terms can also be complex $m(\btheta)=|m(\btheta)|{\rm
  e}^{{\rm i}\phi[\btheta]}$, here we assume a scalar spatially varying
quantity, and will investigate further generalisation in future work.

The E and
B mode decomposition of the spin-2 field $e_{\rm measure}(\btheta)$
can be written in general as a rotation in Fourier space (see GREAT10
handbook) such that 
\ba 
E(\bell)\pm{\rm i}B(\bell)&=&\ell^*\ell^*|\ell|^{-2}\left[e_{1,{\rm
    measure}}(\bell)+{\rm i}e_{2,{\rm measure}}(\bell)\right]\nn
E(\bell)\pm{\rm i}B(\bell)&=&{\rm e}^{\mp2{\rm i}\phi_{\ell}}\left[e_{1,{\rm
    measure}}(\bell)+{\rm i}e_{2,{\rm measure}}(\bell)\right]
\ea
where $e_{\rm measure}(\bell)$ is the Fourier transform of $e_{\rm
  measure}(\btheta)$. 

When creating a power spectrum 
the auto-correlations of the first three terms of equation (\ref{genmc}) have a simple
interpretation, but the fourth term has an effective weight map as a
function of position such that (only focusing on the contribution
from the fourth term) we have that the estimated E and B-mode terms are
\be 
\widetilde E(\bell)\pm{\rm i}\widetilde B(\bell)=\int\frac{{\rm d}^2\ell'}{(2\pi)^2}
{\rm e}^{\mp2{\rm i}\phi_{\ell'}}W_m(\bell-\bell')\left[E(\bell')\pm{\rm
    i}B(\bell')\right],
\ee
where $W_m$ is 2D Fourier transform of the the $m(\btheta)$
field. Equivalently for the E-mode part only we have 
\be
\label{Et}
\widetilde E(\bell)=\int\frac{{\rm d}^2\ell'}{(2\pi)^2} W_m(\bell-\bell')
\left[\cos(2(\phi_{\ell}-\phi_{\ell'}))E(\bell')-\sin(2(\phi_{\ell}-\phi_{\ell'}))B(\bell')\right],
\ee
where this equation has the interpretation of a rotation of E and B to
ellipticity in Fourier space, a convolution with the window/weight
function and then a rotation back to E and B. We now wish to compute
the effect that the weight map has on the E-mode power. 
In Fourier space the auto and cross power are defined as 
\be
\langle
X_i(\bell)X^*_j(\ell')\rangle=(2\pi)^2C^{X_iX_j}_{\ell}\delta^D(\bell-\bell')
\ee
where isotropy of the field is assumed. 
This means that an unbiased estimator can be written in the flat sky limit as an
average over angle in $\ell$-space
\be
\label{cff}
\langle C^{X_iX_j}_{\ell}\rangle=\int\frac{{\rm
    d}\phi_{\ell}}{(2\pi)}\langle X_i(\bell)X^*_j(\bell')\rangle. 
\ee
Hence by taking the correlation function of equation (\ref{Et}) we can
calculate the estimated power spectrum in the presence of a 
systematic weight map. This follows the calculations of Memari (2010), the resulting expressions for the EE
power and BB power are below, and we include the EB expression for
completeness (however in the flat sky limit there is no EE, BB and EB
mixing; there is between EE and BB though)
\ba
\label{MM}
\langle \widetilde C^{EE}_{\ell}\rangle&=&
\int\frac{{\rm d}^2\ell'}{(2\pi)^2}\left\{\int {\rm d}LL\frac{W_{mm}(L)}{\ell\ell'\sin\eta}
\left([1+\cos4\eta]\langle C^{EE}_{\ell'}\rangle+[1-\cos4\eta]\langle
C^{BB}_{\ell'}\rangle\right)\right\}\nn
\langle \widetilde C^{EB}_{\ell}\rangle&=&
\int\frac{{\rm d}^2\ell'}{(2\pi)^2} \left\{\int {\rm
  d}LL\frac{W_{mm}(L)}{\ell\ell'\sin\eta}2\cos4\eta\langle
C^{EB}_{\ell'}\rangle\right\}\nn
\langle \widetilde C^{BB}_{\ell}\rangle&=&
\int\frac{{\rm d}^2\ell'}{(2\pi)^2} \left\{\int {\rm d}LL\frac{W_{mm}(L)}{\ell\ell'\sin\eta}
\left([1-\cos4\eta]\langle C^{EE}_{\ell'}\rangle+[1+\cos4\eta]\langle
C^{BB}_{\ell'}\rangle\right)\right\},
\ea
where the additional $L$-mode forms a triangle with $\ell$ and
$\ell'$, ($|\ell-\ell'|<L<\ell+\ell'$), with 
$\cos\eta=(\ell^2+\ell'^2-L^2)/2\ell\ell'$ and similarly for
$\sin\eta$ and $W_{mm}$ is the angle-average of the modulus squared of the weight function 
\be
W_{mm}(L)\equiv\int\frac{{\rm d}\phi_{\ell}}{(2\pi)}|W_{m}(\bL)|^2.
\ee
In the discrete case we can write equations (\ref{MM}) in a compact
form using \emph{mixing matrices} such that 
\be 
\left(\begin{array}{c}
 \langle \widetilde C^{EE}_{\ell}\rangle \\
 \langle \widetilde C^{BB}_{\ell}\rangle
\end{array}\right)=
\sum_{\ell'}
\left(\begin{array}{cc}
  M^{EE,mm}_{\ell\ell'} & M^{BB,mm}_{\ell\ell'} \\
  M^{BB,mm}_{\ell\ell'} & M^{EE,mm}_{\ell\ell'}
\end{array}\right)\left(\begin{array}{c}
 \langle  C^{EE}_{\ell}\rangle \\
 \langle  C^{BB}_{\ell}\rangle
\end{array}\right),
\ee
where 
\ba
M^{EE,mm}_{\ell\ell'}&\equiv&\frac{\Delta\ell'\ell'}{(2\pi)^2}\sum_L
\Delta L L W_{mm}(L)\frac{1+\cos4\eta}{\ell\ell'\sin\eta}\nn
M^{BB,mm}_{\ell\ell'}&\equiv&\frac{\Delta\ell'\ell'}{(2\pi)^2}\sum_L
\Delta L L W_{mm}(L)\frac{1-\cos4\eta}{\ell\ell'\sin\eta},
\ea
and similarly for the EB power; $\Delta \ell'$ is the separation
  between the discrete $\ell'$ modes. These expressions assume that the
systematic fields are uncorrelated with the shear and intrinsic
ellipticity fields. This may not be the case in real data (e.g. selection effects
over galaxy populations may have particular biases), but for GREAT10
selection effects are not investigated and the biases are quoted as
averages over populations. We leave a
generalisation of this formalism to correlated systematic-ellipticity fields for future
work.  

Using this we can write a power spectrum estimate of the
quantities in equation (\ref{genmc}) (we drop the angle brackets over
$\phi_{\ell}$ for clarity from here) including the $\gamma$I cross term
\ba
\widetilde
C^{EE}_{\ell}&=&(1+2m_{\ell})[C^{EE,\gamma\gamma}_{\ell}+C^{EE,II}_{\ell}+C^{EE,\gamma I}_{\ell}]+{\mathcal A}^{EE}_{\ell}\nn
&+&\sum_{\ell'}(M^{EE,mm}_{\ell\ell'}[C^{EE,\gamma\gamma}_{\ell'}+C^{EE,II}_{\ell'}+C^{EE,\gamma I}_{\ell'}] + M^{BB,mm}_{\ell\ell'}[C^{BB,\gamma\gamma}_{\ell'}+C^{BB,II}_{\ell'}+C^{BB,\gamma I}_{\ell'}])
\ea
where ${\mathcal A}_{\ell}$ is the angle averaged power spectrum of the $c(\btheta)$
variation; here, through isotropy, is it assumed that that the power contains all relevant
information. This could be generalised to 
include non-isotropic variation in all terms i.e. not taking the
angle averages. $m_{\ell}$ 
is the angle averaged Fourier transform of
$m(\btheta)$. Our notation, for example $C^{EE,AB}_{\ell}$, 
refers to the $EE$ power corresponding to correlations between
quantities $A$ and $B$ as a function of $\ell$. 
We do not include terms from the quadratic $q(\btheta)$
contribution. For GREAT10 the $\gamma$ field is E-mode only and the intrinsic
ellipticity field is B-mode only, with no $\gamma$I term, so we have a simpler expression 
\be
\label{genCl}
\widetilde
C^{EE}_{\ell}=(1+2m_{\ell})C^{EE,\gamma\gamma}_{\ell}+{\mathcal A}^{EE}_{\ell}+
\sum_{\ell'}(M^{EE,mm}_{\ell\ell'}C^{EE,\gamma\gamma}_{\ell'} +
M^{BB,mm}_{\ell\ell'}C^{BB,II}_{\ell'}).
\ee
These expressions are general for a wide class of shape measurement biases, and
are trivially extendable, for example to include cross-terms that may appear in
real data (e.g. $\langle c m\rangle$ cross terms) if required. 

Equation (\ref{genCl}) represents in general how shape
measurement inaccuracies in GREAT10 can propagate through to the shear
power spectrum. In the case that the
weight-map is constant ($m(\btheta)=$constant$=m_0$, and
$c(\btheta)=c_0$ with some associated error $\sigma(c)$) the Fourier transform
becomes a delta-function and the mixing matrices become 
$M^{EE,mm}_{\ell\ell}=I_{N_{\ell}}\times m_0^2$ and $M^{BB,mm}_{\ell\ell'}=0$. This leads to 
\be
\label{genCl3}
\widetilde
C^{EE}_{\ell}=C^{EE,\gamma\gamma}_{\ell}+{\mathcal A}+
{\mathcal M}C^{EE,\gamma\gamma}_{\ell}
\ee
where ${\mathcal M}=2m_0+m_0^2$ and 
${\mathcal A}=\sigma(c)^2$ are constant functions of scale. 
In general the mixing matrices are not only dependent on a single
$\ell$ (i.e. diagonal $M_{\ell\ell}$) except in the case that the systematic is isotropic
or constant. Unfortunately this is likely not to be the case in weak
lensing where for example PSF ellipticity and size is often coherent but not constant 
across a field of view. Massey et al (in prep) will discuss requirements
on these parameters ${\mathcal M}$ and ${\mathcal A}$, and how they
relate to uncertainty in PSF parameters. 

We note that this formalism means that we only need to recover the statistical
properties of the varying $m(\btheta)$ field (the power spectrum and
mixing matrix) in order to propagate its impact through to the shear
power spectrum. In addition, as shown in Appendix B, this formalism can also be
used to generate expressions for correlation coefficients between the
systematic $m(\btheta)$ and $c(\btheta)$ fields and any spatially
varying quantity. Given these definitions and formalism we can now proceed to outline the
metrics used in this paper, taking into account some practicalities
such as pixel noise removal. 

\section*{Appendix B: Description of the Evaluation Metrics}
The variable shear nature of the simulations enables a variety of
metrics to be calculated, each of which allow us to infer different
properties of the shape measurement method under scrutiny. In this
paper we define a variety of metrics that we explain in detail in this Section.

\subsection*{B1. Quality factor}
\begin{figure}
  \centerline{\includegraphics[width=0.5\columnwidth,angle=0,clip=]{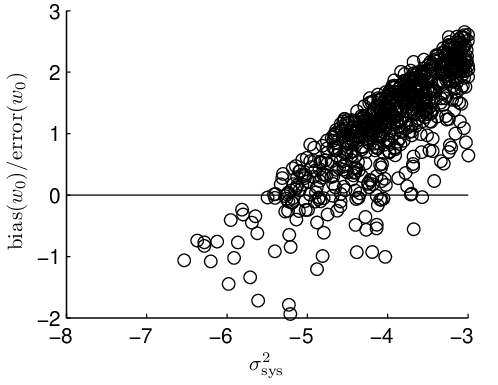}}
 \caption{Monte Carlo realisations of submitted shear power spectrum
   where $\sigma_{\rm sys}^2$ is the denominator in equation (\ref{Q}),
   and the calculated bias in dark energy parameter with respect to
   its error.}
 \label{numerator}
\end{figure}

In general for a variable field we define the power spectrum as the
Fourier transform of the correlation function as described in Appendix
A. We wish to compare the power
reconstructed from the submissions against the true shear power
spectrum and so define a baseline evaluation metric, the quality factor (Q), as 
\be 
\label{Q}
Q=1000\frac{5\times 10^{-6}}{\int \!{\rm d}\!\ln \ell |\widetilde C^{EE}_{\ell}-C^{EE,\gamma\gamma}_{\ell}|\ell^2}.
\ee
The numerator $5\times 10^{-6}$ is calculated by generating Monte Carlo realisations of
a mock submitted power spectra and calculating the bias in the dark energy equation
of state parameter $w_0$ (Linder, 2003) which would occur if such an observation were made (using the
functional form filling formalism described in Kitching et al.,
2008) over a survey of $20,000$ square degrees using the same redshift
distribution as described in Section \ref{Variable shear and intrinsic
  ellipticity fields}. In Figure \ref{numerator} we show the result of this procedure
for GREAT10 (where the numerator in equation \ref{Q} is labelled as $\sigma_{\rm sys}^2$),
where we take a threshold value of bias-to-error ratio of $1$. This is
in fact conservative as shown in Massey et al., (2012, in prep). The
factor of $1000$ normalises the metric such that a good method should
achieve $Q\simeq 1000$. A factor $(1/2\pi)$ could be included in the denominator, but we
absorb this into the factor $5\times 10^{-6}$.
This was the quality factor used in the online leaderboard during the
challenge. 

\subsection*{B2. Pixel noise corrected quality factor}
In general we can express the measured total ellipticity by including a
noise term in equation (\ref{genmc}), where $e_{\rm n}$
is some inaccuracy in this estimator due to stochastic terms in shape
measurement method, or due to pixel noise in the images (finite
signal-to-noise). In the simulations, for ellipticity catalogue
submissions, we averaged over $N_{\rm realisation}$ realisations of
the noise. In this averaging the mean of the noise contribution is assumed to be zero
$\langle e_{\rm n}\rangle=0$ over realisations, but where there is an error on this
mean that remains. By propagating this through to the power spectrum we recover 
\be 
\widetilde C^{EE}_{\ell}\rightarrow\widetilde C^{EE}_{\ell}+\frac{\sigma^2_{\rm n}}{N_{\rm realisation}N_{\rm object}}
\ee
where the noise term is white noise (constant over all scales) with a variance
$\sigma^2_{\rm n}$, which is a sum of the $e_1$ and $e_2$ components. 
The noise term is now averaged over the number of realisations
and the number of objects. For values of $N_{\rm realisation}=200$ and
$N_{\rm object}=10^4$ the expected fractional contribution to the
measured power $\sigma^2_{\rm n}/(N_{\rm realisation}N_{\rm
  object}\langle C_{\ell,{\rm estimated}}\rangle)\approx(\sigma/0.05)^2$. 

The measured power spectra inferred from the ellipticity catalogue
submissions and used in the quality factor ($Q$) defined in equation
(\ref{Q}), therefore includes this noise term. However for an error induced
by noise on ellipticity estimates of $\sigma\ls 0.05$ the impact on
the metric should be subdominant. It is commonly assumed
that such noise terms could be removed in real data (this is trivial
for correlation functions, but is more complex for power spectrum
estimates; that require an estimate of $\sigma_n$ from data -- the full
covariance of the shear estimators, see also e.g. Schneider et al. 2010), and some power
spectrum submissions (see Section \ref{Description of the Methods})
did employ techniques to remove this term from the submitted power
spectrum. Hence we here introduce a quality factor that accounts for
this noise term 
\be 
\label{Qdn}
Q_{\rm dn}=1000\frac{5\times 10^{-6}}{\int {\rm d}\ln l |\widetilde C^{EE}_{\ell}-C^{EE,\gamma\gamma}_{\ell}-
\frac{\langle\sigma^2_{\rm n}\rangle}{N_{\rm realisation}N_{\rm object}}|l^2}
\ee
where $\langle\sigma^2_{\rm n}\rangle$ is an estimated value of the
pixel noise term from the ellipticity catalogue submissions. 

To estimate the value of $\langle\sigma^2_{\rm n}\rangle$ from the
simulations we have to separate the E-mode shear field from the B-mode
only intrinsic ellipticity field, otherwise the variance of the
ellipticities from a submitted entry will be dominated by the variance
of the intrinsic ellipticities. This is done using the rotations described in
Appendix A, here we describe this pedagogically (we also use explicit
Cartesian coordinates $\btheta=(x,y)$ and $\bell=(\ell_x,\ell_y)$ for clarity). We
make a 2D discrete Fourier transform of the submitted ellipticity values such that 
\be 
\label{f1}
\epsilon_{\rm measure}(\ell_x,\ell_y)={\rm FT}[e_{\rm measure}(x,y)]
\ee
where here the measured ellipticity is averaged over all noise
realisations before transformation. We then rotate this field such that  
\be 
\epsilon_{\rm
  rot,measure}(\ell_x,\ell_y)=(\ell^*\ell^*/|\ell^2|)\epsilon_{\rm
  measure}(\ell_x,\ell_y)
\ee
and then inverse Fourier transform to real space 
\be
\label{f3}
\kappa(x,y)+{\rm i}\beta(x,y)= {\rm iFT}[\epsilon_{\rm rot,measure}(\ell_x,\ell_y)]
\ee
where we now have a $\kappa(x,y)$ field which contains E-mode
power only and a $\beta(x,y)$ field that contains B-mode power only. 
The simulations have been set up such that the intrinsic
ellipticity field has B-mode power only, such that we can now take the
$\kappa(x,y)$ map and generate an E-mode only ellipticity catalogue that
should only contain the estimated shear values and the noise term only  
\be 
\label{ishea}
\kappa(x,y)\rightarrow e_{\rm E, measure}(x,y)\approx\hat\gamma(x,y)+e_{\rm n}(x,y),
\ee 
where $\hat\gamma$ is the estimate shear for each position (object) in
field. We do this by following the inverse steps of transformations from
equations (\ref{f1}) to (\ref{f3}), and assume noise is equally
distributed between E and B modes. The expression is only approximate
because of position dependent biases (see Appendix A and next Section), that can mix E
and B modes, but for the majority of methods presented in this paper
this affect seems to be subdominant.
By taking the normal variance of $e_{\rm E, measure}(x,y)$ we find
that 
\be 
\sigma^2_{\rm E, measure}=\sigma^2_{\gamma}+\sigma^2_{\rm n}
\ee
and so our estimate of the noise variance is 
\be 
\sigma^2_{\rm n}=\sigma^2_{\rm E, measure}-\sigma^2_{\gamma}.
\ee
To calculate this we use the true shear values to find 
$\sigma^2_{\gamma}$, this is unrealistic but note that the true individual shear values
are not used directly only to calculate the variance. For
real data, as as done by `fit2-unfold' we expect that noise
estimates from each galaxy will be used to calculate this
correction. Indeed part of the challenge, demonstrable by the
`fit2-unfold' submissions, was to develop optimal estimates for $\sigma^2_{\rm n}$.

\begin{figure}
  \centerline{\includegraphics[width=0.5\columnwidth,angle=0,clip=]{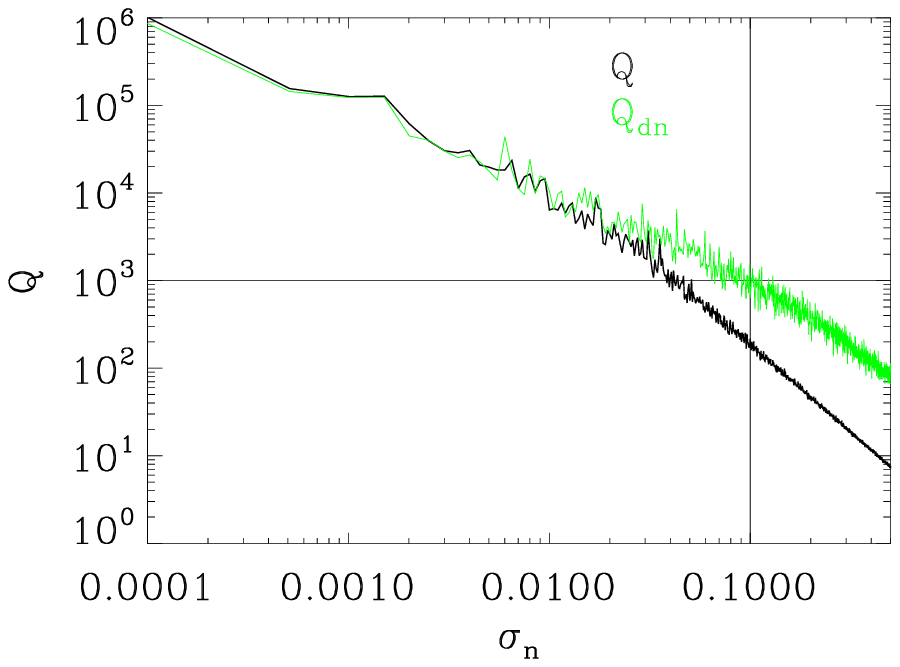}}
 \caption{A simulation of the effect on $Q$ (black line) and $Q_{\rm
     dn}$ (green line) as the noise in a mock submission (containing
   only noise and the true shear values) increases. Lines at $Q=1000$
   and $\sigma_n=0.1$ are to guide the reader.}
 \label{noisefig}
\end{figure}

To test that such a correction works we simulated a submission by
taking the true shear values and adding random normally distributed
numbers to each of the $10$,$000\times 200\times 24$ shear values. We show
results in Figure \ref{noisefig}. We find as expected that as the
noise increases the value of $Q$ (equation \ref{Q}) decreases, but
that including the noise correction (equation \ref{Qdn}) increases the
value. Note that due to the finite size of the simulations any
estimation of $\sigma^2_{\rm n}$ is itself noisy which means the
corrected value of $Q_{\rm dn}<\infty$ even in this ideal case.

\subsection*{B3. One-point estimator shear relations}

As well as metrics that integrate over the measured power spectra we
can also investigate a number of metrics that encapsulate a 
relation between the measured and true shears for individual 
objects. This ties the quality factor metrics to the STEP (Heymans et
al., 2006) $m$ and $c$ values where 
\be 
\hat\gamma_i = (1+m_{ij})\gamma^t_j + c_i 
\ee
where $\gamma^t_i$ is the true shear and $\hat\gamma_i$ is the measured
shear for each component, this is a simplification of
equation (\ref{genmc}), and that used for all constant shear
simulations (with no position dependence). We also add a quadratic
non-linear term to this relation ($q^{1/2}_{ij}\gamma_j|\gamma|_kq^{1/2}_{ki}$)
\be 
\hat\gamma_i = (1+m_{ij})\gamma^t_j + c_i + q^{1/2}_{ij}\gamma_j|\gamma|_kq^{1/2}_{ki}
\ee
that contains $\gamma|\gamma|$, not $\gamma^2$, since we may expect divergent behaviour to more
positive and more negative shear values for each domain
respectively. In general $m_{ij}$ and $q_{ij}$ could be non-diagonal
matrices, however in this paper we assume that they are diagonal and
take an average over the two shear components to give 
\be 
\label{mcq}
\hat\gamma = (1+m)\gamma^t + c + q\gamma|\gamma|
\ee 
where all quantities are averaged over $\gamma_1$ and $\gamma_2$.

In a variable shear simulation
calculating $m$, $c$ and $q$ by regressing $e_{\rm measure}$ and 
$(\gamma+e_{\rm intrinsic})$ would result in a noisy estimator
dominated by intrinsic ellipticity noise. However we 
can calculate $m$, $c$ and $q$ directly by finding the estimated shear
for each galaxy individually, removing the intrinsic ellipticity
contribution (equation \ref{ishea}). This is for
every galaxy a noisy estimate of the shear, we then average these
estimates over bins in $\gamma^t$. This enables the $m$, $c$ and $q$ parameters
to be recovered, and in fact the variable field simulations allows for
a flexible binning as a function of any other spatially varying
quantity (see Appendix E), and an exact removal of shape noise (through the B-mode
intrinsic power). This method of calculating the $m$, $c$ and $q$ 
parameters is a one-point estimate of the shape measurement biases
and makes no assumption about spatially correlated effects. 

\subsection*{B4. Power spectrum relations}
As described in Appendix A we can write an expression for the
estimated power using two linear parameters ${\mathcal M}$
and ${\mathcal A}$, taking into account the pixel noise removal we
have a similar expression 
\be
\label{genCl2}
\left[C^{EE}_{\ell}-C^{EE,\gamma\gamma}_{\ell}-\frac{\langle\sigma^2_{\rm n}\rangle}{N_{\rm realisation}N_{\rm object}}\right]=
{\mathcal M}C^{EE,\gamma\gamma}_{\ell}+{\mathcal A}. 
\ee
This can be related to the $m$ and $c$ parameters 
\ba 
\label{MMA}
{\mathcal M}&\simeq&m^2+2m\approx 2m\nn
{\mathcal A}&\simeq&\sigma(c)^2 
\ea
where $\sigma(c)$ is the variance of the $c$ parameter, 
but only approximately because of the assumption of some form of
spatial variation (constant in this case). 
\begin{figure}
  \centerline{\includegraphics[width=0.7\columnwidth,angle=0,clip=]{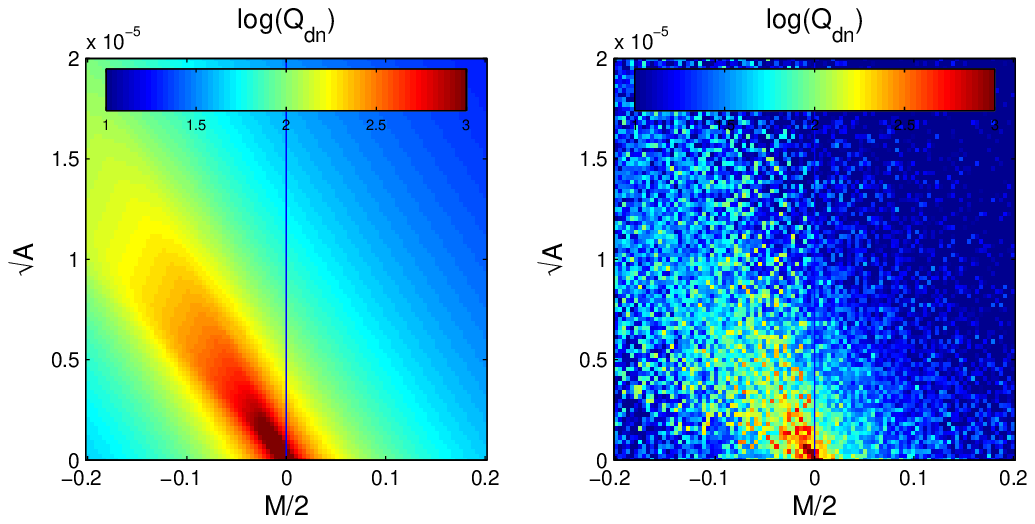}}
  \centerline{\includegraphics[width=0.7\columnwidth,angle=0,clip=]{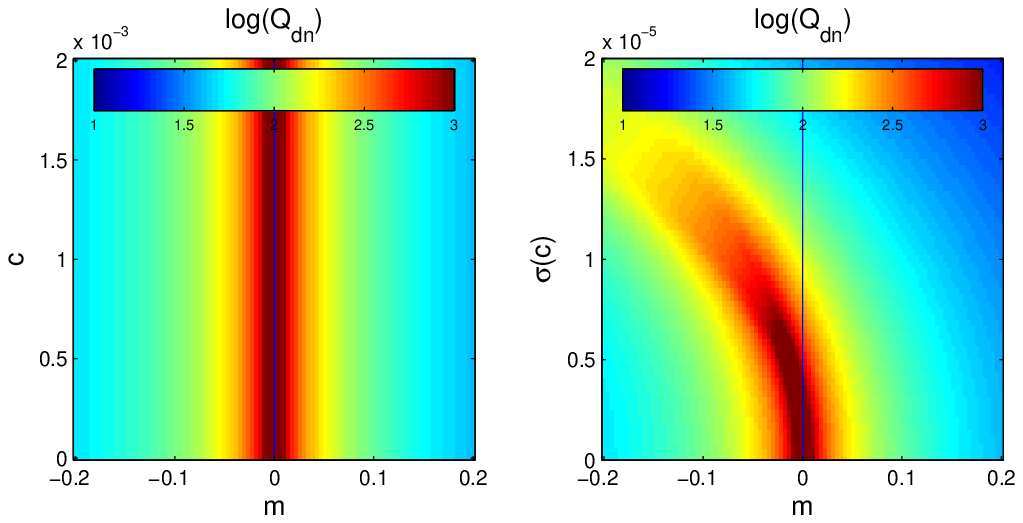}}
 \caption{An exploration of the (${\mathcal M}$, ${\mathcal A}$), ($m$, $c$) and ($m$, $\sigma(c)$) planes, 
   where at each point the quality factor is calculated using a noise free fiducial power spectrum. The colour scale shows the 
   logarithm of the quality factor. This can be compared to Figure \ref{avmc}.}
 \label{QMAmc}
\end{figure}

In Figure \ref{QMAmc} we show how the $Q_{\rm dn}$, ${\mathcal M}$, ${\mathcal A}$ and the point estimators $m$ and $c$ are
related. To create this we explore the (${\mathcal M}$, ${\mathcal A}$) plane 
and using the fiducial power spectrum calculate $Q_{\rm dn}$ 
for each value. We also show a realisation where random components have been added, 
${\mathcal M}(1+R)$ where $R$ is a uniform random number and similarly for ${\mathcal A}$, at each point in parameter 
space to simulate a more realistic submission. We find that there is degenerate line in $Q_{\rm dn}$ 
where an offset ${\mathcal A}$ can be partially 
cancelled by a negative ${\mathcal M}$ yielding the same $Q_{\rm dn}$, and a more straightforward relation for 
${\mathcal M}\geq 0$. As expected the $c$ parameter does not impact
the quality factor but the variance of $c$ does. There  
is a similar degeneracy between $m$, $\sigma(c)$ and $Q_{\rm dn}$ as
with the linear power spectrum parameters, this is as  
expected as in equation (\ref{MMA}), except that for large
negative $m$ the quadratic $m^2$ term begins to become  
important. 

\subsection*{B5. Correlations with spatially varying quantities}

To relax the assumption of constant $m$ and $c$ in power spectrum
analysis we can assume that
each of these is correlated with some spatially varying parameter $X(\btheta)$
\ba 
m(\btheta)&=&m_0+\alpha \left[\frac{X(\btheta)}{X_0}\right]\nn
c(\btheta)&=&c_0+\beta \left[\frac{X(\btheta)}{X_0}\right]
\ea
with correlation coefficients $\alpha$ and $\beta$; $X_0$
is constant reference value to ensure that the units of $\alpha$ and $\beta$
are dimensionless: for ellipticity this is set to unity for PSF size squared
this is the mean PSF size squared. This is a simple
relation and could be made significantly more complex. 

We explain in a correlation function notation how these propagate
through, for pedagogical purposes, but for the full treatment one
should refer to the pseudo-Cl methodology that we present in Appendix
A. A simple correlation function approximation of the measured shear can be written
\ba
\langle\langle e_{\rm measure}\rangle_n\langle e_{\rm measure}\rangle^*_n\rangle&=&
(\alpha/X_0)^2\langle X X^*\rangle[\langle \gamma\gamma^*\rangle+\langle \langle
  e_{\rm intrinsic}\rangle_n\langle e_{\rm
    intrinsic}\rangle^*_n\rangle]\nn
&+&(2(1+m_0)(\alpha/X_0)\langle X\rangle+(1+m_0)^2)[\langle\gamma\gamma^*\rangle+\langle \langle
  e_{\rm intrinsic}\rangle_n\langle e_{\rm
    intrinsic}\rangle^*_n\rangle]+(\beta/X_0)^2\langle XX^*\rangle\nn
\ea 
not including the pixel noise term.
We can also take the cross correlation between the measured
ellipticity and these quantities 
\ba 
\langle\langle e_{\rm measure}\rangle_n X^*\rangle&=&
\left\langle \left((1+m_0+(\alpha/X_0)X)(\gamma+\langle e_{\rm
  intrinsic}\rangle_n) + c_0+(\beta/X_0)X\right)X^*\right\rangle\nn
&=&
(1+m_0)\langle (\gamma+\langle e_{\rm intrinsic}\rangle_n)X^*\rangle +
(\alpha/X_0) \langle X (\gamma+\langle e_{\rm intrinsic}\rangle_n)X^*\rangle
+ (\beta/X_0)\langle X X^*\rangle + c_0\langle X^*\rangle\nn
&\approx& 
(1+m_0)\langle (\gamma+\langle e_{\rm intrinsic}\rangle_n)X^*\rangle +(\beta/X_0)\langle X X^*\rangle+ c_0\langle X^*\rangle
\ea
which results in an expression that is not dependent on $\alpha$ and assuming that third
order correlations and noise-$X$ correlations are zero. 

The corresponding full expressions for the pseudo-Cl power spectrum, including
the noise correction term (which we assume is uncorrelated with all
other terms) are  
\ba 
\left[\widetilde C^{EE}_{\ell}-C^{EE,\gamma\gamma}_{\ell}-\frac{\langle\sigma^2_{\rm n}\rangle}{N_{\rm
      realisation}N_{\rm object}}\right]&=&(m_{\ell}^2+2m_{\ell})C^{EE,\gamma\gamma}_{\ell}+
(\alpha/X_0)^2
\sum_{\ell'}[M^{EE,XX}_{\ell\ell'}C^{EE,\gamma\gamma}_{\ell'}+M^{BB,XX}_{\ell\ell'}C^{BB,II}_{\ell'}]\nn
&+&(\alpha/X_0) (1+m_{\ell})\langle X\rangle C^{EE,\gamma\gamma}_{\ell}+(\beta/X_0)^2C^{XX}_{\ell}\nn
\left[\widetilde C^{EX}_{\ell}-C^{\gamma
    X}_{\ell}-C^{IX}_{\ell}\right]&=&m_{\ell}(C^{\gamma X}_{\ell}+C^{IX}_{\ell})+(\beta/X_0) C^{XX}_{\ell}+c_0\langle X_{\ell}\rangle.
\ea
The second expression has cross-power
spectra on the both sides. The matrices $M^{XX}$ are the mixing
matrices for the spatially varying quantity $X$. In general the
variation of $X$ is not isotropic -- PSF ellipticity for example can have a
preferred direction in an image -- however here we make the assumption of
isotropy in defining the power $C^{XX}_{\ell}$. 

To calculate these from the simulations we find the best fitting $\alpha$
and $\beta$ values (using a minimum least squares estimator over the
$\ell$ range defined in Section \ref{Description of the Methods}) for
$X=$PSF size squared and PSF ellipticity. Because
this calculation is done on sets that are averaged over noise
realisations this can only be calculated for those sets in which
the PSF is fixed for a set (for the PSF correlations). 

The relation to the linear power relations
${\mathcal M}$ and ${\mathcal A}$ is not
straightforward because of the non-diagonal mixing matrix in general. 
Therefore in the results Sections (Section \ref{Results}) we will quote values for these correlation
coefficients $\alpha_e$, $\alpha_{R^2}$, $\beta_e$, $\beta_{R^2}$ for
ellipticity and PSF size squared (the square of the size is the most relevant
quantity for propagated PSF-shear behaviour, see Massey et al. 2012,
in prep and Paulin-Henriksson et al. 2008). Note that $\alpha$ and
$\beta$ are unitless and scaled by a reference value $X_0=[\langle X\rangle]$: for PSF size
correlations this means units of $X_0=3.4^2=11.56$ pixel$^2$, and for
ellipticity correlations the quantities are unitless $X_0=1$. If one
were to expand the bias in terms of a different scaling, a natural
expansion one may use for example is as a function of $R_{\rm
  PSF}/R_{\rm galaxy}$, then a scaling can be applied to results
presented in this paper.

\section*{Appendix C : Simulation Modelling}

In this Section we provide some further details of the variable shear and PSF field, as
well as the local modelling of the galaxies and stars. 

\subsection*{C1. Scaling of the shear field}
We note that in performing the process of sampling the shear field 
discretely and then generating a postage stamp for each sampling 
the inter-postage stamp
separation in the final image has a distance of $\theta_{\rm
  image}/100$ but this is not necessarily related to the pixel scale of the
postage stamps i.e. $\theta_{\rm pixel}\times
48\times100\not=\theta_{\rm image}$ in general. As a result the
number density of the galaxies can be scaled as 
\be 
\frac{n_0}{\rm square \,\,arcmin}=\frac{10^4}{3600\theta^2_{\rm image}}=\frac{2.77}{\theta^2_{\rm
    image}}  
\ee
and the maximum $\ell$ set by the grid-separation of the galaxies
scales as
\be 
\ell_{\rm max}=0.5\frac{2}{\theta^2_{\rm image}/180/100}=\frac{18,000}{\theta_{\rm
    image}}
\ee
where $100$ is the number of grid positions on a side. 
But note that the true underlying simulated shear field is always fully sampled in every case.

For the case of $\theta_{\rm image}=10$ degrees this gives
values of $n_0=0.0277$ and $\ell_{\rm max}=1800$. The images however
can be scaled to match a variety of other configurations, with the caveat that the
absolute value of the shear power is constant, $\theta_{\rm image}=1$
degrees gives a scaling of $n_0=2.77$ and $\ell_{\rm max}=18$,$000$,
and $\theta_{\rm image}=0.5$ degrees gives a scaling of $n_0=11.1$ and
$\ell_{\rm max}=36$,$000$. In each case the absolute amplitude of the calculated shear power
also needs to be scaled. It is fair to then match the simulations to
either of these cases, which span a reasonable expected dynamical range in
number density of objects but with a coupled increase in the maximum
$\ell$-range. The $\ell$ values used for the Q metrics are 
$\ell=(233,415,600,789,977,1162,1350,1538)$, 
these are specified by: i) defining the maximum and minimum
$\ell$-modes, we do not generate $\ell$ modes
above that corresponding to the grid the separation, and avoid the
smallest $\ell$-modes where the 
signal-to-noise is low; ii) choose $8$ bins linearly spaced in $\ell$
between these limits; iii) define a grid in ($\ell_x$,$\ell_y$) for
the power spectrum calculation, defined with
$\Delta\ell=36$; iv) integrate over this grid and take the mean $\ell$
value from the grid points in each of the $8$ $\ell$-bins. The bins
were originally defined
under the assumption that an equivalent accuracy of $Q\gs 1000$ in
each $\ell$-bin independently is desirable; see Figure \ref{noisefig}
where given the size
of the simulation ($200$ noise realisations), and assuming that
$\sigma_n\sim 0.01$ for a good method, we find 
$Q\sim 1000$ at $\sigma_n=0.01$x$\sqrt{(200/8)}=0.05$; although this
is only an estimated number for any given method. 
$8$ $\ell$-bins were also defined for computational speed. We caution
here that accuracy statements will be dependent on the maximum and
minimum $\ell$ ranges, and on the shape of the power spectrum in
general.

We could replace the integrals in the Q factor definitions with
sums for the discrete $\ell$ case where $\int {\rm d}\ell
\rightarrow \sum_{\ell=(233,415,600,789,977,1162,1350,1538)} \Delta
\ell$ but we keep the integral version in the text to maintain a
general expression and for clarity. The power $C^{EE}\ell^2$ is binned, and compared
to the binned equivalent of the true/input power spectrum -- the power spectrum of the
actual realization of the shear field -- calculated in exactly the same
way as the submitted power (one may refer to this as the ``sample'' input power spectrum).

\subsubsection*{$\ell$ integration}
Here we briefly discuss a technical issue with regard to the $\ell$
integral accuracy used for the $Q$ factor calculation. 
The Q value is defined via
\begin{equation}
\label{ll1}
\frac{Q_{N}}{Q} = \int_{\log\ell_{\rm min}}^{\log\ell_{\rm max}} d(\log\ell)\, f(\ell) = \int_{\ell_{\rm min}}^{\ell_{\rm max}} \frac{d\ell}{\ell} \, f(\ell)
\end{equation}
with $Q_N = 0.005$ and
\begin{equation}
f(\ell) \equiv |\widetilde C_{\ell}^{EE} - C_{\ell}^{EE,\gamma\gamma}| \ell^2 \; .
\end{equation}
We can write equation (\ref{ll1}) without any approximations as
\begin{equation}
\frac{Q_{N}}{Q} = \sum_{i=1}^{N_{\rm bins}} I_i \,\,\,\,\,\,\, {\rm
  with} \,\,\,\, I_i \equiv \int_{\ell_{i-1}}^{\ell_i} \frac{d\ell}{\ell} \, f(\ell) \; .
\end{equation}
For concreteness, we assume equally spaced bins that are linear in $\ell$: $\ell_i \equiv \ell_{min} + i \Delta\ell$
with $i = 1,2,\ldots, N_{\rm bins}$ and $\Delta\ell = (\ell_{\rm max}-\ell_{\rm min})/N_{\rm bins}$.
We calculate the integral over the difference in the power using Monte Carlo integration of the average
value of $\ell^2 C_{\ell}^{EE,\gamma\gamma}$ for $\ell_{i-1} < \ell
\le \ell_i$ based on the ellipticities associated with a single
realization of $\widetilde C_{\ell}^{EE}$, and similarly for $\ell^2
\widetilde C_{\ell}^{EE}$. Therefore, we have a quantity that is
related to $I_i$ that can be written 
\begin{equation}
\widetilde I_i \simeq \frac{1}{\Delta\ell}\,\int_{\ell_{i-1}}^{\ell_{i}} d\ell\, f(\ell) \; .
\end{equation}
Working to second order in $\Delta\ell$ to evaluate different schemes
for estimating the value of equation (\ref{ll1}) we have:
\begin{equation}
f_i(\ell) = f_{i-1/2} + f'_{i-1/2} (\ell - \ell_{i-1/2}) + \frac{1}{2} f''_{i-1/2} (\ell - \ell_{i-1/2})^2 + {\cal O}(\Delta\ell)^3
\end{equation}
with
\begin{equation}
\ell_{i-1/2} \equiv \ell_i - \frac{\Delta\ell}{2} \quad, \quad f_{i-1/2} \equiv f_{i-1/2}(\ell_{i-1/2}) \quad , \quad \mathrm{etc}\ldots
\end{equation}
then
\begin{equation}
I_i = \frac{f_{i-1/2}}{\ell_{i-1/2}}+ \frac{\Delta\ell^2}{\ell_{i-1/2}^3}\,\left[
f_{i-1/2} - f'_{i-1/2} \ell_{i-1/2} + \frac{1}{2} f''_{i-1/2} \ell_{i-1/2}^2 \right]
 +  {\cal O}(\Delta\ell)^3 \; .
\end{equation}
and
\begin{equation}
\widetilde I_i \simeq f_{i-1/2} + \frac{\Delta\ell^2}{24} f''_{i-1/2} +  {\cal O}(\Delta\ell)^3 \; .
\end{equation}
We are now in a position to calculate the numerical approximation
errors inherent in different schemes for combining values of $\widetilde I_i$ to estimate the value of equation (\ref{ll1}). 
\\

\noindent {\bf Linear scheme}: A straightforward implementation of the
integration over $\ell$ in equation (\ref{ll1}) in terms of a finite
sum yields $1/\ell_{i-1/2}$ weights and is accurate to second order: 
\begin{equation}
\label{ll2}
\sum_{i=1}^{N_{\rm bins}} \frac{1}{\ell_{i-1/2}}\,\widetilde I_i \simeq \frac{Q_{N}}{Q} + \frac{\Delta\ell^2}{12}
\sum_{i=1}^{N_{\rm bins}} \frac{f'_{i-1/2} \ell_{i-1/2} - f_{i-1/2}}{\ell_{i-1/2}^3} \; .
\end{equation}

\noindent {\bf Log scheme}: We can also implement the integration over
$\log\ell$ (first equality in  equation, \ref{ll1}) as a
straightforward finite sum approximation, 
which implies $\log(\ell_i/\ell_{i-1})/\Delta\ell$ weights and is also
formally accurate to second order: 
\begin{equation}
\label{ll3}
\sum_{i=1}^{N_{\rm bins}} \log(\ell_i/\ell_{i-1})/\Delta\ell\,\widetilde I_i  \simeq \frac{Q_{N}}{Q} + \frac{\Delta\ell^2}{12}
\sum_{i=1}^{N_{\rm bins}} \frac{f'_{i-1/2}}{\ell_{i-1/2}^2} \; .
\end{equation}

Comparing the two schemes above, both are accurate to second-order
(there are further scheme that are only accurate to first
order). In order to compare the two methods, we need to assume
something about how the error in each bin $\widetilde I_i \simeq
f_{i-1/2}$ grows with $\ell$, and then compare
\begin{equation}
\frac{f'_{i-1/2} \ell_{i-1/2} - f_{i-1/2}}{\ell_{i-1/2}^3} \,\,\,\,\,\,\,
{\rm with} \,\,\,\,
\frac{f'_{i-1/2}}{\ell_{i-1/2}^2} \; .
\end{equation}
Suppose that the leading term in the Taylor expansion of $f(\ell)$ is
$c \ell^n$, then we can calculate the leading behavior for the ratio
of equations (\ref{ll2}) and (\ref{ll3}) explicitly as 
\begin{equation}
\frac{f'_{i-1/2} \ell_{i-1/2} - f_{i-1/2}}{\ell_{i-1/2}^3} \cdot \frac{\ell_{i-1/2}^2}{f'_{i-1/2}} = \frac{n-1}{n} \; .
\end{equation}
Therefore, we conclude that the linear scheme is generally more
accurate, and that the log scheme is only competitive in the unlikely
scenario that $f(\ell)$ depends very strongly on $\ell$. Since we find
empericilly that $f(\ell) \propto \ell^2$ (i.e., that
$|\tilde{C}_{\ell}^{EE} - C_{\ell}^{EE,\gamma\gamma}|$ is
approximately constant over bins), $n=2$ is a good approximation and
the linear scheme is then roughly twice as accurate than the log
scheme. 

\subsection*{C2. The galaxy models} 

Here we describe how the individual galaxies are modelled. Each galaxy
is composed of a bulge and a disk defined as radial
intensity profiles with 
\ba 
I(r)&=&I_{i}{\rm exp}\left[{-\left(K\frac{r}{r_i}\right)^{1/n}}\right]\nn
\ea
where $K=2n-0.331$ with $n=4$ for the bulge and $n=1$ for the disks
and $i=\{b,d\}$ for bulge and disk.
Both are Sersic profiles (the second simply a
exponential). The intensity is normalised to match the signal-to-noise
and the scale radii for the disk and bulge, $r_d$ and $r_b$
respectively, are in general free parameters, 
fiducial values these were set to be $r_b=2.3$ and $r_d=4.8$ pixels. 
In Bridle et al. (2010), and for the
code used for this challenge, the value of radii $r$ are the
half-light radius for both bulges and disks. 
The disk exponential scale length and half-light scale radii differ by that factor $1.669$.

In most
sets the size distribution over objects was a compact Gaussian, with a
variance 
of $\sigma_R=0.01$
\be 
p(r)\propto {\rm exp}\left[-\frac{(r-r_b)^2}{2\sigma_R^2}\right].
\ee
and similarly for the disk distribution. 
In three sets (see Section \ref{Simulation Structure}) the galaxy size 
varied for each galaxy in the set, in this case the functional 
form for the signal-to-noise variation was a Rayleigh distribution 
\be 
P(r)\propto \frac{r}{\sigma_R^2}{\rm exp}\left[-\frac{(r-r_b)^2}{\sigma_R^2}\right],
\ee
where $\sigma_R=2.0$ for these sets, and the $r_b$ and $r_d$ are the
fiducial values. There is a caveat that the sizes referred to here (and in the GREAT08 
simulations) refer to the \emph{pre-sheared} radii of the objects, as such there is 
a ellipticity-size correlation that was present in the simulations.

The bulge
and disk in general can be mis-centered, however in all but
two sets the bulge and disk profiles were co-centered. Object 
positions were centered in each postage stamp with a Gaussian error
position with a standard deviation of $0.5$ pixels. This means that the
distribution of centroids is not uniform across
pixels but (unrealistically) clustered symmetrically towards the
center; this is one of the simplifying aspects of GREAT10 designed to
militate against biases causes by centroiding errors in methods.
 
The bulge-to-disk fraction was 50\% for the majority of sets i.e. the
flux in the bulge and disk was equal. In those sets in which this
varied we used a uniform distribution of bulge-to-disk ratios over the range
$b/d=[0.3,0.95]$, to avoid very low and very high fractions. 

\begin{figure}
  \centerline{\includegraphics[width=0.6\columnwidth,angle=0,clip=]{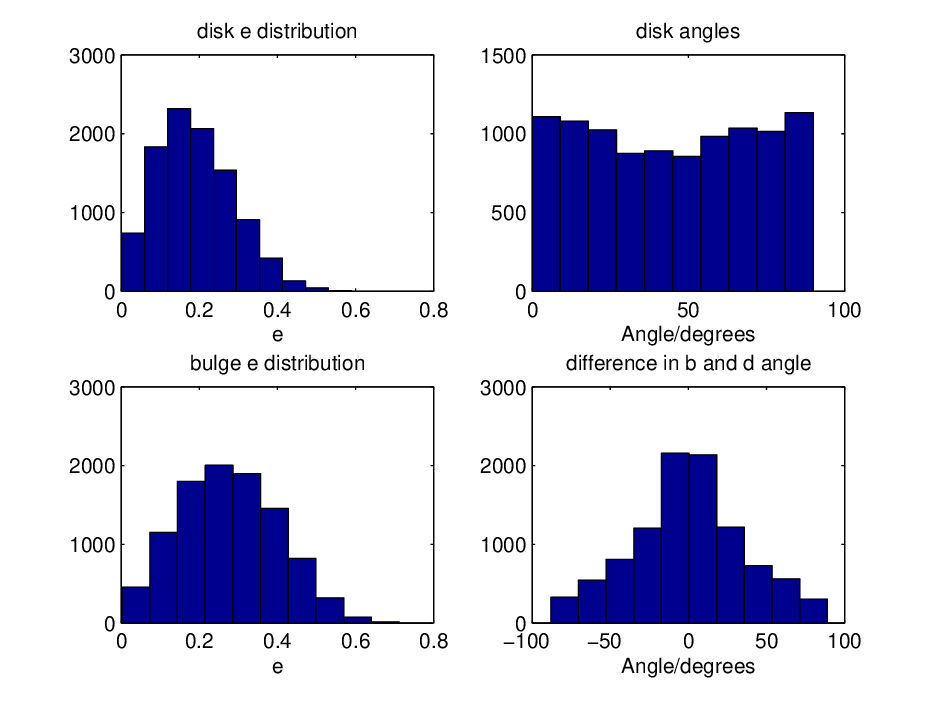}}
 \caption{The distributions of bulge and disk ellipticities for a
   typical image within the fiducial set. Left panels show the
   distribution of ellipticities for bulge and disk. The top right
   panel shows the uniform distribution of disk position angles, and
   the bottom right panel shows the difference between the bulge and
   disk positions angles.}
 \label{pefig}
\end{figure}

The bulge and disk components of
the galaxies in the simulations had different intrinsic ellipticity
distributions, each described by 
\be 
\label{pe}
P_i(e)=e\cos\left(\frac{\pi e}{2}\right){\rm
  exp}\left[{-2\left(\frac{e}{B_i}\right)^C_i}\right]
\ee
where $B=0.09$ and $C=0.577$ for the bulges and $B=0.19$ and $C=0.702$
for the disks (these values are taken from the APM survey, Crittenden
et al. 2001). To remove any very highly elliptical galaxies from the
sample we truncated this distribution at $e=0.8$. This model was
slightly more complex than in Bridle et al. (2010) by allowing for
non-coelliptical profiles (i.e. the bulge and disk were allowed to
have different ellipticities). This was done so that the ellipticity
distributions in equation (\ref{pe}) were conserved. As an example we
show the distribution of the disk and bulge angles in Figure \ref{pefig}.

The signal-to-noise was implemented by calculating the noise-free model
flux by integrating over the galaxy model and then adding a constant
Gaussian noise with a variance of unity and rescaling the galaxy model
to yield the correct signal-to-noise. The signal-to-noise was
scaled to match the default SExtractor (Bertin \& Arnouts 1996) {\tt
  flux\_auto/flux\_err\_auto} parameter combination. 
The galaxy signal-to-noise distribution was a compact Gaussian in the majority of
sets, with a variance of $\sigma_S=0.1$, centered on $(S/N)_i=20$ for
the fiducial set
\be 
p(S/N)\propto {\rm exp}\left[-\frac{(S/N-(S/N)_i)^2}{2\sigma_S^2}\right].
\ee
In three sets (see Section \ref{Simulation Structure}) the signal to
noise varied for each galaxy in the set with a functional 
form for the signal-to-noise variation that was a Rayleigh distribution 
\be 
P(S/N)\propto \frac{S/N}{\sigma_S^2}{\rm exp}\left[-\frac{(S/N-(S/N)_i)^2}{\sigma_S^2}\right],
\ee
where $(S/N)_i=20$ and $\sigma_S=5.0$ for these sets. 

\subsection*{C3. The PSF models} 
\label{The PSF Models} 
The PSF model consisted of a static component that modelled the local
PSF functional form and a spatially varying kernel that mapped the
parameters of this local model across the image plane. The local
functional form was a Moffat profile 
\be
I(r)=\left[1+\left(\frac{r}{r_d}\right)^2\right]^{-\beta},
\ee 
the scale radius $r_d$ was a variable quantity across each image,
related to the FWHM, the
power $\beta=3$ for all images. After generating a circular PSF, it
was made into an elliptical shape by distortion using the shear matrix
given in Kitching et al. (2011) such that there were three parameters
which locally describe the PSF $(r_d,e_1,e_2)$. 
Where similarly to the galaxies the size was the \emph{pre-sheared} size of the 
PSF.

\begin{figure}
  \centerline{\includegraphics[width=0.6\columnwidth,angle=0,clip=]{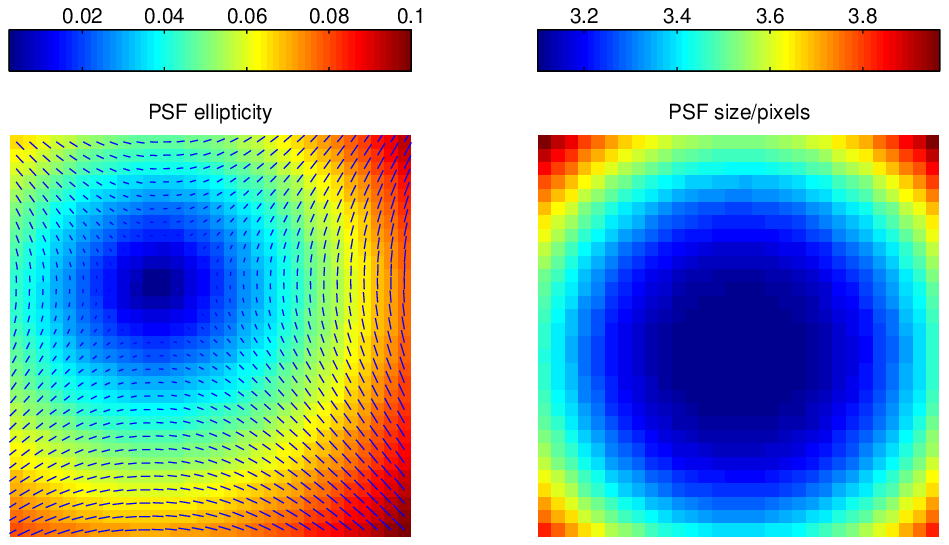}}
  \centerline{\includegraphics[width=0.6\columnwidth,angle=0,clip=]{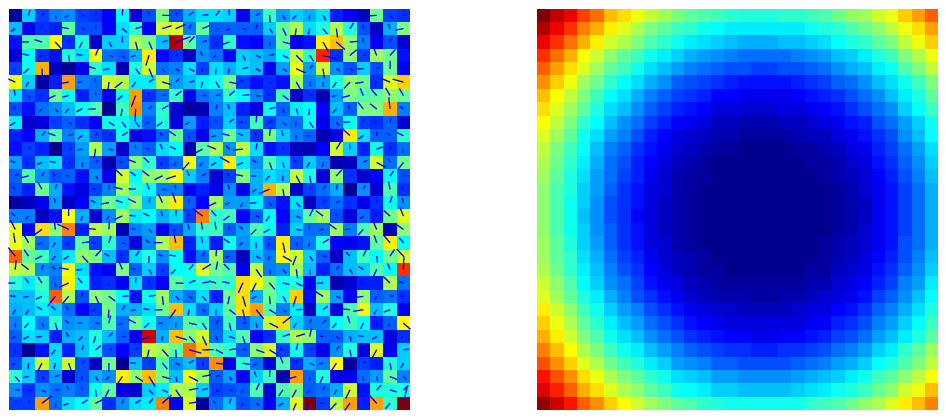}}
 \caption{Each panel shows an entire simulated image, showing the 
   typical PSF pattern for an image in a set (image $100$ in set $1$) 
   with no random Kolmogorov component (upper panels) 
   and for an image in a set (image $100$ in set $19$) with a random Kolomogorov 
   component (lower panels). The $100$x$100$ grid has been downsampled to $30$x$30$ in these
   panels for clarity. The left panels show the amplitude of the ellipticity in the colour scale, and the 
   orientation of the PSF denoted by the whiskers. The right hand panels show the size of the PSF in the colour 
   scale in unit of pixels. In each image in a set these patterns changed, except in those sets where the PSF spatial 
   variation was fixed (see Appendix D).}
 \label{psf12}
\end{figure}

The PSF spatial variation consisted of three components 
\begin{itemize} 
\item 
{\bf Static Component.} These were spatially constant across the image
and consisted of i) a Gaussian smoothing kernel that added to the PSF
size, this had a variance of $0.1$ present in all images, ii) a static
additive ellipticity component of $0.05$ in $e_{1,{\rm PSF}}$ and
$e_{2,{\rm PSF}}$ to simulate tracking error. 
\item 
{\bf Deterministic Component.} This was to simulate the impact of the
telescope on the PSF size and ellipticity. We used the Jarvis, Schecter
and Jain (2008) model to simulate this with fiducial parameters 
$(a_0= 0.014$, $a_1= 0.0005$, $d_0=-0.006$, $d_1= 0.001$,
$c_0=-0.010)$, which is dominated by primary astigmatism ($a_0$), 
primary de-focus ($d_0$) and coma ($c_0$).
\item 
{\bf Random Component.} To simulate the random turbulent effect of the
atmosphere in some of the sets we additionally included a random
Gaussian field in the ellipticity only with a Kolmogorov power spectrum of 
$C_{\ell}=\ell^{-11/6}$ (see Rowe, 2010 and Heymans et al., 2012 for discussion on 
this kind of power spectrum PSF variation seen in optical weak lensing images). 
\end{itemize}
In Figure \ref{psf12} we show a typical PSF pattern for an image in a set with no random Kolmogorov 
variation and one in which there is a random Kolmogorov component. 
As described in Section \ref{Description of the Competition}
participants were provided with the PSF as an exact functional form,
consisting of tabulated numbers for $(r_d, e_1, e_2)$ at the position of each galaxy and as a pixelated stellar
image.

\section*{Appendix D : Set Description}
In the Table below we provide the parameter values that define each
set in the GREAT10 Galaxy Challenge simulations. 
\begin{landscape}
\begin{table*}
\begin{center}
\begin{tabular}{|l|c|c|c|c|c|c|c|c|c|c|c|}
\hline
&Set Name& Fixed& S/N& S/N Dist.& $r_b$/pix.& $r_d$/pix. & B/D Fraction&B-D
Offset/pix.$^2$& $r$ Dist.&KM Power\\
\hline
$1$ & Fiducial & --  & $20$& Gaussian& $2.3$& $4.8$& $0.5$&$0.0$& Gaussian &None\\ 
$2$ & Fiducial & {\bf PSF} & $20$& Gaussian& $2.3$& $4.8$& $0.5$&$0.0$& Gaussian &None\\ 
$3$ & Fiducial & {\bf Int}& $20$& Gaussian& $2.3$& $4.8$& $0.5$&$0.0$& Gaussian &None\\ 
$4$ & Low S/N & -- &  \boldmath$10$& Gaussian& $2.3$& $4.8$& $0.5$&$0.0$& Gaussian &None\\ 
$5$ & Low S/N & {\bf PSF} &  \boldmath$10$& Gaussian& $2.3$& $4.8$& $0.5$&$0.0$& Gaussian &None\\ 
$6$ & Low S/N & {\bf Int} & \boldmath$10$& Gaussian& $2.3$& $4.8$& $0.5$&$0.0$& Gaussian &None\\ 
$7$ & High S/N & -- & \boldmath$40$& Gaussian& $2.3$& $4.8$& $0.5$&$0.0$& Gaussian &None\\ 
$8$ & High S/N  & {\bf PSF}& \boldmath$40$& Gaussian& $2.3$& $4.8$& $0.5$&$0.0$& Gaussian &None\\ 
$9$ & High S/N  & {\bf Int}& \boldmath$40$& Gaussian& $2.3$& $4.8$& $0.5$&$0.0$& Gaussian &None\\ 
$10$ & Smooth S/N & --&$20$& {\bf Rayleigh}& $2.3$& $4.8$& $0.5$&$0.0$& Gaussian &None\\ 
$11$ & Smooth S/N & {\bf PSF}&$20$& {\bf Rayleigh}& $2.3$& $4.8$& $0.5$&$0.0$& Gaussian &None\\ 
$12$ & Smooth S/N & {\bf Int}&$20$& {\bf Rayleigh}& $2.3$& $4.8$& $0.5$&$0.0$& Gaussian &None\\ 
$13$ & Small Galaxy & -- &$20$& Gaussian& \boldmath$1.8$& \boldmath$2.6$& $0.5$&$0.0$& Gaussian &None\\ 
$14$ & Small Galaxy & {\bf PSF}&$20$& Gaussian& \boldmath$1.8$& \boldmath$2.6$& $0.5$&$0.0$& Gaussian &None\\ 
$15$ & Large Galaxy & -- &$20$& Gaussian& \boldmath$3.4$& \boldmath$10.0$& $0.5$&$0.0$& Gaussian &None\\ 
$16$ & Large Galaxy & {\bf PSF}&$20$& Gaussian& \boldmath$3.4$& \boldmath$10.0$& $0.5$&$0.0$& Gaussian &None\\ 
$17$ & Smooth Galaxy & -- &$20$& Gaussian& $2.3$& $4.8$& $0.5$&$0.0$& {\bf Rayleigh} &None\\ 
$18$ & Smooth Galaxy & {\bf PSF}&$20$& Gaussian& $2.3$& $4.8$& $0.5$&$0.0$& {\bf Rayleigh} &None\\ 
$19$ & Kolmogorov & -- &$20$& Gaussian& $2.3$& $4.8$& $0.5$&$0.0$& Gaussian &{\bf Yes}\\ 
$20$ & Kolmogorov & {\bf PSF}&$20$& Gaussian& $2.3$& $4.8$& $0.5$&$0.0$& Gaussian &{\bf Yes}\\ 
$21$ & Uniform b/d & --&$20$& Gaussian& $2.3$& $4.8$& \boldmath$[0.3,0.95]$&$0.0$& Gaussian &None\\ 
$22$ & Uniform b/d  & {\bf PSF}&$20$& Gaussian& $2.3$& $4.8$& \boldmath$[0.3,0.95]$&$0.0$& Gaussian &None\\ 
$23$ & Offset b/d & --&$20$& Gaussian& $2.3$& $4.8$& $0.5$&\boldmath$0.5$& Gaussian &None\\ 
$24$ & Offset b/d  & {\bf PSF}&$20$& Gaussian& $2.3$& $4.8$& $0.5$&\boldmath$0.5$& Gaussian &None\\ 
\hline
\end{tabular}
\caption{A summary of the variables that define each set in the GREAT10
  Galaxy Challenge simulations. The variables in bold are those that distinguish each
  set from the fiducial one. The third columns lists those fields that were 
  fixed over each image in each set. Columns 4 and 9 list the
  distribution used for the signal-to-noise and galaxy sizes
  respectively. Column 8 shows the variance of the offset between the
  bulge and disk components in pixels squared.}
\label{sets}
\end{center}
\end{table*}
\end{landscape}

\section*{Appendix E : Description of the Methods}
\label{Description of the Methods}

Here we briefly summarise the methods that took part in the
challenge. We encourage the reader to refer to the methods' own
papers for more details. 

For each method we show 3 figures these are 
\begin{enumerate}
\item
A reconstruction of the
shear power spectrum for each set comparing the submitted power, true
power and pixel noise corrected power, and the ${\mathcal M}$,
${\mathcal A}$ and $Q_{\rm dn}$ values for all sets.
\item 
The measured minus
true shear on an object-by-object basis as a function of the true
shear $\gamma^t$, the PSF ellipticity and size, the bulge-to-disk
angle and fraction and the bulge size; for $\gamma^t$ the gradient and
offset of this fit is are $m$ and $c$, in all cases we make 10 bins
the variable quantity. We also show a value for $q$, a non-linear
shear response for each metric keeping $m$ and $c$ fixed at their best
fit values (see equation \ref{mcq}). 
\item
The $m$ and $c$ values as a function of PSF ellipticity and size, the bulge-to-disk
angle and fraction and the bulge size. In all cases we make 10 bins
the variable quantity. 
\end{enumerate}
Because these figures contain a wealth of information for the latter two we
plot the gradient and offset values for a linear fit through the
points and display these values in the figures. In the top righthand
corner of each of the subplots we show the difference in the reduced
$\chi^2$ between the best linear fit and the best constant fit
(gradient equal to zero) 
$\Delta\chi^2=\chi^2({\rm gradient},{\rm offset})-\chi^2({\rm
  offset})$; this can be used as an indicator of the significance of
any linearly varying behaviour. 

For power spectrum
submissions the later two plots (concerned with individual one-point 
shear biases) will not be shown. 

We have also provided postscripts of all Figures in this Appendix
online here {\tt http://great.roe.ac.uk/data/galaxy\_article\_figures}.

\newpage
\subsection*{E1. ARES : Peter Melchior}
Comparing the results of DEIMOS and KSB, we found several sets where
the ellipticities measured with either method strongly and
consistently disagreed, with relative deviations of up to 25\%. With
additional simulations we investigated when such discrepancies between
KSB and DEIMOS occur, and concluded that mainly very small, i.e. badly
resolved, galaxies are responsible for large relative deviations, with
KSB having a too weak and DEIMOS a too strong response to galactic
ellipticities. Hence, a linear combination of the shear estimates of
KSB and DEIMOS appeared advantageous. With the results of our
simulations, a weighting scheme was defined that aims to minimise the
mean squared error on the ellipticity of each galaxy. For GREAT10, the
weight for each set was adjusted independently.
\begin{figure*}
  {\includegraphics[width=\columnwidth,angle=0,clip=]{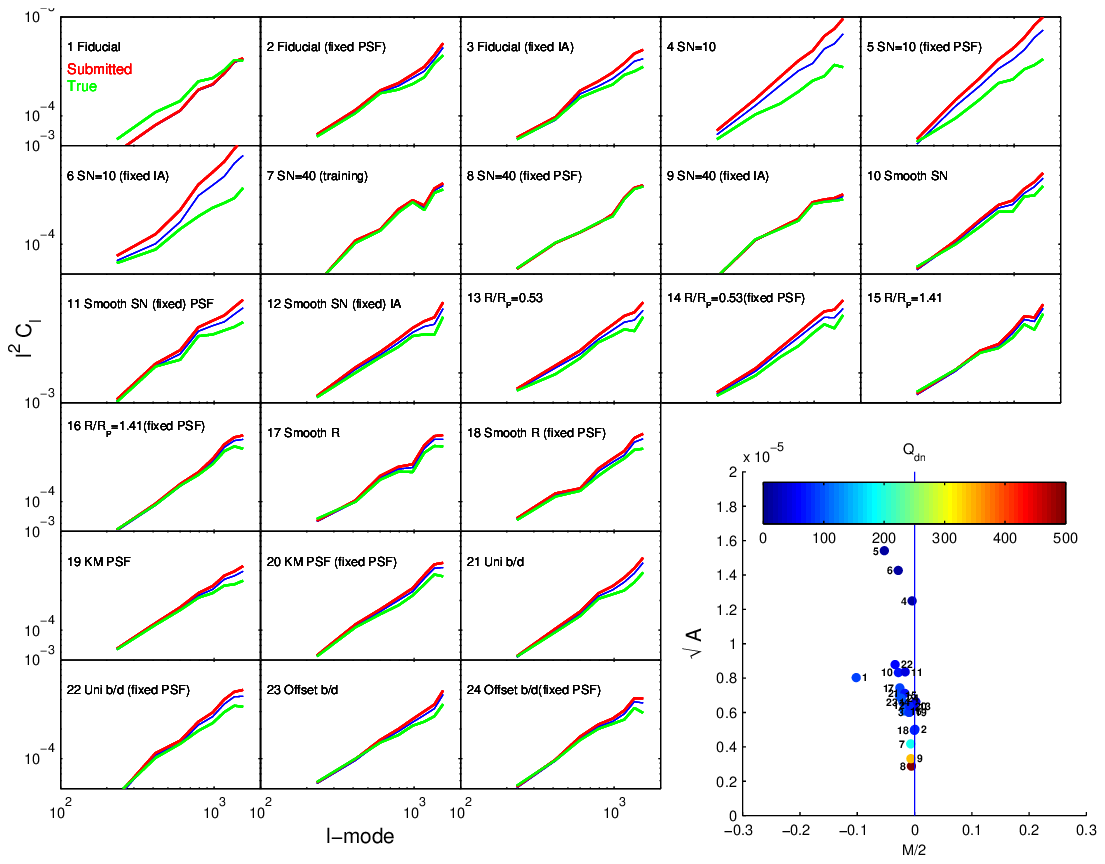}}
 \caption{The true shear power (green) for each set and the shear power for
   the `ARES 50/50' submission (red), we also show the `denoised'
   power spectrum (blue) for each set (where this is indistinguishable
   from the raw submission a red line is only legible). The y-axes 
   are $C_{\ell}\ell^2$ and the x-axis is $\ell$. In
   the bottom righthand corner we show the ${\mathcal M}/2$,
   $\sqrt{{\mathcal A}}$ and 
   the colour scale represents the logarithm of
   the quality factor. The small numbers next to each point label the
   set number.}
 \label{ares}
\end{figure*}
\begin{figure*}
  {\includegraphics[width=0.75\columnwidth,angle=0,clip=]{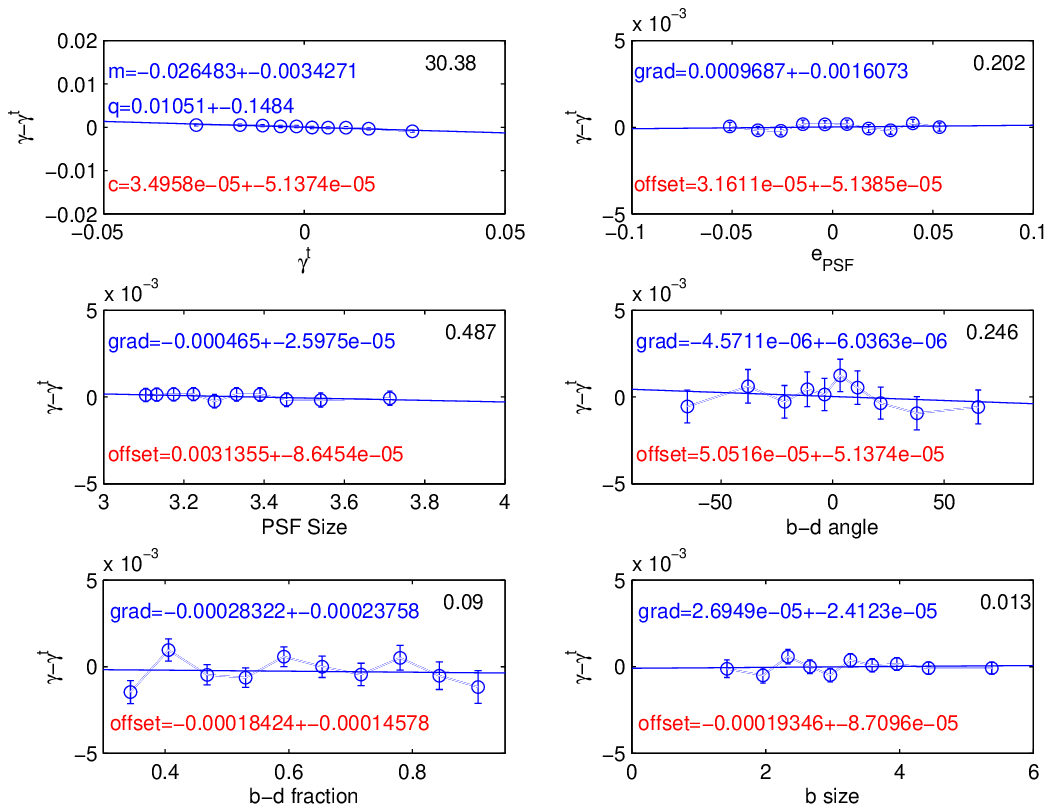}}
 \caption{The measured minus true shear for the `ARES 50/50' submission
   as a function of the true shear, PSF ellipticity, PSF FWHM, galaxy
   bulge-to-disk offset angle, galaxy bulge-to-disk fraction and
   galaxy bulge size. For each dependency we fit a linear function
   with a gradient and offset, for the top left hand panel this is
   the STEP $m$ and $c$ values, additionally for the shear
   dependency we include a quadratic term separately $q$. The top right hand corners show 
   $\Delta\chi^2=\chi^2({\rm gradient},{\rm offset})-\chi^2({\rm offset})$.}
  {\includegraphics[width=\columnwidth,angle=0,clip=]{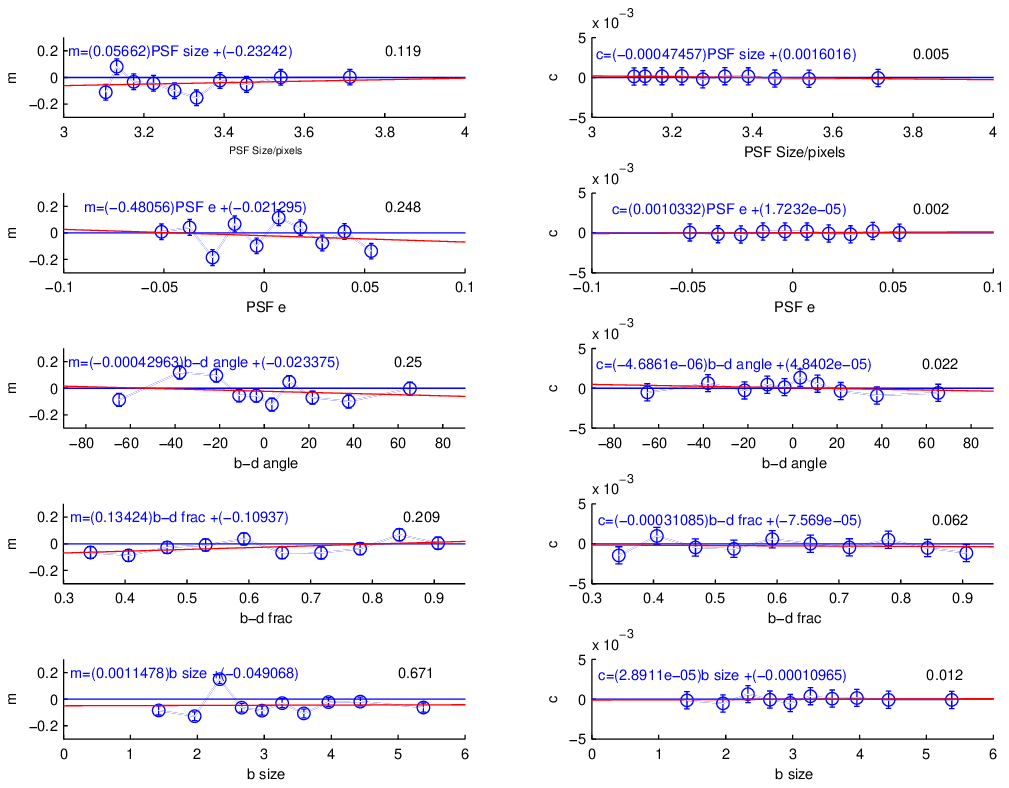}}
 \caption{The STEP $m$ and $c$ values for the `ARES 50/50' submission
   as a function of PSF FHWM and ellipticity, galaxy
   bulge-to-disk offset angle, galaxy bulge-to-disk fraction and
   galaxy bulge size. For each variable we plot the a linear relation
   to the behaviour of $m$ and $c$. We do not explicitly quote errors
   on all parameters for clarity, the average errors on $m$ and $c$ are
   $\simeq 0.005$ and $5\times 10^{-5}$ respectively. The top right hand corners show 
   $\Delta\chi^2=\chi^2({\rm gradient},{\rm offset})-\chi^2({\rm offset})$.}
\end{figure*}
\newpage

\subsection*{E2. cat-unfold: David Kirkby, Daniel Margala}
See fit-unfold description.
\begin{figure*}
  {\includegraphics[width=\columnwidth,angle=0,clip=]{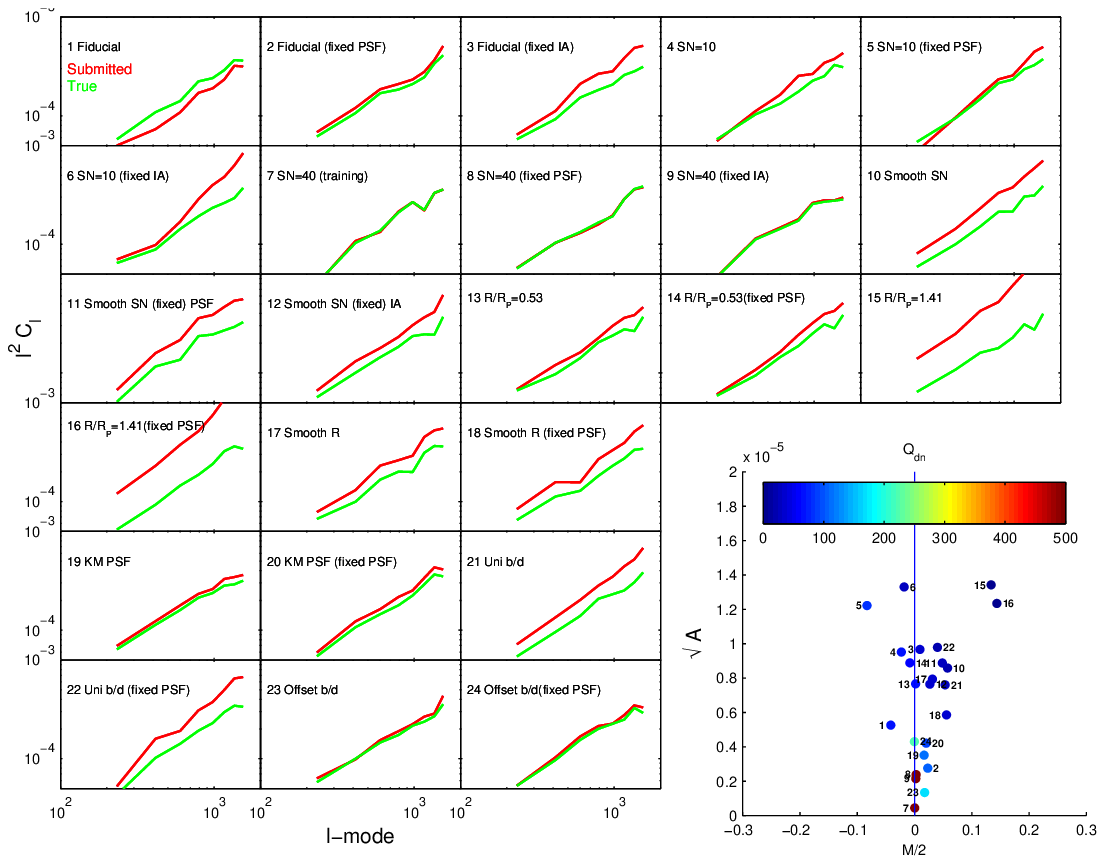}}
 \caption{The true shear power (green) for each set and the shear power for
   the `cat2-unfold' submission (red). 
   The y-axes 
   are $C_{\ell}\ell^2$ and the x-axis is $\ell$. In
   the bottom righthand corner we show the ${\mathcal M}/2$,
   $\sqrt{{\mathcal A}}$ and 
   the colour scale represents the logarithm of
   the quality factor. The small numbers next to each point label the
   set number.}
 \label{cat2unfold}
\end{figure*}
\newpage

\subsection*{E3. DEIMOS : Peter Melchior, Massimo Viola, Julia Young, Kenneth Patton}

DEIMOS (Melchior et al., 2011) measures the second-order moments of the light distribution
using an elliptical Gaussian weight function, whose width is adjusted
such as to maximise the S/N of the measurement. 
The centroid of the galaxy and ellipticity of the weight function is
iteratively matched to the apparent (i.e. PSF-convolved) galaxy (the method has first been
described by Bernstein \& Jarvis, 2002). The application of the weight
function to the image is then corrected by considering higher-order
moments. These corrections become increasingly accurate with
increasing width of the weight function, or the correction
order. For GREAT10 we used correction order of 4 to 8,
i.e. considering the effect of weighting on the moments of order 6 to
10. This correction scheme has been shown to introduce very small
biases on the order of 1\%, mostly for very small galaxies. After the deweighting, we deconvolve the galactic
                        moments from the moments of the PSF, for
                        which we have established an exact and
                        analytic approach. The PSF has been measured
                        with a weight function of the same width as
                        the galaxy, but the ellipticity of the weight
                        function was allowed to match the ellipticity
                        of the PSF. From the deconvolved moments we
                        determine the complex ellipticity $\epsilon$,
                        which theoretically provides an unbiased
                        estimator of the gravitational shear and thus
                        does not need any susceptibility or
                        responsivity corrections.

The only free parameter is the choice of the correction order, which
we varied from 4 to 8 (e.g. ``DEIMOS C6''), and the range of weight function widths. No
model of either galaxy or PSF is employed. The pixel values are taken
at center-pixel positions, an interpolation to sub-pixel resolution is
not applied.
\begin{figure*}
  {\includegraphics[width=\columnwidth,angle=0,clip=]{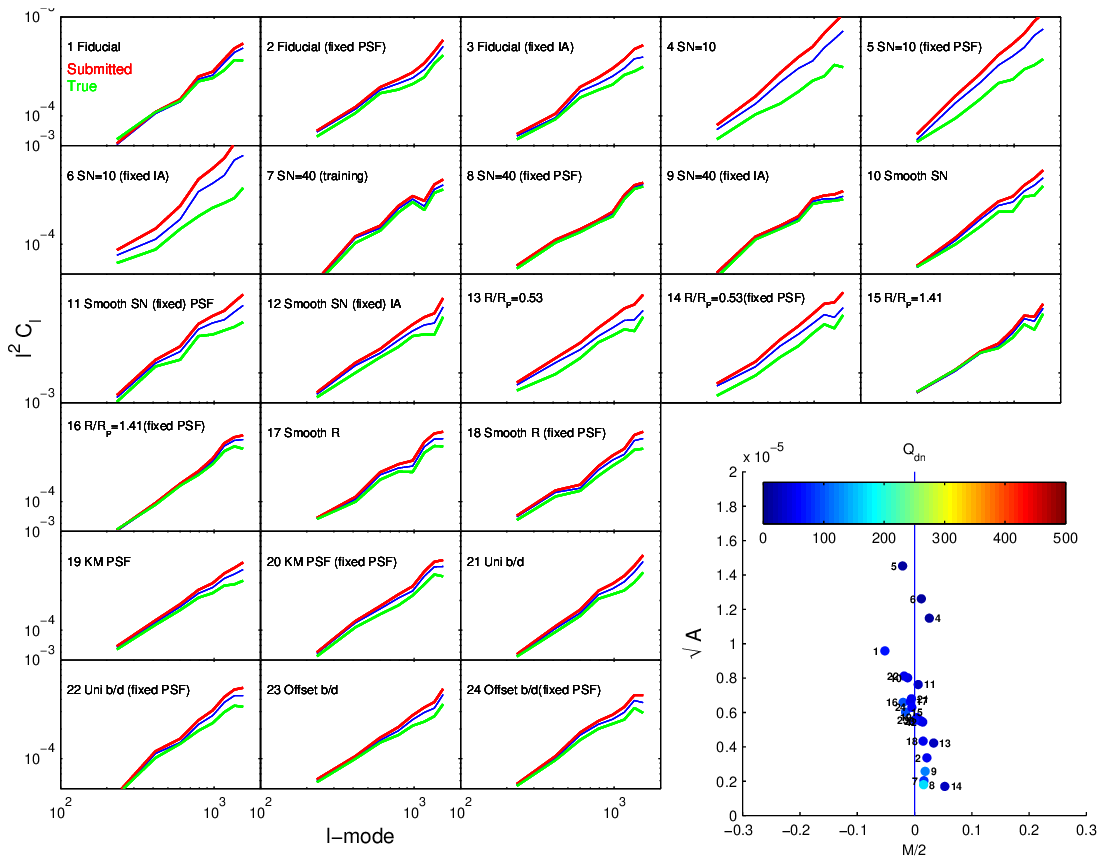}}
 \caption{The true shear power (green) for each set and the shear power for
   the `DEIMOS C6' submission (red), we also show the `denoised'
   power spectrum (blue) for each set (where this is indistinguishable
   from the raw submission a red line is only legible). 
   The y-axes 
   are $C_{\ell}\ell^2$ and the x-axis is $\ell$. In
   the bottom righthand corner we show the ${\mathcal M}/2$,
   $\sqrt{{\mathcal A}}$ and 
   the colour scale represents the logarithm of
   the quality factor. The small numbers next to each point label the
   set number.}
 \label{deimos}
\end{figure*}
\begin{figure*}
  {\includegraphics[width=0.75\columnwidth,angle=0,clip=]{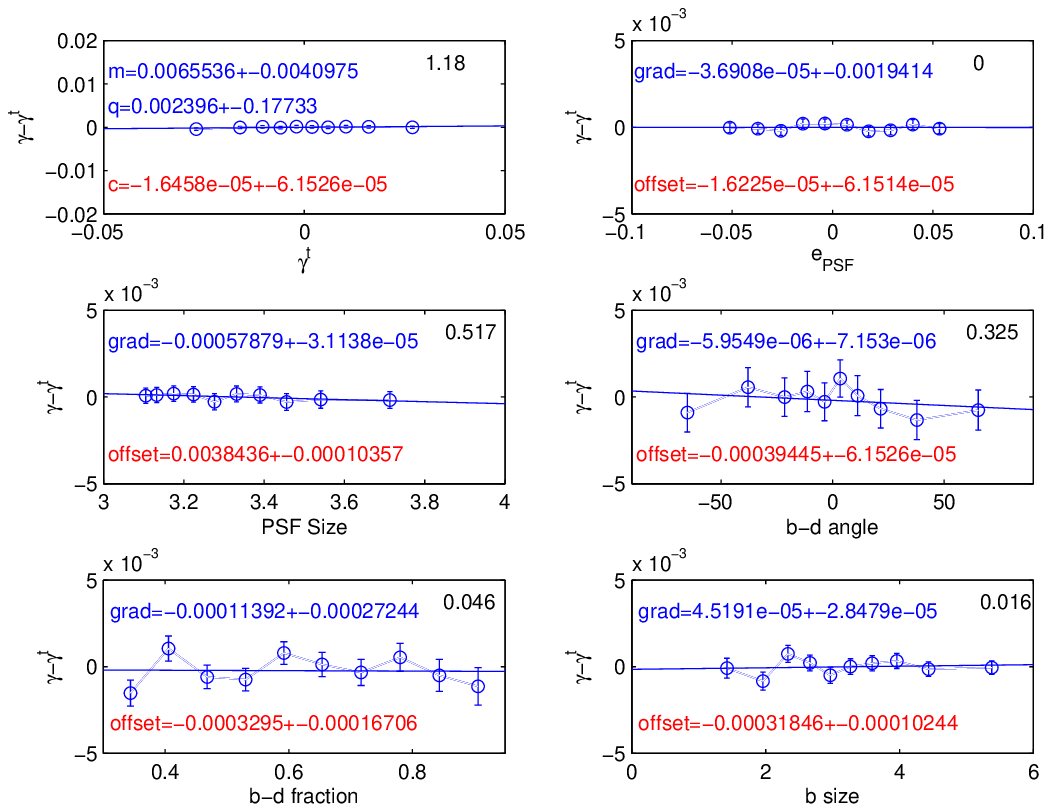}}
 \caption{The measured minus true shear for the `DEIMOS C6' submission
   as a function of the true shear, PSF ellipticity, PSF FWHM, galaxy
   bulge-to-disk offset angle, galaxy bulge-to-disk fraction and
   galaxy bulge size. For each dependency we fit a linear function
   with a gradient and offset, for the top left hand panel this is
   the STEP $m$ and $c$ values, additionally for the shear
   dependency we include a quadratic term separately $q$. The top right hand corners show 
   $\Delta\chi^2=\chi^2({\rm gradient},{\rm offset})-\chi^2({\rm offset})$.}
  {\includegraphics[width=\columnwidth,angle=0,clip=]{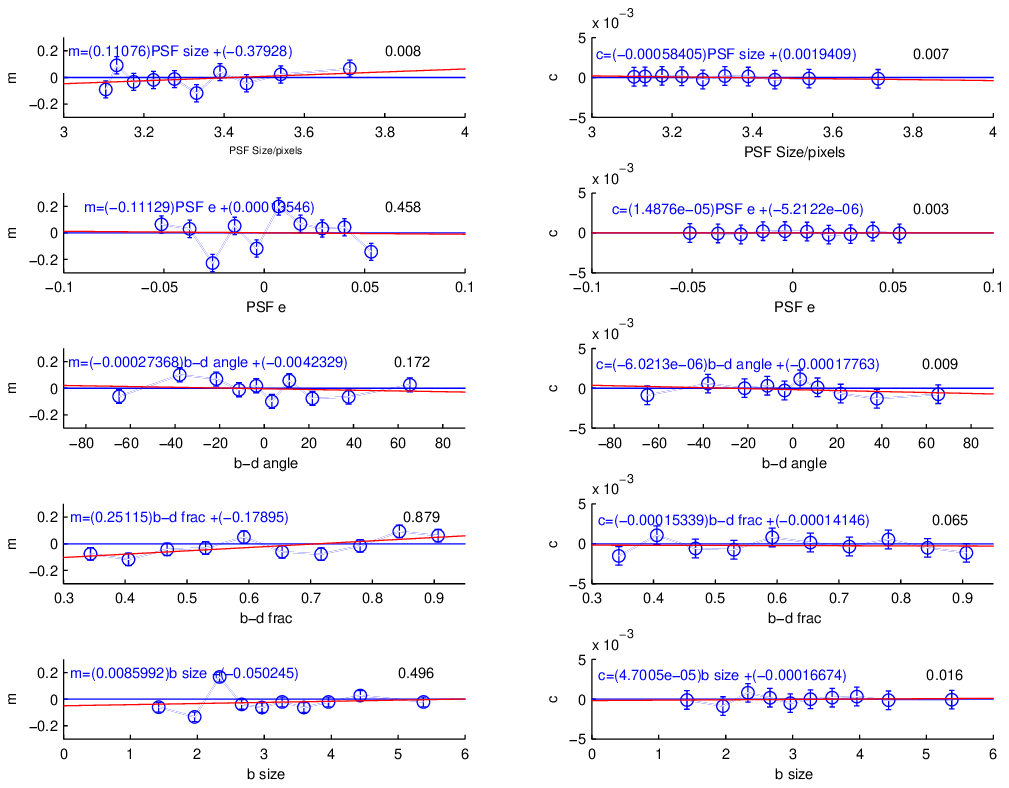}}
 \caption{The STEP $m$ and $c$ values for the `DEIMOS C6' submission
   as a function of PSF FHWM and ellipticity, galaxy
   bulge-to-disk offset angle, galaxy bulge-to-disk fraction and
   galaxy bulge size. For each variable we plot the a linear relation
   to the behaviour of $m$ and $c$. We do not explicitly quote errors
   on all parameters for clarity, the average errors on $m$ and $c$ are
   $\simeq 0.005$ and $5\times 10^{-5}$ respectively. The top right hand corners show 
   $\Delta\chi^2=\chi^2({\rm gradient},{\rm offset})-\chi^2({\rm offset})$.}
\end{figure*}
\newpage

\subsection*{E4. fit-unfold, cat-unfold, shapefit : David Kirkby,
  Daniel Margala}

Each of these names refer to different submissions from the same
underlying software. fit-unfold and cat-unfold were power spectrum
submissions. The DeepZot analysis pipeline consists of four layers of software,
implemented as C++ libraries, that were used for both the GREAT10
Galaxy Challenge and the MDM Challenge (Kitching et al. in prep). 
The first layer provides a
uniform interface to the GREAT10 and MDM datasets. The next layer
performs PSF and galaxy shape estimation using a maximum likelihood
model-fitting method. A half-trace approximation KSB method is also
implemented for comparison with earlier work and to provide a fast
bootstrap of the model fit. The model-fitting code incorporates an
optimised image synthesis engine and uses the MINUIT minimisation
library to calculate full covariance matrices. The third layer
provides supervised machine learning when a suitable training set is
available, and is based on the TMVA package. The best results in the
MDM Challenge were obtained with a 13-input neural network that
derives ellipticity corrections from a combination of model-fitted
parameters, covariance matrix elements, and KSB results. The final
layer of the DeepZot software pipeline performs power-spectrum
estimation and uses the model-fit errors to determine and subtract the
variance due to shape measurement errors. The main computational
bottleneck in the DeepZot pipeline is the model fit, that currently
requires about 500ms per galaxy on a single Intel Xeon core for a
typical fit to a 19-parameter galaxy model in which seven parameters
are floating and a full covariance matrix is obtained.
\begin{figure*}
  {\includegraphics[width=\columnwidth,angle=0,clip=]{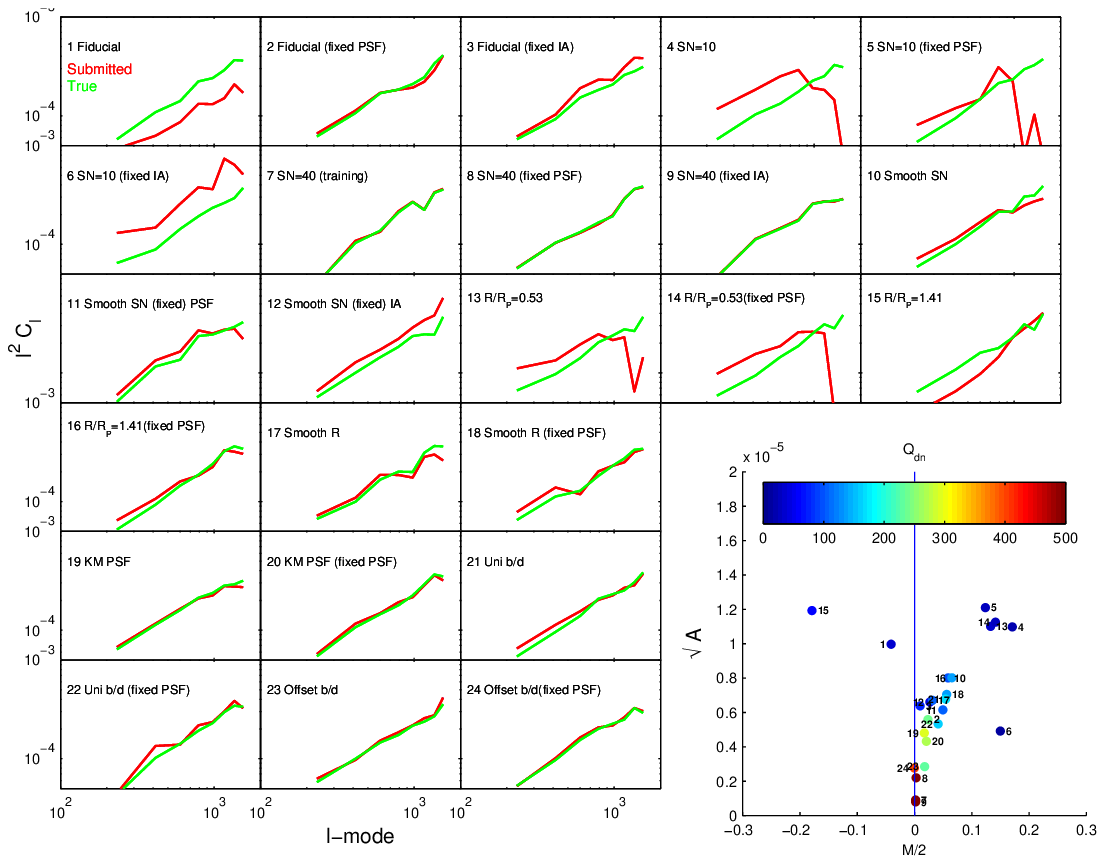}}
 \caption{The true shear power (green) for each set and the shear power for
   the `fit2-unfold' submission (red). 
   The y-axes 
   are $C_{\ell}\ell^2$ and the x-axis is $\ell$. In
   the bottom righthand corner we show the ${\mathcal M}/2$,
   $\sqrt{{\mathcal A}}$ and 
   the colour scale represents the logarithm of
   the quality factor. The small numbers next to each point label the
   set number.}
 \label{fit2unfold}
\end{figure*}
\newpage

\subsection*{E5. gfit : Marc Gentile,  Frederic Courbin, Guldariya Nurbaeva}
The \emph{gfit} shear measurement method is a simple forward model fitting method 
where the underlying galaxy is modelled using a 7-parameter S\'ersic profile. 
The model parameters are the S\'ersic index and radius $(n, r_e)$, 
the galaxy 2-component ellipticity $(e_1,e_2$), the centroid $(x_c, y_c)$ 
and the flux intensity $(I_0)$ at $r = 0$. The galaxy and PSF centroids were 
estimated using {\tt SExtractor} (Bertin \& Arnouts, 1996).

For GREAT10, gfit used a different minimiser than that based on \emph{Levenberg-Marquardt} 
previously used in GREAT08. The minimiser was developed at the 
Laboratory of Astrophysics of EPFL (LASTRO) with GREAT10 in mind. It has proven more 
robust and more accurate when fitting low SNR images.

The `gfit den cs' version of gfit submitted in GREAT10 involved an
experimental implementation of the new \emph{DWT-Wiener} wavelet-based denoising method, also developed at 
LASTRO. DWT-Wiener proved very successful in all other methods we submitted in 
the Galaxy challenge (TVNN, MegaLUT). In the case of gfit, the $Q$ 
factor was boosted by an estimated factor of 1.5. More details about the 
DWT-Wiener method can be found in Nurbaeva, Courbin et al., (2011).
\begin{figure*}
  {\includegraphics[width=\columnwidth,angle=0,clip=]{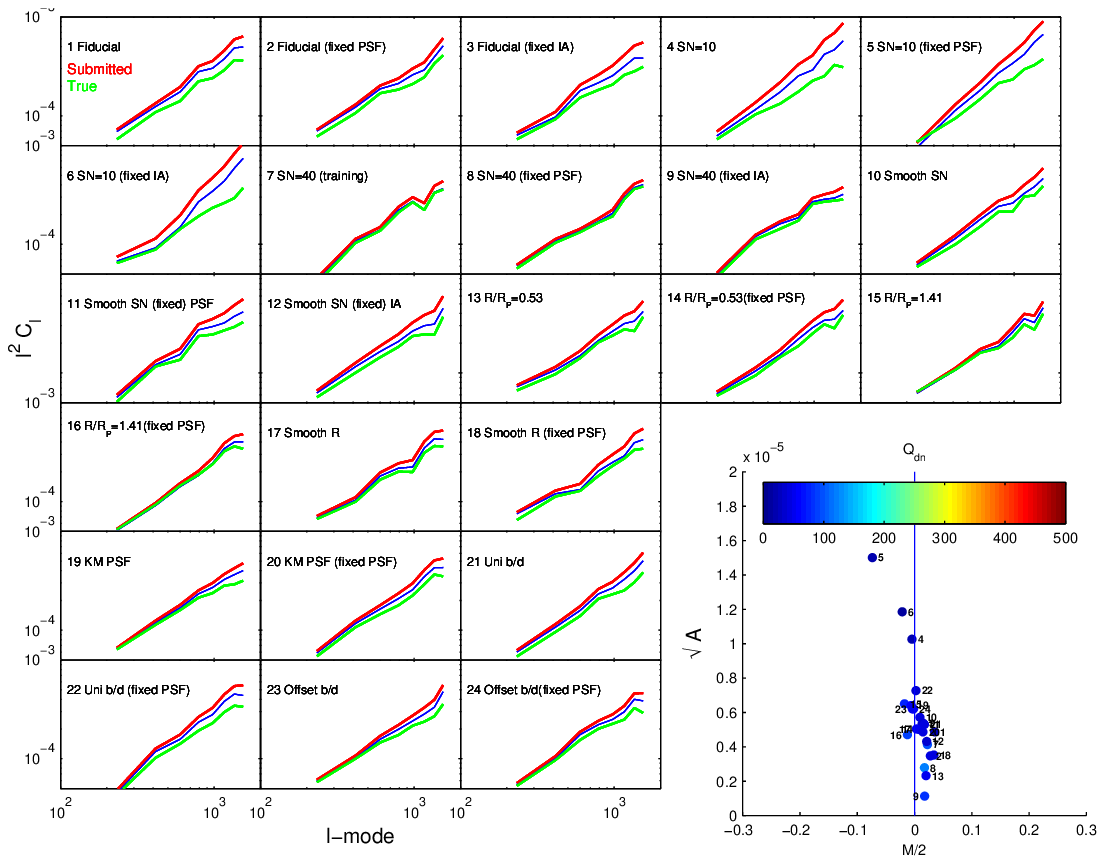}}
 \caption{The true shear power (green) for each set and the shear power for
   the `gfit' submission (red), we also show the `denoised'
   power spectrum (blue) for each set (where this is indistinguishable
   from the raw submission a red line is only legible). 
   The y-axes 
   are $C_{\ell}\ell^2$ and the x-axis is $\ell$. In
   the bottom righthand corner we show the ${\mathcal M}/2$,
   $\sqrt{{\mathcal A}}$ and 
   the colour scale represents the logarithm of
   the quality factor. The small numbers next to each point label the
   set number.}
 \label{gfit}
\end{figure*}
\begin{figure*}
  {\includegraphics[width=0.75\columnwidth,angle=0,clip=]{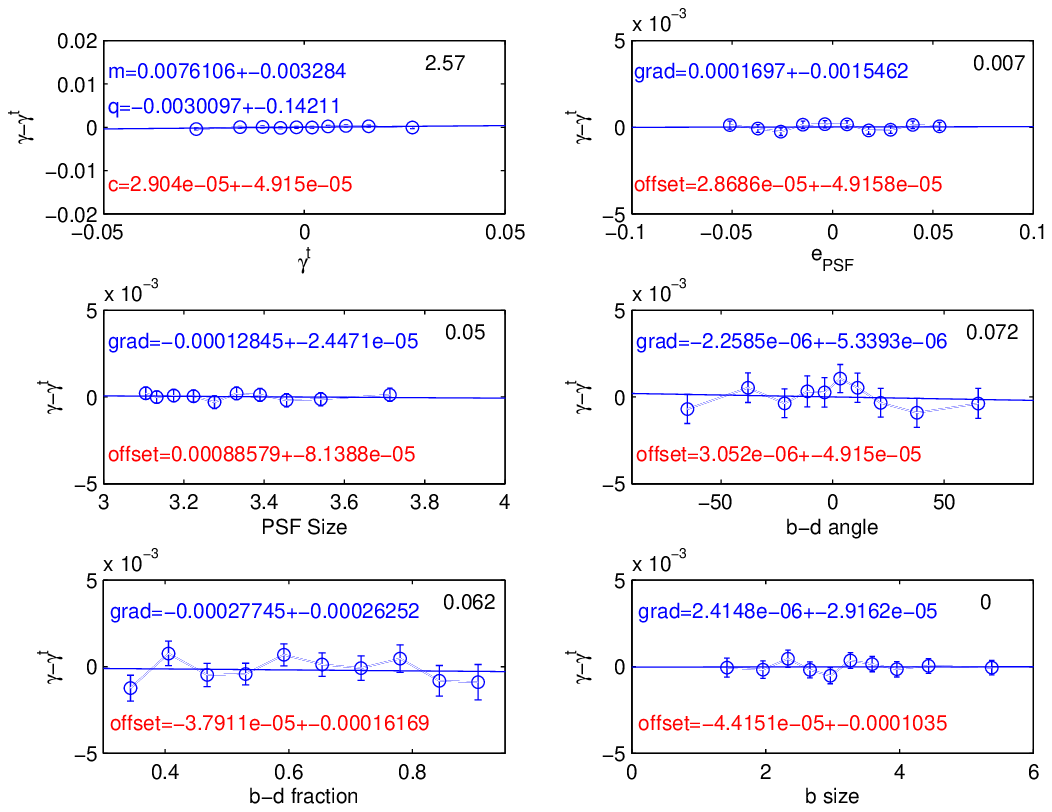}}
 \caption{The measured minus true shear for the `gfit' submission
   as a function of the true shear, PSF ellipticity, PSF FWHM, galaxy
   bulge-to-disk offset angle, galaxy bulge-to-disk fraction and
   galaxy bulge size. For each dependency we fit a linear function
   with a gradient and offset, for the top left hand panel this is
   the STEP $m$ and $c$ values, additionally for the shear
   dependency we include a quadratic term separately $q$. The top right hand corners show 
   $\Delta\chi^2=\chi^2({\rm gradient},{\rm offset})-\chi^2({\rm offset})$.}
  {\includegraphics[width=\columnwidth,angle=0,clip=]{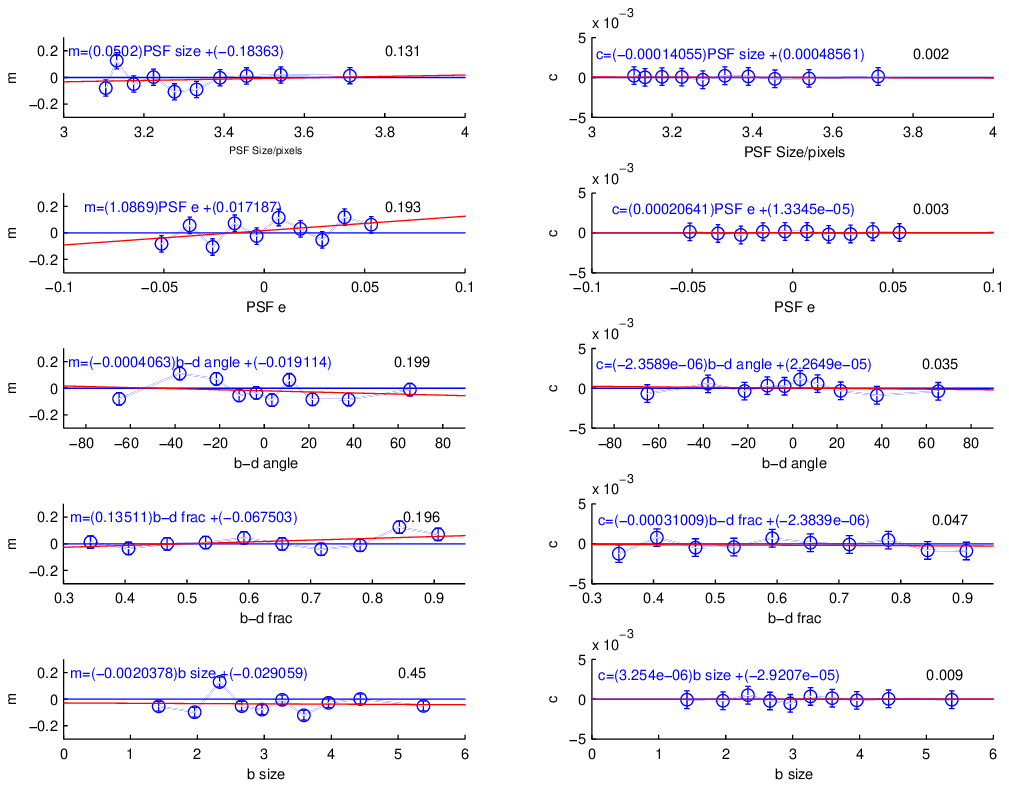}}
 \caption{The STEP $m$ and $c$ values for the `gfit' submission
   as a function of PSF FHWM and ellipticity, galaxy
   bulge-to-disk offset angle, galaxy bulge-to-disk fraction and
   galaxy bulge size. For each variable we plot the a linear relation
   to the behaviour of $m$ and $c$. We do not explicitly quote errors
   on all parameters for clarity, the average errors on $m$ and $c$ are
   $\simeq 0.005$ and $5\times 10^{-5}$ respectively. The top right hand corners show 
   $\Delta\chi^2=\chi^2({\rm gradient},{\rm offset})-\chi^2({\rm offset})$.}
\end{figure*}
\newpage

\subsection*{E6. im3shape : Sarah Bridle, Tomasz Kacprzak, Barney
  Rowe, Lisa Voigt, Joe Zuntz}

{\tt im3shape} fitted a sum of co-elliptical and co-centered Sersic
profiles. In this implementation two Sersic profiles were used with 
the Sersic indices fixed to be $1$ (disk-like) and $4$ (bulge-like) and a
bulge to disk scale radius ratio set to $0.9$. The functional form
for the PSF was provided, and the convolution was performed on a grid three
times the pixel resolution in each direction, with additional
integration in the central pixels of the galaxy model image. 
The maximum likelihood point was used, with a $\chi^2$ evaluated from the full
48$\times$48 postage stamp. The output ellipticity
$(a-b)/(a+b)$ was used as our shear estimate, but with a correction
for noise bias for the submissions marked ``NBC''.
For the noise bias correction a noisy simulated image was produced of a
fiducial galaxy using the machinery in the {\tt im3shape} code. 
Simulations were also produced in which the ellipticity was increased by
$0.1$ in one or other direction.
A straight line was fitted to the output shear estimates relative to
the input ellipticity to measure multiplicative and additive errors
and it was verified that the multiplicative and additive errors were zero in
the absence of noise.
For submissions marked ``NBC0'' two different kinds of
noisy simulations were performed and used these to correct the shear estimates of the
corresponding GREAT10 image sets for (i) Moffat PSF and fiducial GREAT10
SNR (ii) Moffat PSF and lowest GREAT10 SNR. For NBC1 the following combinations were used
(i) Moffat PSF, fiducial GREAT10 SNR, PSF FWHM 3.3 pixels, bulge scale
radius 4.3 pixels (ii) as previous but PSF FWHM 3.1 (iii) as previous but PSF FWHM 3.6
(iv) Moffat PSF, fiducial GREAT10 SNR, PSF FWHM 3.3 pixels, bulge
scale radius 2.3 pixels (v) as previous but bulge scale radius 8 pixels
(vi) Moffat PSF, low GREAT10 SNR, PSF FWHM 3.3 pixels, bulge scale
radius 4.3 pixels (vii) as previous but PSF FWHM 3.1 (viii) as previous but PSF FWHM 3.6.
The optimiser used to find the location of maximum likelihood in
the model parameter space was ``PRAXIS'' (short for Principal AXIS) by
Richard Brent, that is freely available from Netlib at
{\tt http://www.netlib.org/opt/}. The code is specifically written to make
it easy to interchange optimisers and alternatives are also under
investigation. For more information please refer to Zuntz et al. (in prep) for
details about the {\tt im3shape} code in general and Kacprzak et al. (in
prep) for details of the noise bias calibration.
\begin{figure*}
  {\includegraphics[width=\columnwidth,angle=0,clip=]{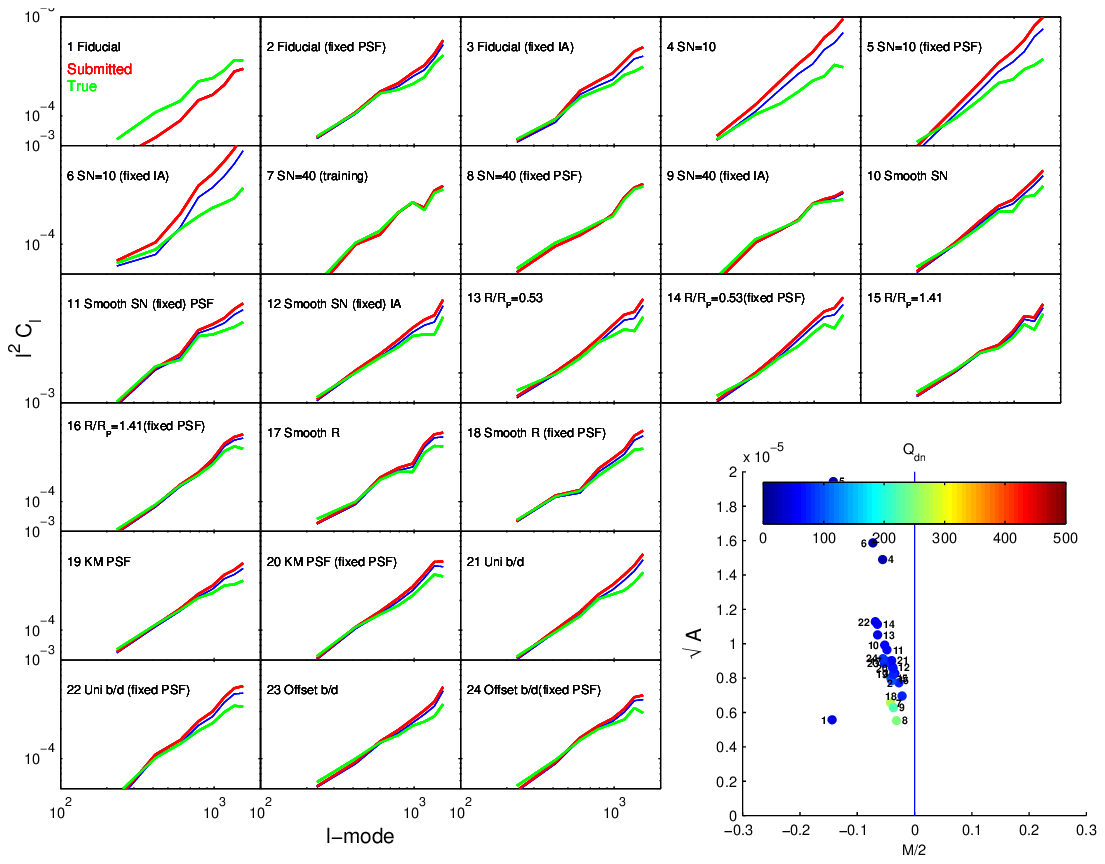}}
 \caption{The true shear power (green) for each set and the shear power for
   the `im3shape NCB0' submission (red), we also show the `denoised'
   power spectrum (blue) for each set (where this is indistinguishable
   from the raw submission a red line is only legible).
   The y-axes 
   are $C_{\ell}\ell^2$ and the x-axis is $\ell$. In
   the bottom righthand corner we show the ${\mathcal M}/2$,
   $\sqrt{{\mathcal A}}$ and 
   the colour scale represents the logarithm of
   the quality factor. The small numbers next to each point label the
   set number.}
 \label{im3shape}
\end{figure*}
\begin{figure*}
  {\includegraphics[width=0.75\columnwidth,angle=0,clip=]{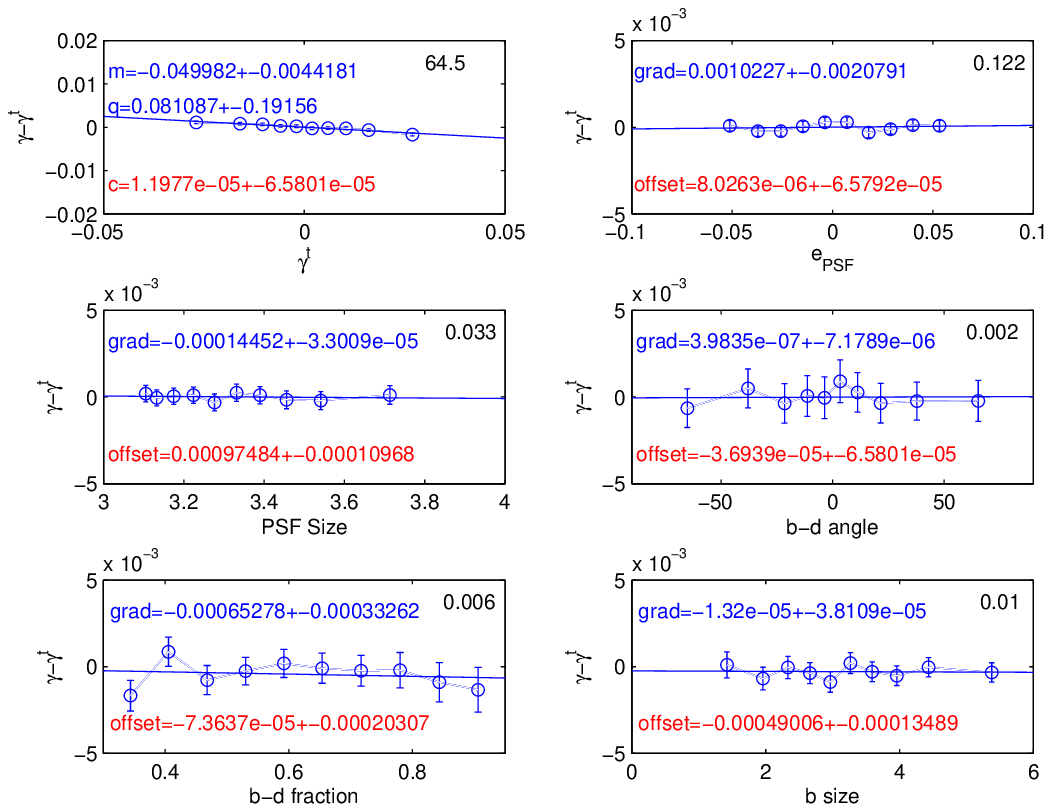}}
 \caption{The measured minus true shear for the `im3shape NCB0' submission
   as a function of the true shear, PSF ellipticity, PSF FWHM, galaxy
   bulge-to-disk offset angle, galaxy bulge-to-disk fraction and
   galaxy bulge size. For each dependency we fit a linear function
   with a gradient and offset, for the top left hand panel this is
   the STEP $m$ and $c$ values, additionally for the shear
   dependency we include a quadratic term separately $q$. The top right hand corners show 
   $\Delta\chi^2=\chi^2({\rm gradient},{\rm offset})-\chi^2({\rm offset})$.}
  {\includegraphics[width=\columnwidth,angle=0,clip=]{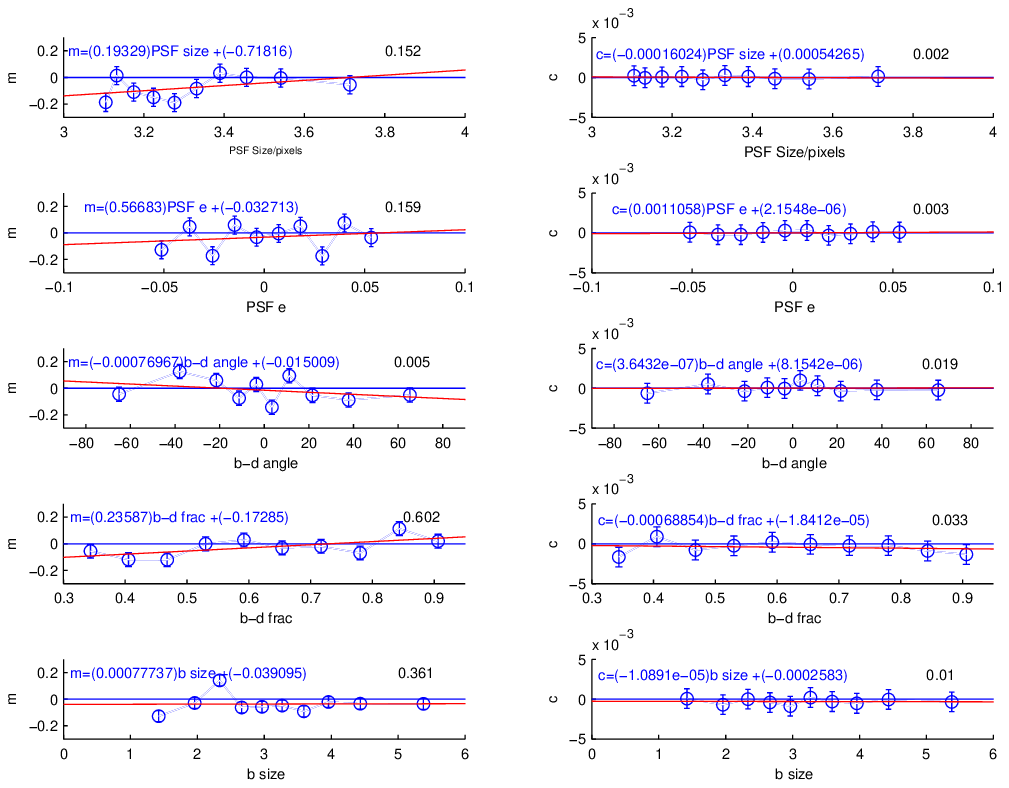}}
 \caption{The STEP $m$ and $c$ values for the `im3shape NCB0' submission
   as a function of PSF FHWM and ellipticity, galaxy
   bulge-to-disk offset angle, galaxy bulge-to-disk fraction and
   galaxy bulge size. For each variable we plot the a linear relation
   to the behaviour of $m$ and $c$. We do not explicitly quote errors
   on all parameters for clarity, the average errors on $m$ and $c$ are
   $\simeq 0.005$ and $5\times 10^{-5}$ respectively. The top right hand corners show 
   $\Delta\chi^2=\chi^2({\rm gradient},{\rm offset})-\chi^2({\rm offset})$.}
\end{figure*}
\newpage

\subsection*{E7. KSB : Julia Young, Peter Melchior}

The original KSB approach was implemented with the `trace-trick',
where the inversion of $P^{sm}$ is achieved by replacing the entire
2x2 matrix by 1/2 of its trace. This approach is employed in several
studies, and it has recently recently been shown (Viola et al., 2011) that is
provides the most unbiased shear estimates for a variety of
observational condition.

To determine galaxy centroid and the width of the circular Gaussian
weight function, the same iterative method employed in DEIMOS was used: 
determine the centroid such that the first moments vanish, and
the size of the weight function such as to maximise S/N. 
For the final shear estimate, we did not apply additional fudge factors or responsivity corrections.
\begin{figure*}
  {\includegraphics[width=\columnwidth,angle=0,clip=]{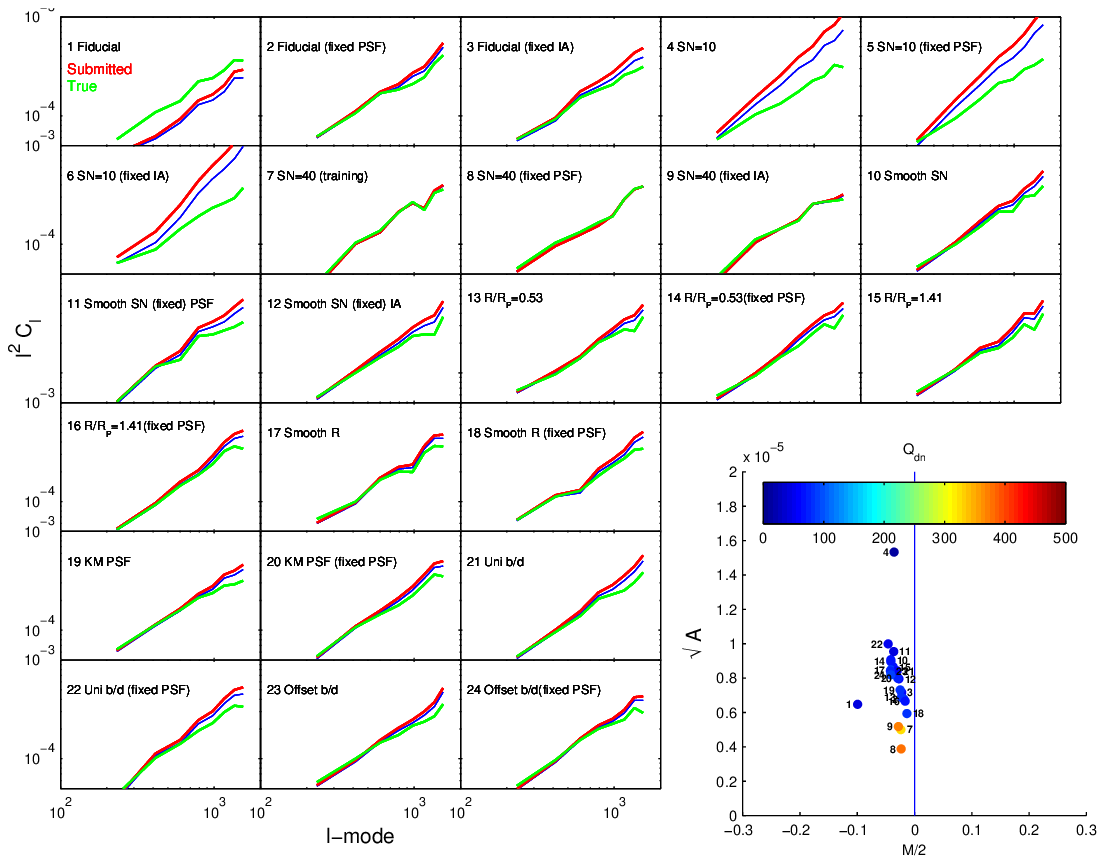}}
 \caption{The true shear power (green) for each set and the shear power for
   the `KSB' submission (red), we also show the `denoised'
   power spectrum (blue) for each set (where this is indistinguishable 
   from the raw submission a red line is only legible). 
   The y-axes 
   are $C_{\ell}\ell^2$ and the x-axis is $\ell$. In
   the bottom righthand corner we show the ${\mathcal M}/2$,
   $\sqrt{{\mathcal A}}$ and 
   the colour scale represents the logarithm of
   the quality factor. The small numbers next to each point label the
   set number.}
 \label{ksb}
\end{figure*}
\begin{figure*}
  {\includegraphics[width=0.75\columnwidth,angle=0,clip=]{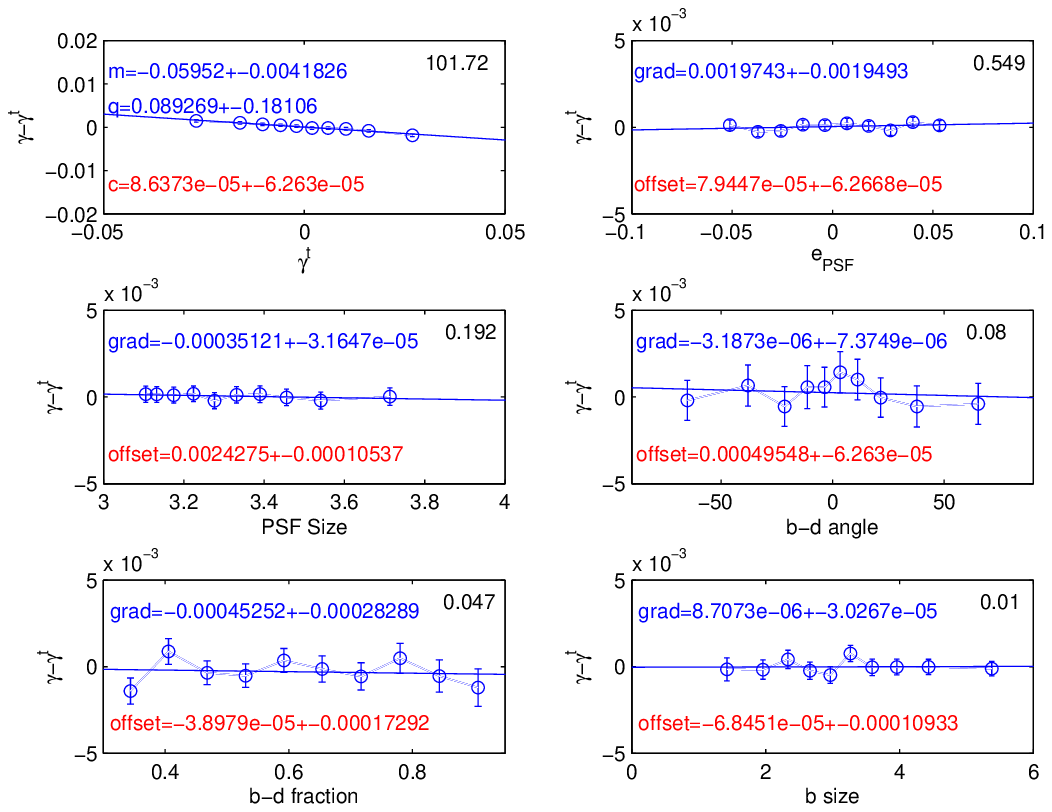}}
 \caption{The measured minus true shear for the `KSB' submission
   as a function of the true shear, PSF ellipticity, PSF FWHM, galaxy
   bulge-to-disk offset angle, galaxy bulge-to-disk fraction and
   galaxy bulge size. For each dependency we fit a linear function
   with a gradient and offset, for the top left hand panel this is
   the STEP $m$ and $c$ values, additionally for the shear
   dependency we include a quadratic term separately $q$. The top right hand corners show 
   $\Delta\chi^2=\chi^2({\rm gradient},{\rm offset})-\chi^2({\rm offset})$.}
  {\includegraphics[width=\columnwidth,angle=0,clip=]{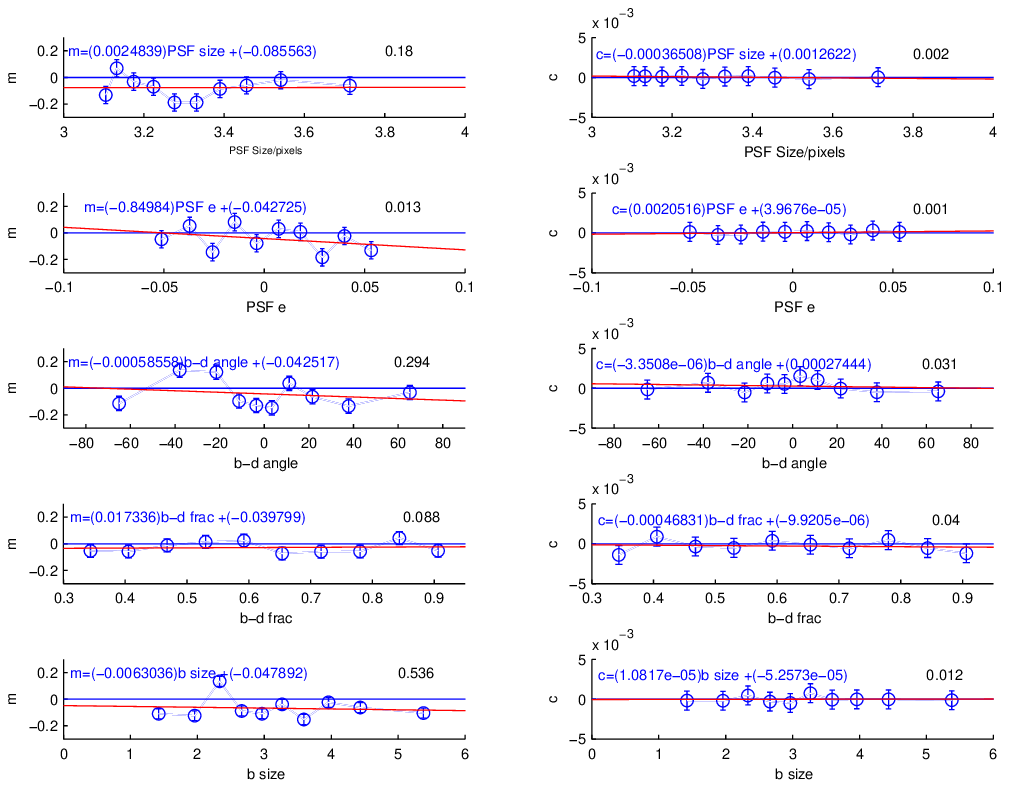}}
 \caption{The STEP $m$ and $c$ values for the `KSB' submission
   as a function of PSF FHWM and ellipticity, galaxy
   bulge-to-disk offset angle, galaxy bulge-to-disk fraction and
   galaxy bulge size. For each variable we plot the a linear relation
   to the behaviour of $m$ and $c$. We do not explicitly quote errors
   on all parameters for clarity, the average errors on $m$ and $c$ are
   $\simeq 0.005$ and $5\times 10^{-5}$ respectively. The top right hand corners show 
   $\Delta\chi^2=\chi^2({\rm gradient},{\rm offset})-\chi^2({\rm offset})$.}
\end{figure*}
\newpage

\subsection*{E8. KSB f90 : Catherine Heymans}
KSB f90 is a benchmark implementation of the longstanding KSB+ method
(Kaiser, Squires \& Broadhurst 1995, Luppino \& Kaiser 1996 and
Hoekstra et al 1998). This code is identical to that used in the `CH'
analysis of STEP1 and GREAT08 (Heymans et al 2006a, Bridle et al 2010)
and can therefore be viewed as a benchmark to compare the different
simulations. KSB f90 is publicly available and
can be downloaded from {\tt http://www.roe.ac.uk/$\sim$heymans/KSBf90}.  The code has
been used to analyse the GEMS and STAGES HST surveys (Heymans et al
2005, Heymans et al 2008). The accuracy of KSB f90 has a strong S/N
dependence as shown in this paper yielding an incorrect redshift
scaling of the lensing signal in real data. For this reason, whilst
KSB f90 has been shown to perform well on average
and for signal-to-noise $>20$, author C. Heymans advises not to use this shape
measurement method for low signal-to-noise data.
\begin{figure*}
  {\includegraphics[width=\columnwidth,angle=0,clip=]{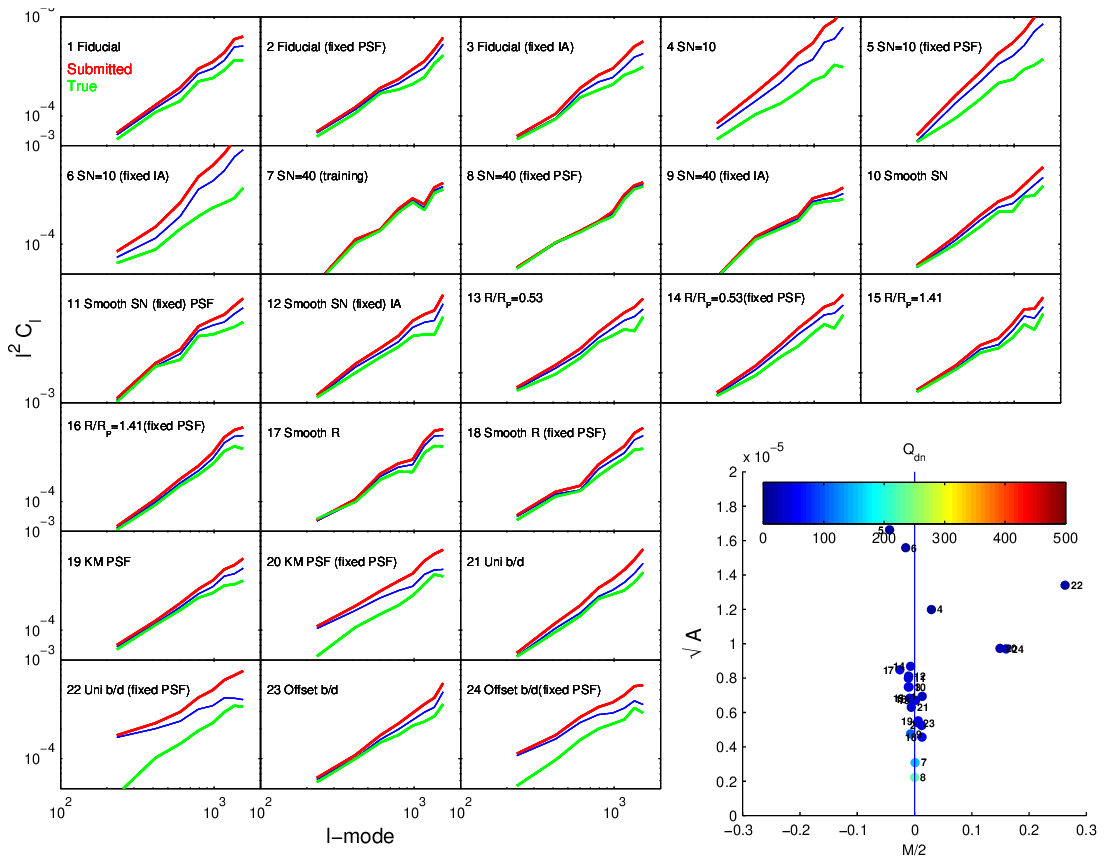}}
 \caption{The true shear power (green) for each set and the shear power for
   the `KSB f90' submission (red), we also show the `denoised'
   power spectrum (blue) for each set (where this is indistinguishable
   from 
   the raw submission a red line is only legible). 
   The y-axes 
   are $C_{\ell}\ell^2$ and the x-axis is $\ell$. In
   the bottom righthand corner we show the ${\mathcal M}/2$,
   $\sqrt{{\mathcal A}}$ and 
   the colour scale represents the logarithm of
   the quality factor. The small numbers next to each point label the
   set number.}
 \label{ksb}
\end{figure*}
\begin{figure*}
  {\includegraphics[width=0.75\columnwidth,angle=0,clip=]{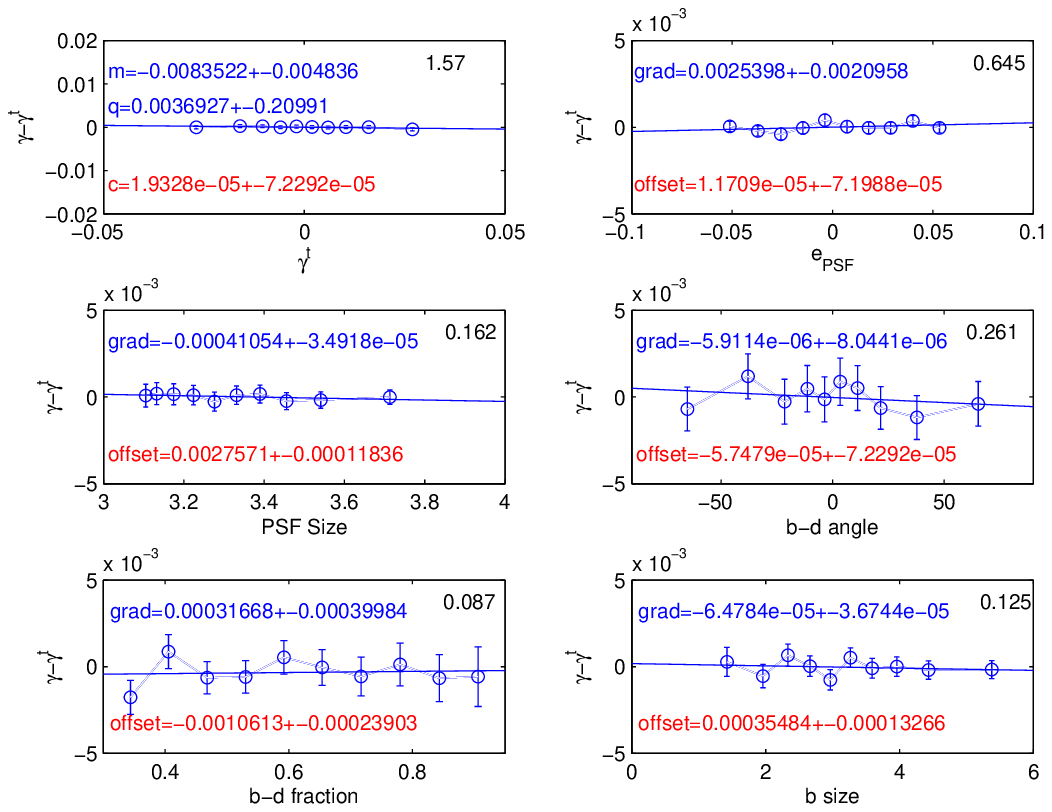}}
 \caption{The measured minus true shear for the `KSB f90' submission
   as a function of the true shear, PSF ellipticity, PSF FWHM, galaxy
   bulge-to-disk offset angle, galaxy bulge-to-disk fraction and
   galaxy bulge size. For each dependency we fit a linear function
   with a gradient and offset, for the top left hand panel this is
   the STEP $m$ and $c$ values, additionally for the shear
   dependency we include a quadratic term separately $q$. The top right hand corners show 
   $\Delta\chi^2=\chi^2({\rm gradient},{\rm offset})-\chi^2({\rm offset})$.}
  {\includegraphics[width=\columnwidth,angle=0,clip=]{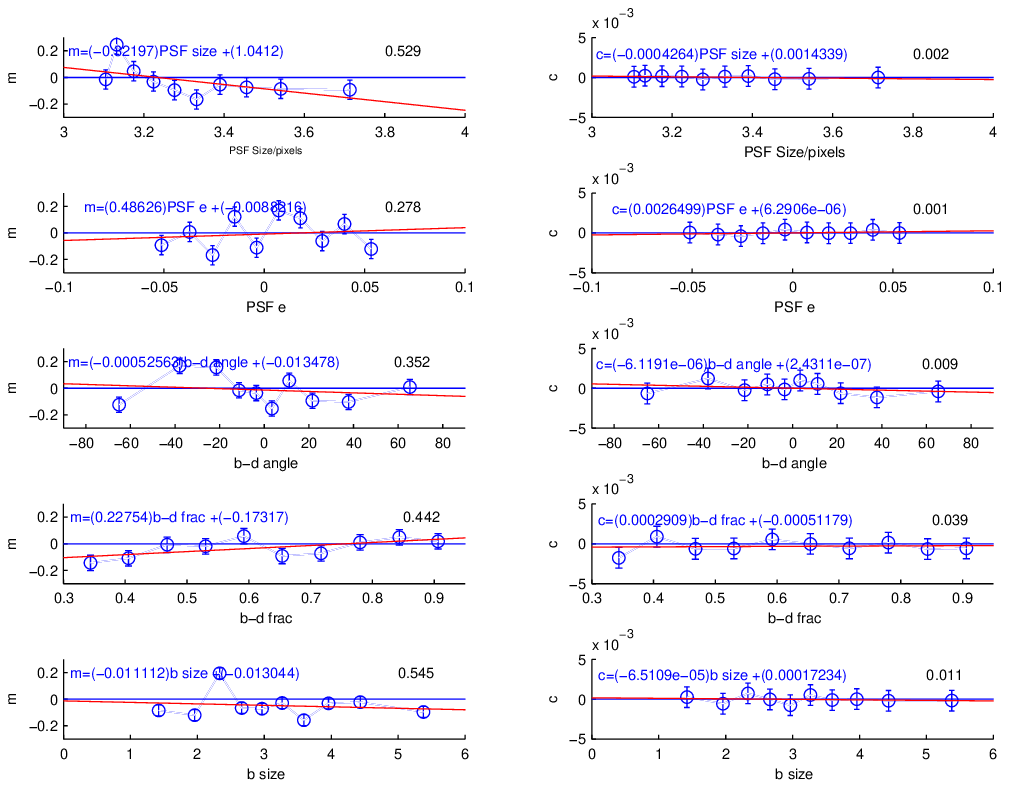}}
 \caption{The STEP $m$ and $c$ values for the `KSB f90' submission
   as a function of PSF FHWM and ellipticity, galaxy
   bulge-to-disk offset angle, galaxy bulge-to-disk fraction and
   galaxy bulge size. For each variable we plot the a linear relation
   to the behaviour of $m$ and $c$. We do not explicitly quote errors
   on all parameters for clarity, the average errors on $m$ and $c$ are
   $\simeq 0.005$ and $5\times 10^{-5}$ respectively. The top right hand corners show 
   $\Delta\chi^2=\chi^2({\rm gradient},{\rm offset})-\chi^2({\rm offset})$.}
\end{figure*}
\newpage

\subsection*{E9. MegaLUT : Malte Tewes, Nicolas Cantale, Frederic Courbin}

MegaLUT is a fast empirical method to correct ellipticity measurements
of galaxies for the distortions by the PSF. It uses a straightforward
classification scheme, namely a lookup table (LUT), built by
supervised learning. In the scope of our submissions to GREAT10, the
successive steps of MegaLUT can be summarised as follows: 1. Simulate
a large number of realistic galaxy and PSF stamps and store the
sheared galaxy ellipticities prior to the PSF convolution. This leads
to a learning sample of images. 2. Run a shape measurement algorithm
on the galaxies and PSFs of this learning sample and create a lookup
table that connects the measured galaxy and PSF shapes to the known
galaxy ellipticities stored in the first step. 3. For a given
galaxy/PSF pair in the GREAT10 data, run the same shape measurement
algorithms as in step 2. Query the lookup table to identify the
galaxy/PSF pairs of the learning sample that have similar measured
shapes. The galaxy ellipticities of these selected pairs, {\it as
 stored at step 1}, yield our estimate of the galaxy ellipticity
prior to the convolution by the PSF. The complex problem of PSF
correction is therefore reduced to a simple and fast array indexing
operation.  

For the final submission `MegaLUTsim2.1 b20', we denoised the galaxy
and PSF images with wavelet filtering, and built simple threshold
masks. The shapes were then measured using second order moments of the
masked light distributions. The lookup table was generated from 2.1
million simulated galaxy/PSF pairs. 
\begin{figure*}
  {\includegraphics[width=\columnwidth,angle=0,clip=]{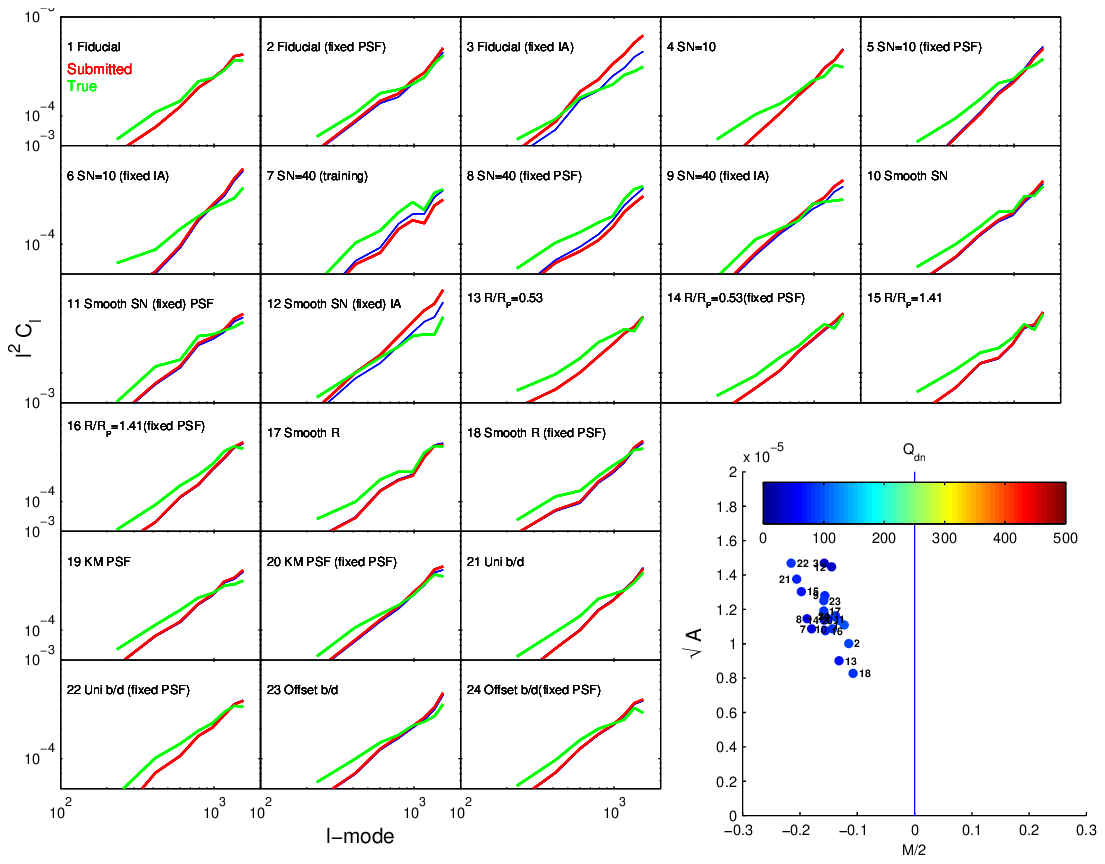}}
 \caption{The true shear power (green) for each set and the shear power for
   the `MegaLUTsim2.1 b20' submission (red), we also show the `denoised'
   power spectrum (blue) for each set (where this is indistinguishable
   from the raw submission a red line is only legible). 
   The y-axes 
   are $C_{\ell}\ell^2$ and the x-axis is $\ell$. In
   the bottom righthand corner we show the ${\mathcal M}/2$,
   $\sqrt{{\mathcal A}}$ and 
   the colour scale represents the logarithm of
   the quality factor. The small numbers next to each point label the
   set number.}
 \label{megalut}
\end{figure*}
\begin{figure*}
  {\includegraphics[width=0.75\columnwidth,angle=0,clip=]{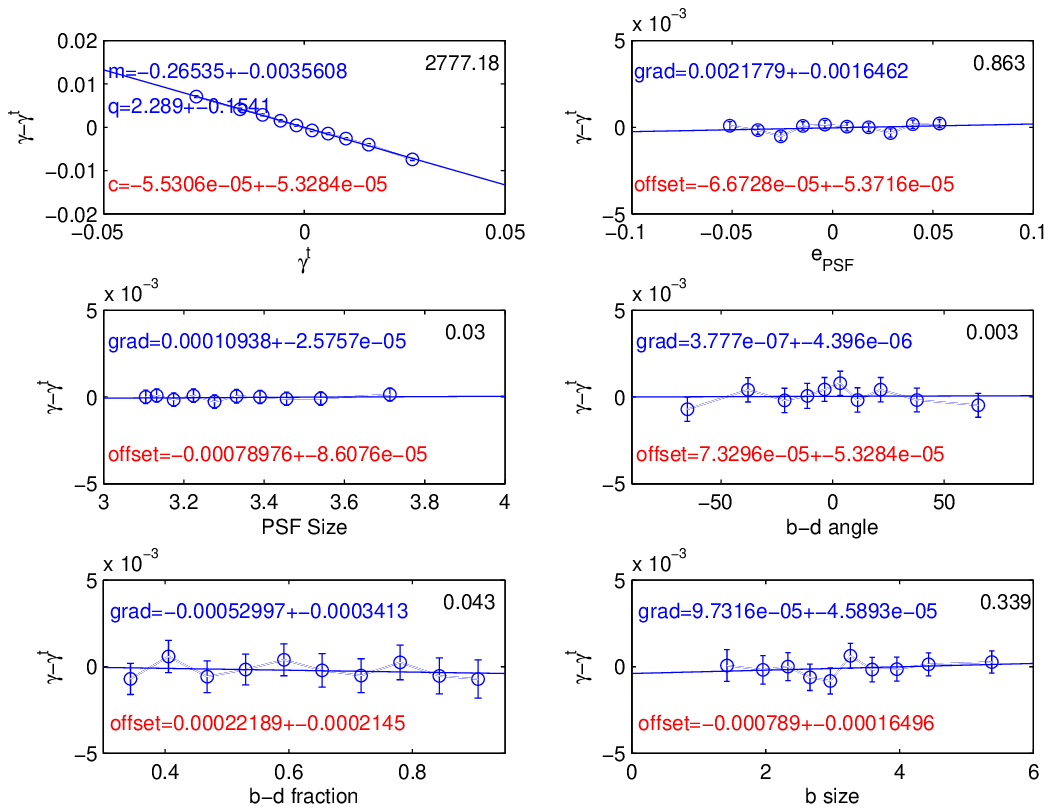}}
 \caption{The measured minus true shear for the `MegaLUTsim2.1 b20' submission
   as a function of the true shear, PSF ellipticity, PSF FWHM, galaxy
   bulge-to-disk offset angle, galaxy bulge-to-disk fraction and
   galaxy bulge size. For each dependency we fit a linear function
   with a gradient and offset, for the top left hand panel this is
   the STEP $m$ and $c$ values, additionally for the shear
   dependency we include a quadratic term separately $q$. The top right hand corners show 
   $\Delta\chi^2=\chi^2({\rm gradient},{\rm offset})-\chi^2({\rm offset})$.}
  {\includegraphics[width=\columnwidth,angle=0,clip=]{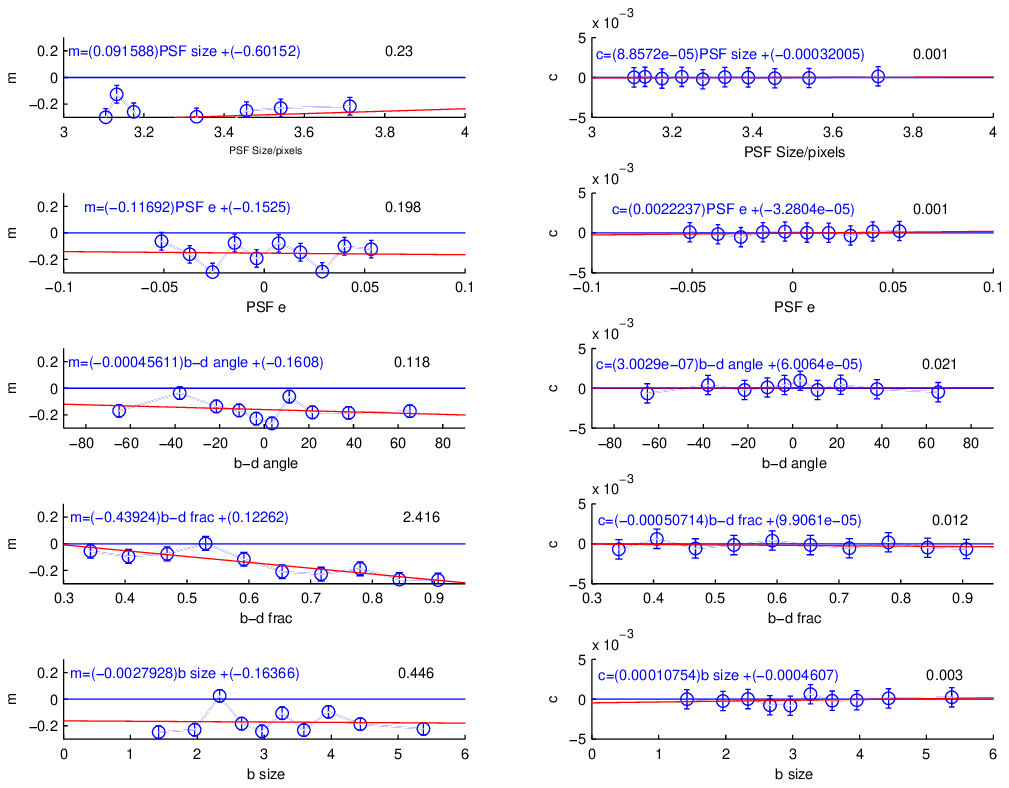}}
 \caption{The STEP $m$ and $c$ values for the `MegaLUTsim2.1 b20' submission
   as a function of PSF FHWM and ellipticity, galaxy
   bulge-to-disk offset angle, galaxy bulge-to-disk fraction and
   galaxy bulge size. For each variable we plot the a linear relation
   to the behaviour of $m$ and $c$. We do not explicitly quote errors
   on all parameters for clarity, the average errors on $m$ and $c$ are
   $\simeq 0.005$ and $5\times 10^{-5}$ respectively. The top right hand corners show 
   $\Delta\chi^2=\chi^2({\rm gradient},{\rm offset})-\chi^2({\rm offset})$.}
\end{figure*}
\newpage

\subsection*{E10. method4,5,7 : Micheal Hirsch, Stefan Harmeling}

In a series of submissions named method0x with $x\in\{1,..,7\}$ 
the effect of taking higher order pixel correlations on the accuracy
of shear measurement was tested. In method01 the shear was measured by subtracting the
quadrupole moments of the auto-correlated images of the galaxy and
corresponding PSF images. The assumption of uncorrelated noise is
confirmed by the fact that the auto-correlation is highly peaked at
zero shift. To get rid of this peak which impedes accurate moment
estimation, a rough estimate of the noise variance was obtained by
computing the variance of pixels with negative intensity values only
(assuming Gaussian noise with zero mean) which was then subtracted
from the central pixel. As in any other KSB-type method,
noise affects moment estimation and has to be accounted for by some
weighting scheme. To this end both galaxy and star images were
modulated 
by a Gaussian with fixed variance and zero centroid. By noticing that
a pixel-wise modulation corresponds to a convolution in Fourier space,
a correction for the induced error due to the modulation could be
removed by
subtracting the measured quadrupole moment and the fixed variance of
the Gaussian distribution used for weighting in the Fourier domain. 
In method04, we went one step further by computing the auto-correlation
of the auto-correlated galaxy or star image, otherwise pursuing the
the same approach as described above. By this the images are even further
smoothed and are still centered such that inaccuracies in centroid
estimation are not an issue in our approach. All other methods are
variants of the above where the empirical moment estimation with a
Gaussian weighting scheme was replaced by a model fitting approach 
(method02), introduced an additional denoising step (method05), did
empirical moment estimation without additional weighting (method03)
and accounted for the PSF by a Wiener deconvolution of the galaxy
images before moment estimation (method07).
\begin{figure*}
  {\includegraphics[width=\columnwidth,angle=0,clip=]{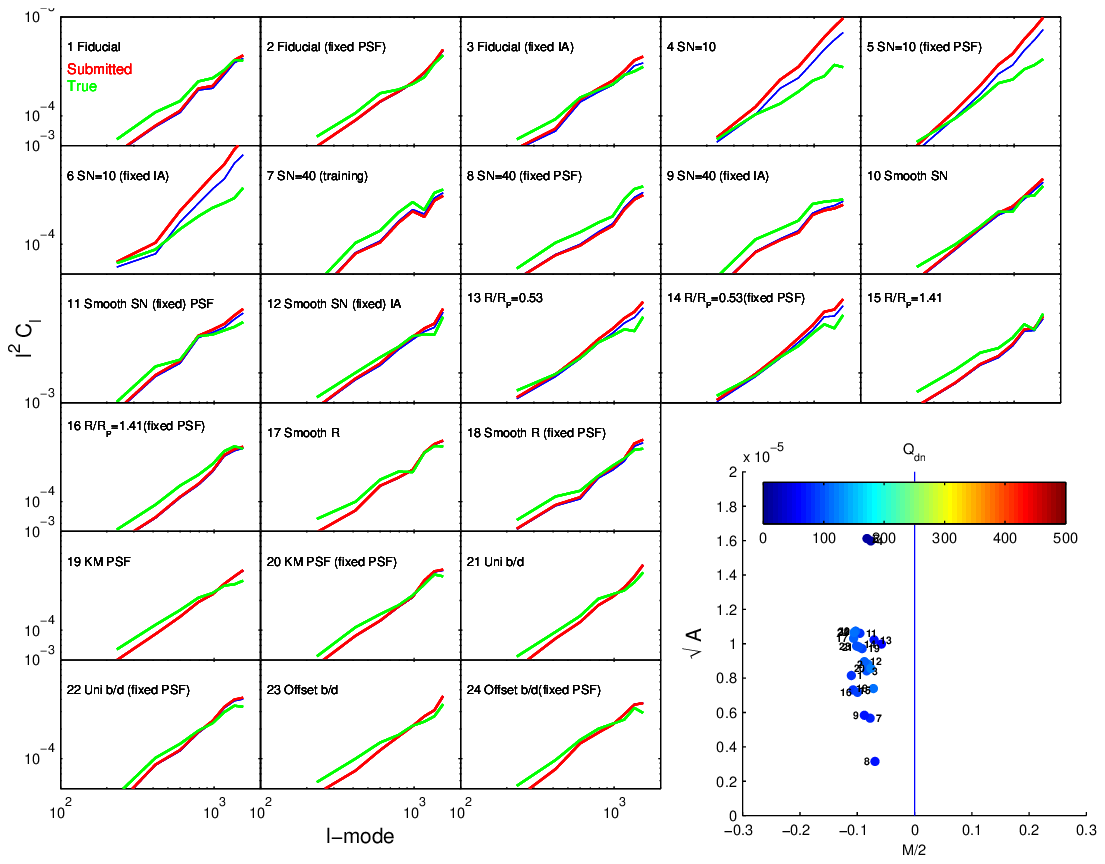}}
 \caption{The true shear power (green) for each set and the shear power for
   the `method 4' submission (red), we also show the `denoised'
   power spectrum (blue) for each set (where this is indistinguishable 
   from the raw submission a red line is only legible). 
   The y-axes 
   are $C_{\ell}\ell^2$ and the x-axis is $\ell$. In
   the bottom righthand corner we show the ${\mathcal M}/2$,
   $\sqrt{{\mathcal A}}$ and 
   the colour scale represents the logarithm of
   the quality factor. The small numbers next to each point label the
   set number.}
 \label{method4}
\end{figure*}
\begin{figure*}
  {\includegraphics[width=0.75\columnwidth,angle=0,clip=]{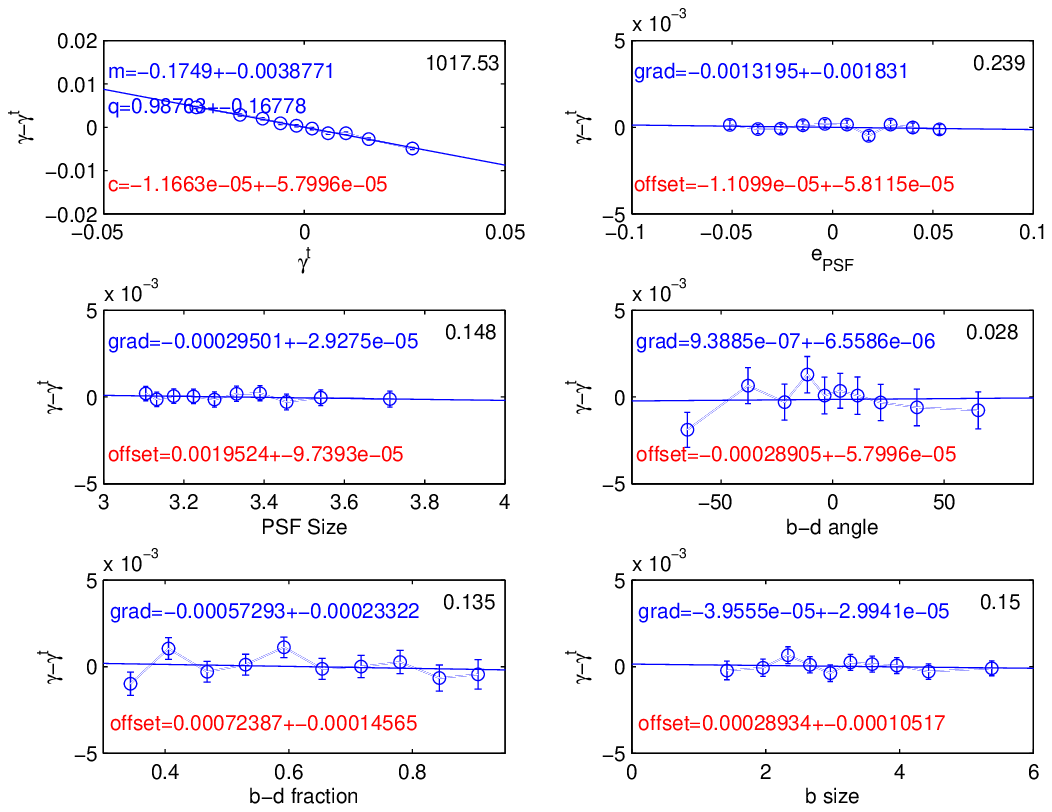}}
 \caption{The measured minus true shear for the `method 4' submission
   as a function of the true shear, PSF ellipticity, PSF FWHM, galaxy
   bulge-to-disk offset angle, galaxy bulge-to-disk fraction and
   galaxy bulge size. For each dependency we fit a linear function
   with a gradient and offset, for the top left hand panel this is
   the STEP $m$ and $c$ values, additionally for the shear
   dependency we include a quadratic term separately $q$. The top right hand corners show 
   $\Delta\chi^2=\chi^2({\rm gradient},{\rm offset})-\chi^2({\rm offset})$.}
  {\includegraphics[width=\columnwidth,angle=0,clip=]{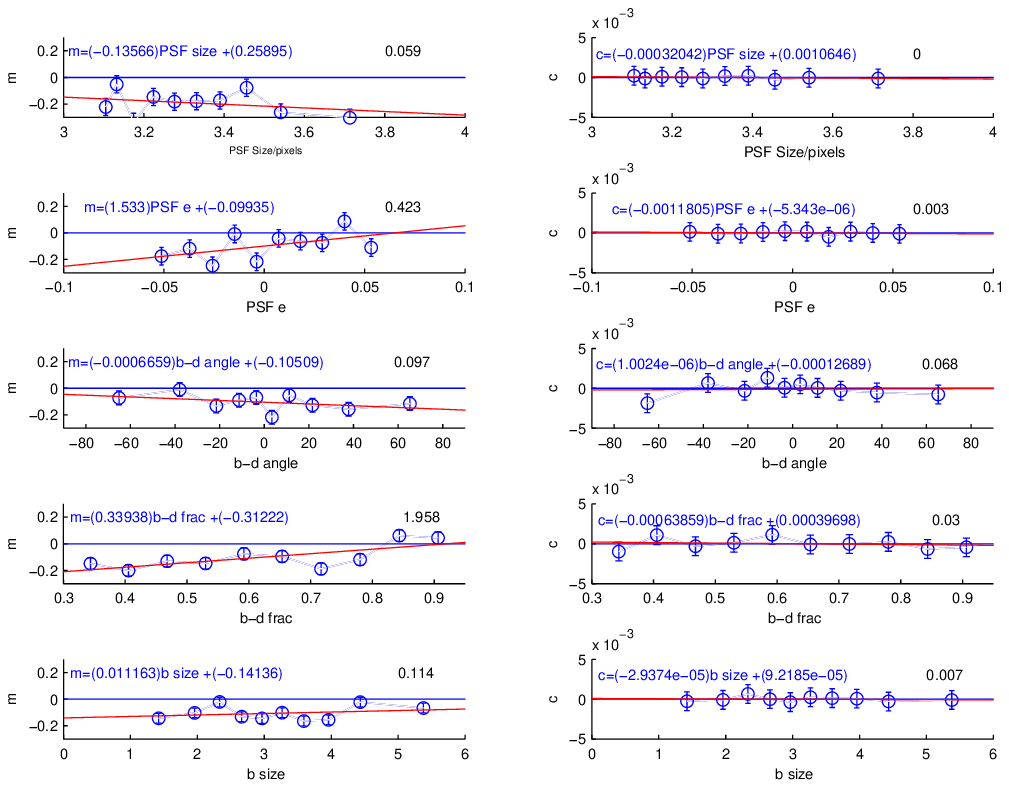}}
 \caption{The STEP $m$ and $c$ values for the `method 4' submission
   as a function of PSF FHWM and ellipticity, galaxy
   bulge-to-disk offset angle, galaxy bulge-to-disk fraction and
   galaxy bulge size. For each variable we plot the a linear relation
   to the behaviour of $m$ and $c$. We do not explicitly quote errors
   on all parameters for clarity, the average errors on $m$ and $c$ are
   $\simeq 0.005$ and $5\times 10^{-5}$ respectively. The top right hand corners show 
   $\Delta\chi^2=\chi^2({\rm gradient},{\rm offset})-\chi^2({\rm offset})$.}
\end{figure*}
\newpage

\subsection*{E11. shapefit: David Kirkby, Daniel Margala}
See fit-unfold description.
\begin{figure*}
  {\includegraphics[width=\columnwidth,angle=0,clip=]{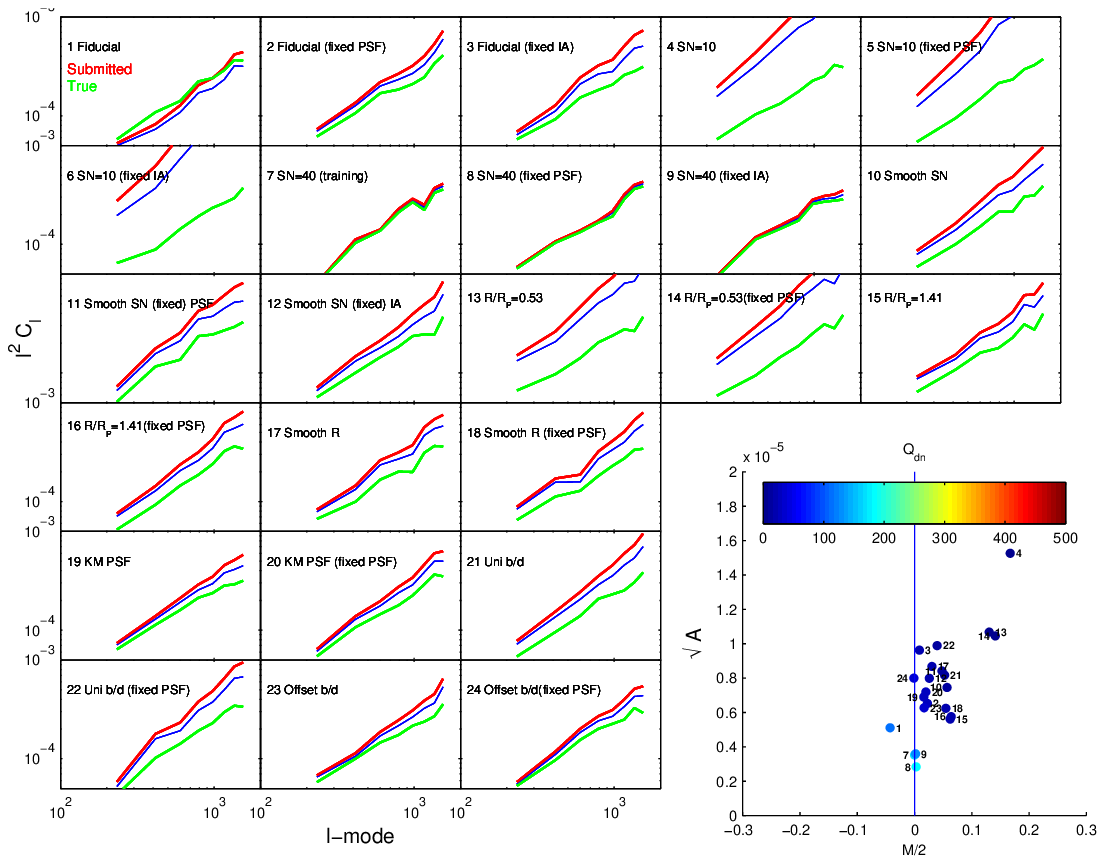}}
 \caption{The true shear power (green) for each set and the shear power for
   the `shapefit' submission (red), we also show the `denoised'
   power spectrum (blue) for each set (where this is indistinguishable 
   from the raw submission a red line is only legible). 
   The y-axes 
   are $C_{\ell}\ell^2$ and the x-axis is $\ell$. In
   the bottom righthand corner we show the ${\mathcal M}/2$,
   $\sqrt{{\mathcal A}}$ and 
   the colour scale represents the logarithm of
   the quality factor. The small numbers next to each point label the
   set number.}
 \label{shapefit}
\end{figure*}
\begin{figure*}
  {\includegraphics[width=0.75\columnwidth,angle=0,clip=]{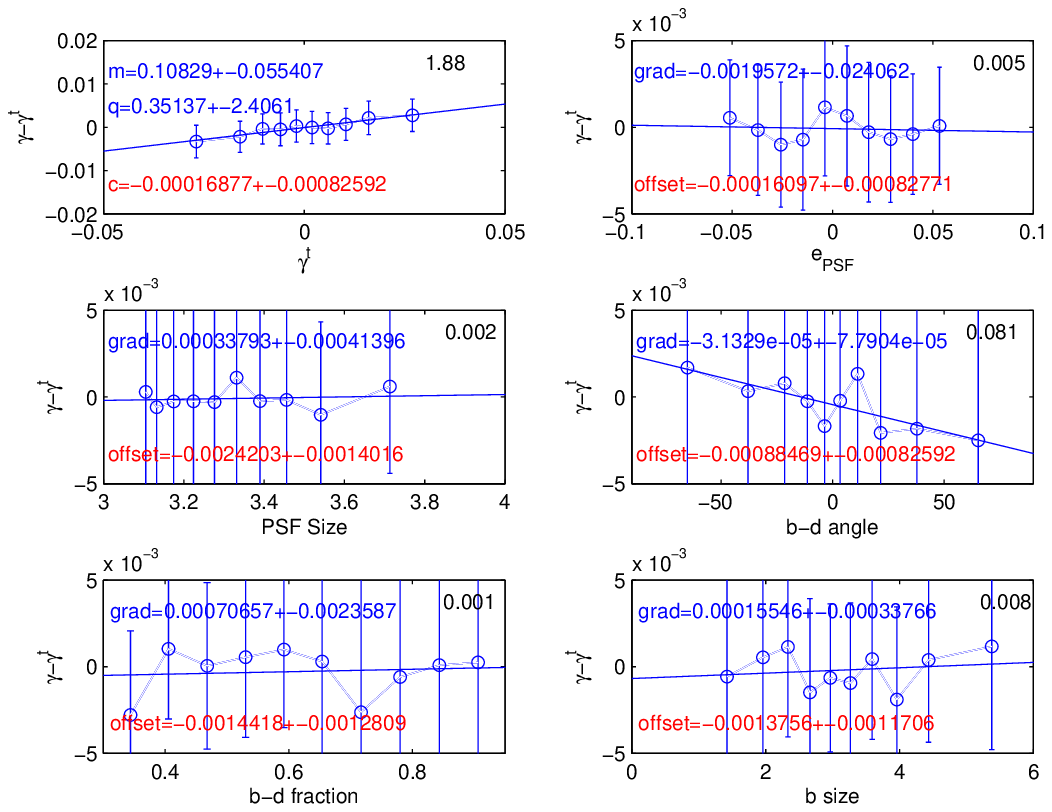}}
 \caption{The measured minus true shear for the `shapefit' submission
   as a function of the true shear, PSF ellipticity, PSF FWHM, galaxy
   bulge-to-disk offset angle, galaxy bulge-to-disk fraction and
   galaxy bulge size. For each dependency we fit a linear function
   with a gradient and offset, for the top left hand panel this is
   the STEP $m$ and $c$ values, additionally for the shear
   dependency we include a quadratic term separately $q$. The top right hand corners show 
   $\Delta\chi^2=\chi^2({\rm gradient},{\rm offset})-\chi^2({\rm offset})$.}
  {\includegraphics[width=\columnwidth,angle=0,clip=]{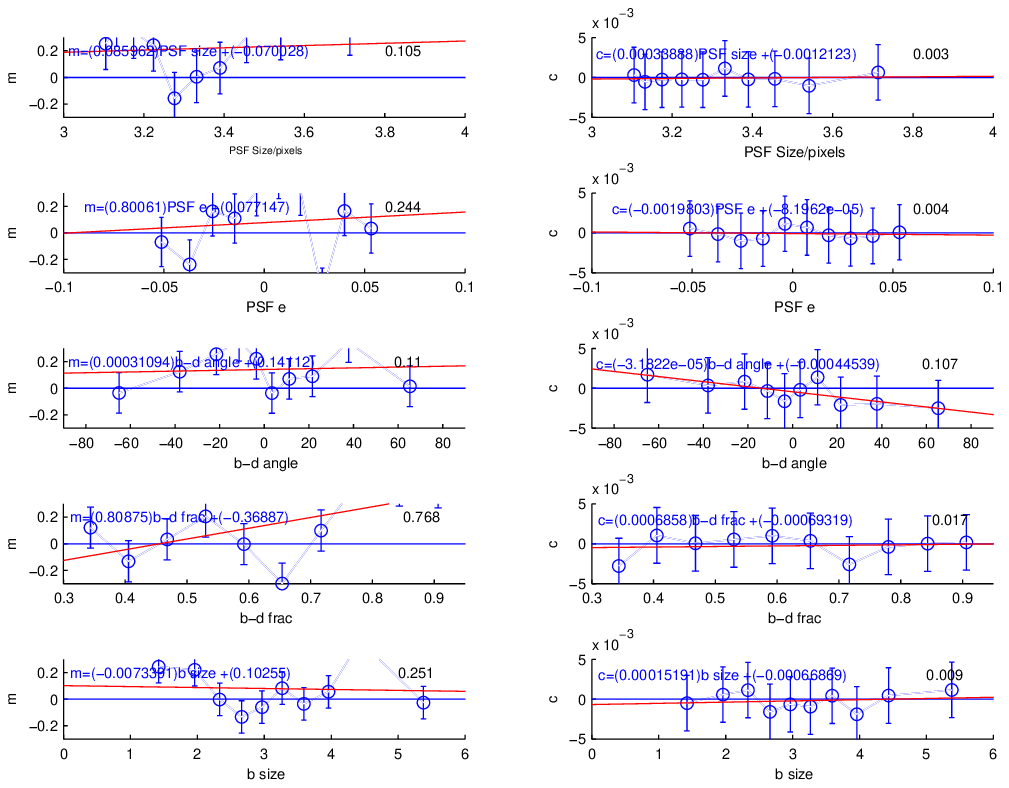}}
 \caption{The STEP $m$ and $c$ values for the `shapefit' submission
   as a function of PSF FHWM and ellipticity, galaxy
   bulge-to-disk offset angle, galaxy bulge-to-disk fraction and
   galaxy bulge size. For each variable we plot the a linear relation
   to the behaviour of $m$ and $c$. We do not explicitly quote errors
   on all parameters for clarity, the average errors on $m$ and $c$ are
   $\simeq 0.005$ and $5\times 10^{-5}$ respectively. The top right hand corners show 
   $\Delta\chi^2=\chi^2({\rm gradient},{\rm offset})-\chi^2({\rm offset})$.}
\end{figure*}
\newpage

\subsection*{E12. TVNN: Guldariya Nurbaeva, Frederic Courbin, Malte Tewes, Marc Gentile}

The methods NN23 func, NN19 and NN21, submitted to GREAT10, 
were variants of the Total
Variation Neural Network (TVNN) method, that is a deconvolution
technique based on the combination of a Hopfield neural network
(Hopfield, 1982) with the Total Variation model proposed by Rudin, Osher
and Faterni (Rudin, 1992). In the Total Variation model, the noise in
the image is assumed to follow a Gaussian distribution. 

The deconvolution process is carried out by minimising the energy
function of the Hopfield Neural Network. This energy function is
composed of the PSF, expressed as a Toeplitz matrix, and of a
regularisation term to minimise the noise. The latter is a Sobel
high-pass operator. The deconvolution itself is done in an iterative
way where at each step, the neurons of the network are updated so as
to minimise the energy function. 

Galaxy ellipticities are then estimated from quadrupole moments
computed on the 2D auto-correlation function (ACF) of the deconvolved
image. The advantages of using the ACF are 1- high signal-to-noise
shape measurement, 2- invariance of the ellipticity measurement with
respect to data (Waerbeke, 1997; Miralda-Escude, 1991) 

In our submissions, the number after the acronym NN stands for the
size of the input data stamps, i.e., NN23 considers images with 23
pixels on a side. This is the first time full-deconvolution of the
data is used to carry out shape measurements.  
\begin{figure*}
  {\includegraphics[width=\columnwidth,angle=0,clip=]{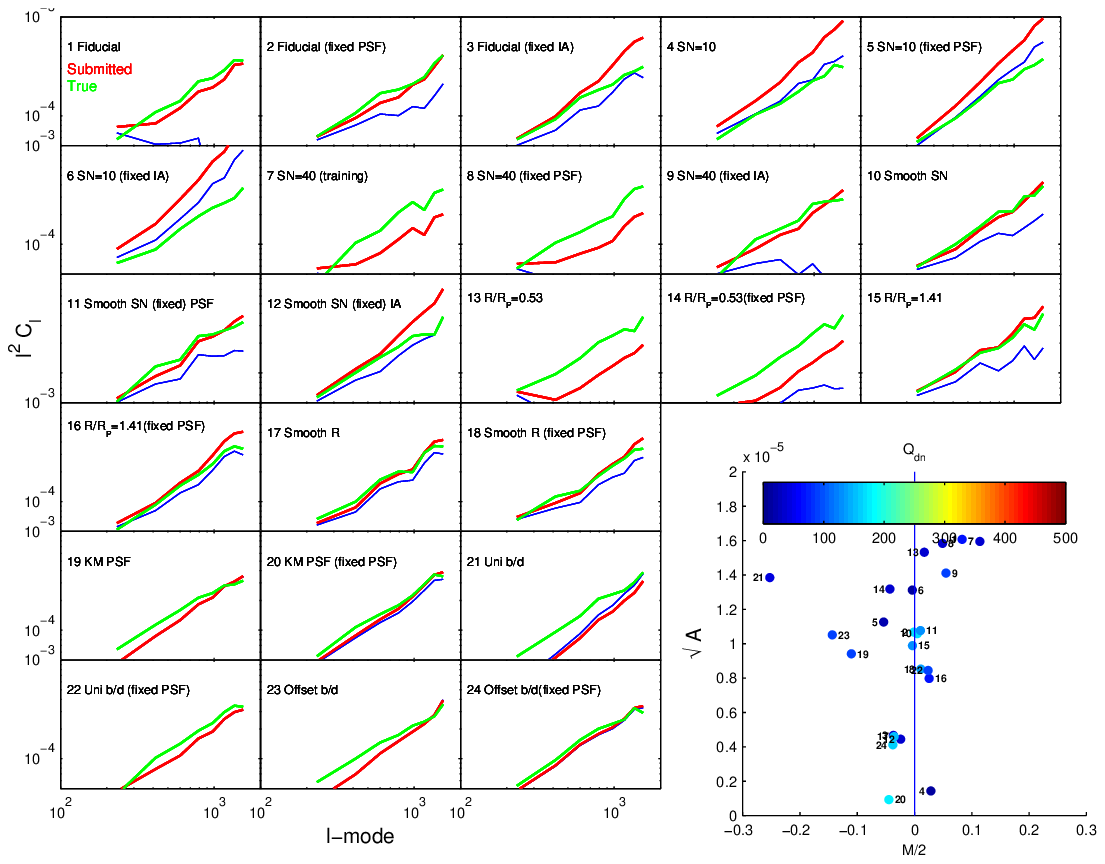}}
 \caption{The true shear power (green) for each set and the shear power for
   the `NN23' submission (red), we also show the `denoised'
   power spectrum (blue) for each set (where this is indistinguishable 
   from the raw submission a red line is only legible). 
   The y-axes 
   are $C_{\ell}\ell^2$ and the x-axis is $\ell$. In
   the bottom righthand corner we show the ${\mathcal M}/2$,
   $\sqrt{{\mathcal A}}$ and 
   the colour scale represents the logarithm of
   the quality factor. The small numbers next to each point label the
   set number.}
 \label{NN23}
\end{figure*}
\begin{figure*}
 {\includegraphics[width=0.75\columnwidth,angle=0,clip=]{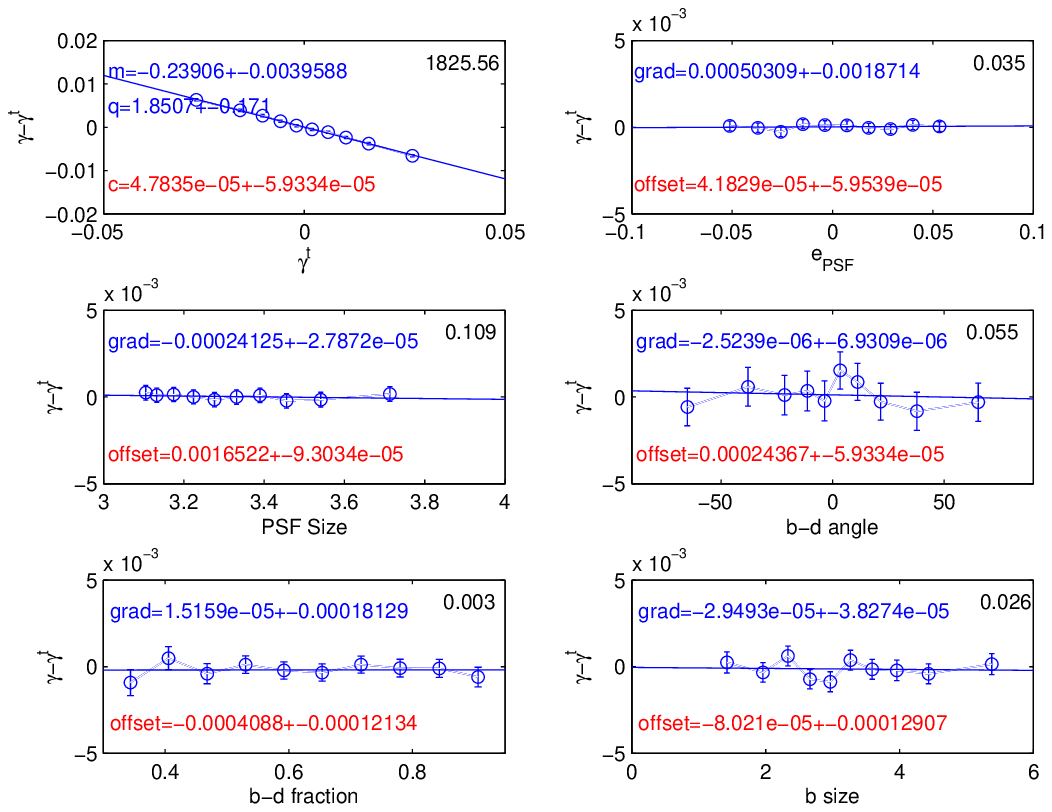}}
 \caption{The measured minus true shear for the `NN23' submission
   as a function of the true shear, PSF ellipticity, PSF FWHM, galaxy
   bulge-to-disk offset angle, galaxy bulge-to-disk fraction and
   galaxy bulge size. For each dependency we fit a linear function
   with a gradient and offset, for the top left hand panel this is
   the STEP $m$ and $c$ values, additionally for the shear
   dependency we include a quadratic term separately $q$. The top right hand corners show 
   $\Delta\chi^2=\chi^2({\rm gradient},{\rm offset})-\chi^2({\rm offset})$.}
  {\includegraphics[width=\columnwidth,angle=0,clip=]{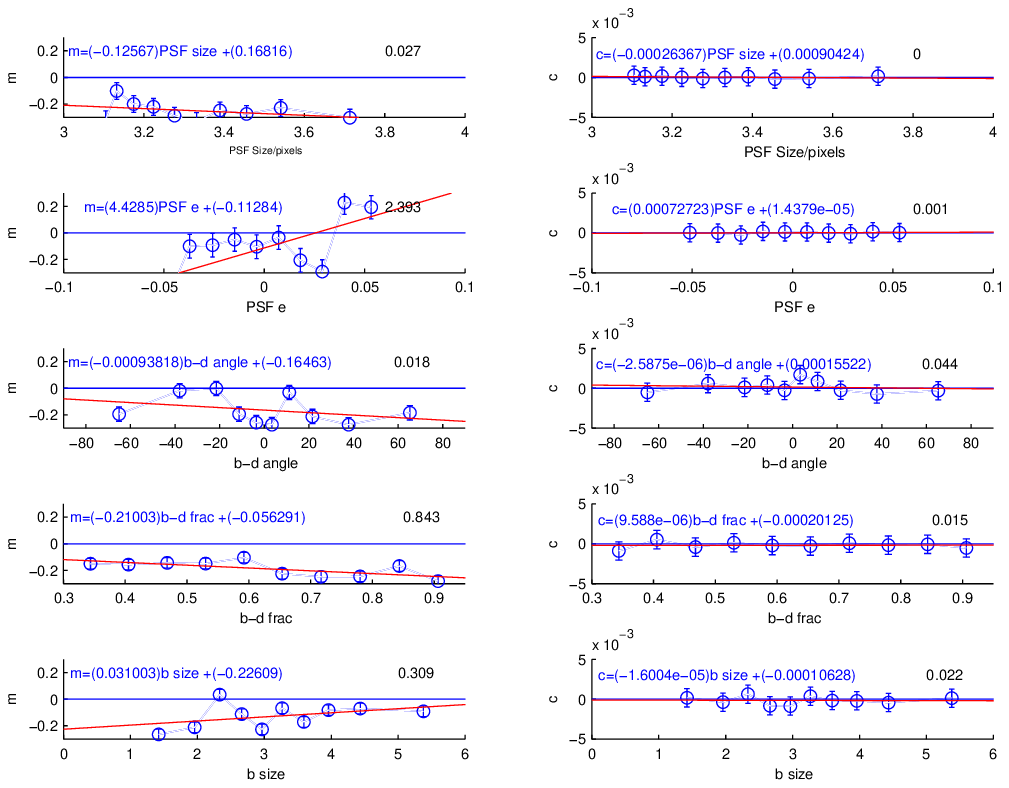}}
 \caption{The STEP $m$ and $c$ values for the `NN23' submission
   as a function of PSF FHWM and ellipticity, galaxy
   bulge-to-disk offset angle, galaxy bulge-to-disk fraction and
   galaxy bulge size. For each variable we plot the a linear relation
   to the behaviour of $m$ and $c$. We do not explicitly quote errors
   on all parameters for clarity, the average errors on $m$ and $c$ are
   $\simeq 0.005$ and $5\times 10^{-5}$ respectively. The top right hand corners show 
   $\Delta\chi^2=\chi^2({\rm gradient},{\rm offset})-\chi^2({\rm offset})$.}
\end{figure*}
\newpage

\section*{Appendix F: Simulations}
Inevitably, with a simulation the size of the GREAT10 Galaxy Challenge,
there were several points in which the data or interpretation of the
data/competition instructions were inadvertently misinterpreted by
participants. We list these here: 
\begin{enumerate} 
\item 
Approximately $1\%$ of the data were found to contain image glitches and were
replaced during the challenge as a patch to the data. 
\item 
The functional PSFs used a convention in $(x$,$y)$ coordinate and
ellipticity for which some methods had to make the following
transformations $e_2\rightarrow -e_2$, $x\rightarrow y$ and
$y\rightarrow x$, $r_{\rm PSF}\rightarrow r_{\rm
  PSF}/(1+e_1^2+e_2^2)$. This convention warning was listed in the header of every
functional PSF description during the challenge. 
\item 
An additional two sets contained ``pseudo-Airy'' PSFs using the functional
form of Kuijken (2006). However there was a misinterpretation by some
participants between the functional PSF
description  and the PSF {\rm FITS} images
generated using the photon-shooting method used in the GREAT10
code. This arose because in the photon-shooting method photons
at large $r$ are generated using a uniform distribution from 0 to 1
and then their values replaced by a reciprocal; but the PDF of such a
process yields a variation of $1/r^2$ not $1/r$, that when modulated
by the function gives $1/r^4$ (not $1/r^3$, given in equation 21,
Kuijken, 2006; the same equation that 
was provided to participants).
This was identified during the challenge and all participants were informed,
and the code used to produce the PSFs made public\footnote{\tt 
http://great.roe.ac.uk/data/code/sm/} on $7^{\rm th}$ February 2011
($7$ months before the challenge deadline), 
however we have not included the results from these sets in this paper
because several submissions were affected. 
\end{enumerate}
Each of these issues were addressed during the challenge, however the 
nature of the participation rate (see Section
\ref{astrocrowdsourcing}, all submissions were made in the final 3 weeks) meant that some methods did not have time to
create alternative submissions before the official challenged
closed. The challenge was extended by one week, into a post-challenge
submission period, but those methods submitted during this time could
not officially `win' the competition, in the event none of these
additional submissions improved on the winning score

When using the GREAT08/GREAT10 code we note a number of issues
that should be taken into account in its description in Bridle et al.,
(2010). The signal to noise used in Bridle et al.,
(2010) is approximately half the standard definition used in this article. Equation (A8) makes the area of the galaxy
invariant under the primary ellipticity transformation (but not under the cosmological shear
transformation), whereas equation (A9) does not make the PSF area invariant under the
ellipticity transformation.  Also the sense of the transformation in
these equations of $g$ for galaxies and $e$ for PSFs is different; the
PSF shear is in the opposite direction to the cosmic shear.
Finally, we also note that there were two typos in Appendix A of Bridle
et al., (2010). These were 1) in
equation (A5) the left top corner of the matrix should be $r/\sqrt(q)$ and 2)
equation (A8) should be the transpose of which it reads.

\end{document}